\documentclass[
  twocolumn,
  10pt,
  prd,
  nofootinbib,
  preprintnumbers,
  balancelastpage,
  superscriptaddress
]{revtex4-2}
\usepackage[utf8]{inputenc}
\usepackage{flushend}

\usepackage{amsmath,amssymb}
\usepackage[english]{babel}
\usepackage{mathrsfs}
\usepackage{siunitx}
\usepackage{mhchem}

\makeatletter
\g@addto@macro\bfseries{\boldmath}
\makeatother

\usepackage[dvips]{graphicx}
\usepackage[dvipsnames]{xcolor}
\definecolor{CiteBlue}{RGB}{45,52,151}
\usepackage[
    colorlinks=true,
    linkcolor=CiteBlue,
    urlcolor=CiteBlue,
    citecolor=CiteBlue
]{hyperref}

\usepackage[capitalise]{cleveref}
\crefname{section}{Sec.}{Secs.}
\Crefname{section}{Section}{Sections}

\usepackage{bm}
\newcommand{\bb}[1]{\bm{\mathrm{#1}}}

\newcommand{\hc}{\mathrm{h.c.}}
\newcommand{\du}{\mathrm{d}}
\newcommand{\BR}{\operatorname{BR}}
\newcommand{\LL}{\mathrm{LL}}
\newcommand{\RL}{\mathrm{RL}}
\newcommand{\LR}{\mathrm{LR}}
\newcommand{\RR}{\mathrm{RR}}
\newcommand{\SM}{\mathrm{SM}}
\newcommand{\lft}{\mathrm{L}}
\newcommand{\rgt}{\mathrm{R}}

\newcommand{\refcite}[1]{Ref.~\cite{#1}}
\newcommand{\refscite}[1]{Refs.~\cite{#1}}

\begin{document}
\preprint{MIT-CTP/5534}

\title{UV physics from IR features: new prospects from top flavor violation}

\author{Wolfgang Altmannshofer}
\email{waltmann@ucsc.edu}
\affiliation{Santa Cruz Institute for Particle Physics and Department of Physics, University of California, Santa Cruz, CA 95064, USA}

\author{Stefania Gori}
\email{sgori@ucsc.edu}
\affiliation{Santa Cruz Institute for Particle Physics and Department of Physics, University of California, Santa Cruz, CA 95064, USA}

\author{Benjamin V. Lehmann}
\email{benvlehmann@gmail.com}
\affiliation{Center for Theoretical Physics, Massachusetts Institute of Technology, Cambridge, MA 02139, USA}

\author{Jianhong Zuo}
\email{jzuo6@ucsc.edu}
\affiliation{Santa Cruz Institute for Particle Physics and Department of Physics, University of California, Santa Cruz, CA 95064, USA}

\begin{abstract}\ignorespaces
    New physics in the rare top decays $t \to q \ell^+\ell^-$ is currently very weakly constrained. We show that in a large class of Standard Model extensions, existing experimental constraints on new physics in flavor-conserving processes imply strong indirect bounds on new physics contributions to flavor-violating processes of the form $t \to q \ell^+\ell^-$. These indirect bounds arise from basic principles of quantum field theory together with a few generic conditions on the UV structure of the theory, and are roughly an order of magnitude stronger than the present experimental bounds on the same processes. These constraints provide a theoretically motivated target for experimental searches for $t \to q \ell^+\ell^-$: violation of these bounds would exclude a large class of new physics models, and would provide nontrivial insight into the UV behavior of the new physics.
\end{abstract}

\maketitle

\section{Introduction}\label{sec:intro}

Flavor-changing neutral current processes involving the top quark are strongly suppressed in the Standard Model (SM) by a loop factor, by small CKM matrix elements, and by an effective Glashow--Iliopoulos--Maiani cancellation. Moreover, these rare top decays compete with the unsuppressed 2-body decay mode $t \to b W$, resulting in extremely small branching ratios to final states like $q \gamma$, $q Z$, and $q h$, in the ballpark of \numrange{e-15}{e-12} for $q=c$ and \numrange{e-14}{e-17} for $q=u$ \cite{Aguilar-Saavedra:2004mfd, Altmannshofer:2019ogm}. The SM predictions for the branching ratios of rare three-body decays of the form $t \to q \ell^+ \ell^-$ are smaller yet by approximately two orders of magnitude. Such tiny branching ratios are far below foreseeable experimental sensitivities. The observation of rare top decays at current or planned colliders would therefore be an unambiguous sign of new physics.

Such new physics contributions can typically be accommodated in a single effective field theory (EFT) in which the SM Lagrangian comprises the renormalizable part. The null results in the searches for new physics particles at the LHC may suggest that there is a significant mass gap between the electroweak scale and the new physics scale, meaning that new physics effects can be parametrized by a set higher-dimensional operators in a model-independent fashion. This basis of operators, known as the Standard Model EFT (SMEFT) \cite{Grzadkowski:2010es}, has become a very popular framework to discuss physics beyond the SM (BSM). Existing constraints on rare top decays and other top flavor-violating processes have already been translated to constraints on the Wilson coefficients of SMEFT operators \cite{Drobnak:2008br, Durieux:2014xla, Forslund:2018qcp, Chala:2018agk, Shi:2019epw, Afik:2021jjh}. Similarly, measurements of top quark properties, and electroweak precision measurements in general, can be used to constrain the Wilson coefficients of flavor-conserving operators involving top quarks \cite{Degrande:2010kt, Zhang:2010dr, Buckley:2015lku, Schulze:2016qas, Maltoni:2016yxb, Degrande:2018fog, deBeurs:2018pvs, Farina:2018lqo, Hartland:2019bjb, Durieux:2019rbz, Brivio:2019ius, Stolarski:2020cvf, Ellis:2020unq, Afik:2021xmi, Miralles:2021dyw, Ethier:2021bye, Liu:2022vgo, Barman:2022vjd}.

Formulating the new physics contributions in a single EFT presents an opportunity to invoke generic constraints on the form of the SMEFT itself. It has long been known that the possible values of Wilson coefficients in an EFT are restricted not only by experimental results, but also by basic principles of quantum field theory, including unitarity, locality and analyticity \cite{Adams:2006sv}. In the bottom-up approach to constructing an EFT, one typically considers all operators that respect the relevant symmetries and treats the corresponding Wilson coefficients as free parameters that are naively allowed to take arbitrary values. However, not all such low energy scenarios have UV completions. Infrared consistency constraints imply, for example, that certain operators need to have positive Wilson coefficients.

Such constraints have been studied in several cases that are relevant for collider physics, as in the context of anomalous triple and quartic gauge boson couplings and vector boson scattering at the LHC and future colliders \cite{Distler:2006if, Vecchi:2007na, Zhang:2018shp, Bi:2019phv, Remmen:2019cyz, Ellis:2019zex, Ellis:2020ljj, Yamashita:2020gtt, Ellis:2022zdw}, fermion--antifermion scattering to dibosons \cite{Bellazzini:2018paj, Gu:2020ldn}, or dilepton production \cite{Fuks:2020ujk, Li:2022rag}. It has also been realized that positivity bounds can have implications for flavor-changing operators \cite{Remmen:2020vts,Bonnefoy:2020yee,Remmen:2020uze}. The results of \refcite{Remmen:2020uze} are of particular interest in context of rare top decays. Under a small set of assumptions, the Wilson coefficients of certain flavor-violating dimension-six four-fermion operators are bounded by combinations of the Wilson coefficients of flavor-conserving operators.

In many cases, flavor-violating operators lead to very prominent signatures that can be searched for with high precision. Thus, indirect constraints on flavor-violating Wilson coefficients obtained from measurements of flavor-conserving processes are typically subdominant to direct experimental constraints from searches for flavor-violating processes. Rare top decays are a notable exception to this rule. As we show in this paper, the theoretical relations obtained in \refcite{Remmen:2020uze}, combined with experimental constraints on new physics in quark flavor conserving processes, imply that the rare leptonic top decays $t \to q \ell^+ \ell^-$ are expected to be approximately an order of magnitude below the current experimental sensitivities in a broad class of new physics models. These expectations provide clear experimental targets for future searches for $t \to q \ell^+ \ell^-$ at the LHC and future colliders. Experimental observation of $t \to q \ell^+ \ell^-$ above the theoretical expectations would imply that the new physics violates some of the assumptions made in \refcite{Remmen:2020uze}. Such a measurement would exclude a broad class of new physics models and would imply that the UV structure of the new physics falls into one of a few narrow classes that evades the theoretical constraints.

This paper is organized as follows: In \cref{sec:EFT}, we introduce the EFT setup for the rare top decays and specify the four-fermion operators that we consider in our analysis. In \cref{sec:theory_bounds}, we summarize the results from \refcite{Remmen:2020uze}. We spell out the theoretical restrictions on the Wilson coefficients and the implications if they were found to be violated. In \cref{sec:exp_constraints}, we discuss the most relevant experimental constraints. We include existing constraints on flavor-violating rare top decays and single top production, as well as constraints on the flavor-conserving operators from dilepton production at the LHC, atomic parity violation, rare $B$ decays, $Z$ decays, and M{\o}ller scattering. \Cref{sec:numerics} contains our main numerical results. Based on the constraints from flavor conserving interactions, we derive the maximal rates for rare top decays and compare them with the existing and expected experimental sensitivities. We discuss the implications of these results and conclude in \cref{sec:conclusions}.

\section{EFT Setup}
\label{sec:EFT}

We consider the following set of dimension-six operators that are subject to sum rules derived in \refcite{Remmen:2020uze} and that are relevant for rare flavor-changing three-body decays of top quarks $t \to u \ell^+ \ell^-$ and $t \to c \ell^+ \ell^-$:
\begin{multline} \label{eq:Heff}
    \mathcal L_\text{eff} =
    \frac{1}{\Lambda^2} \Bigl(
        C_{ijkl}^{\LL 1} Q_{ijkl}^{\LL1} + C_{ijkl}^{\LL2} Q_{ijkl}^{\LL2}
        + C_{ijkl}^{\RR} Q_{ijkl}^{\RR}
        \\
        + C_{ijkl}^{\LR} Q_{ijkl}^{\LR} + C_{ijkl}^{\RL} Q_{ijkl}^{\RL}
    \Bigr).
\end{multline}
The definitions of the various operators are 
\begin{align}
    \label{eq:Q1}
    Q_{ijkl}^{\LL1} &=
        (\bar \ell_i \gamma_\mu \ell_j)(\bar q_k \gamma^\mu q_l),
    \\
    \label{eq:Q2}
    Q_{ijkl}^{\LL2} &=
        (\bar \ell_i \gamma_\mu \tau^I \ell_j)
        (\bar q_k \gamma^\mu \tau^I q_l),
    \\
    \label{eq:Q3}
    Q_{ijkl}^{\RR} &=
        (\bar e_i \gamma_\mu e_j)(\bar u_k \gamma^\mu u_l),
    \\
    \label{eq:Q4}
    Q_{ijkl}^{\LR} &=
        (\bar \ell_i \gamma_\mu \ell_j)(\bar u_k \gamma^\mu u_l),
    \\
    \label{eq:Q5}
    Q_{ijkl}^{\RL} &=
        (\bar e_i \gamma^\mu e_j)(\bar q_k \gamma_\mu q_l),
\end{align}
where $\tau^I$ denotes the Pauli matrices for $I=1,2,3$; $\ell_i$ are the three generations of left-handed lepton doublets; $e_i$ the right-handed charged lepton singlets; $q_i$ are the left-handed quark doublets; and $u_i$ the right-handed quark singlets of up-type.
In the effective Lagrangian of \cref{eq:Heff}, a sum over the flavor indices $i,j,k,l$ is understood. Note that our notation differs from the standard SMEFT notation of \refcite{Grzadkowski:2010es} that is often used in the literature. In the above Lagrangian, we do not include additional four-fermion operators with scalar or tensor currents or dipole operators. Such operators can also lead to rare top decays, but they are not subject to the sum rules derived in \refcite{Remmen:2020uze}.

The sum rules are based on analyticity of the $S$-matrix, partial wave unitarity, and assumptions about the high momentum behavior of forward scattering amplitudes. If the new physics in the UV is dominated by either scalars or vectors, the sum rules imply definite signs of certain combinations of Wilson coefficients. (See also \refscite{Azatov:2021ygj, Zhang:2021eeo, Remmen:2022orj}.) Defining $C_{\alpha \beta} = \alpha_i \alpha_j^* \beta_k^* \beta_l C_{ijkl}$, where $\alpha$ and $\beta$ are arbitrary directions in flavor space, the constraints can be summarized in the following way:
\begin{multline}
    \label{eq:constraint_s}
    s\left(
        C^{\LL1}_{\alpha \beta} \pm \tfrac{1}{4} C^{\LL2}_{\alpha \beta}
    \right) > 0
    ,\\
    s\,C^{\RR}_{\alpha \beta} > 0
    ,~~
    s\,C^{\LR}_{\alpha \beta} < 0
    ,~~
    \textnormal{and}~~
    s\,C^{\RL}_{\alpha \beta} < 0
    ,
\end{multline}
where $s=+1$ ($-1$) if the UV contributions to the Wilson coefficients are dominated by scalars (vectors). If both scalars and vectors contribute at a comparable level to the Wilson coefficients, cancellations may occur, so the sign of $s$ cannot be established from first principles.

Many operators that lead to rare top decays are strongly constrained by flavor-changing neutral current processes involving $B$ mesons \cite{Fox:2007in}. In particular, in the presence of the operators in \cref{eq:Heff} that contain left-handed quark fields, $SU(2)_\lft$ gauge invariance implies that sizable rates of $t \to c (u) \ell^+ \ell^-$ decays are directly related to large new physics effects in $b \to s (d) \ell^+ \ell^-$ or $b \to s (d) \nu\bar\nu$ decays. Barring tuned cancellations, the strong experimental constraints on the rare $B$ decays exclude rare top decays at an experimentally accessible level. As we will discuss in \cref{sec:rareB}, for operators that contain right-handed up-type quarks, constraints from rare $B$ decays only arise at the loop level, leaving much more room for sizable rare top branching ratios. In the following, we will thus focus on the operators $Q^{\LR}$ and $Q^{\RR}$. Furthermore, we will only consider operators with electrons or muons. Operators with taus could be included in an analogous way, but are much more challenging to probe experimentally. The interactions that we are interested in can be split into lepton-flavor--conserving and lepton-flavor--violating operators. The ones that conserve lepton flavor are explicitly given by
\begin{widetext}
\begin{multline}
    \label{eq:L_LFC}
    \Lambda^2 \mathcal L_\text{eff}^\text{LFC} =
    C_{eett}^{\LR} \bigl[ (\bar \nu_e \gamma_\alpha P_\lft \nu_e)
    + (\bar e \gamma_\alpha P_\lft e) \bigr] (\bar t \gamma^\alpha P_\rgt t)
    + C_{eett}^{\RR} (
        \bar e \gamma_\alpha P_\rgt e)(\bar t \gamma^\alpha P_\rgt t)
    \\
    + C_{eecc}^{\LR} \bigl[ (\bar \nu_e \gamma_\alpha P_\lft \nu_e)
    + (\bar e \gamma_\alpha P_\lft e) \bigr] (\bar c \gamma^\alpha P_\rgt c)
    + C_{eecc}^{\RR} (
        \bar e \gamma_\alpha P_\rgt e)(\bar c \gamma^\alpha P_\rgt c)
    \\
    + \Bigl( C_{eect}^{\LR} \bigl[ (\bar \nu_e \gamma_\alpha P_\lft \nu_e)
    + (\bar e \gamma_\alpha P_\lft e) \bigr] (\bar c \gamma^\alpha P_\rgt t)
    + C_{eect}^{\RR} (\bar e \gamma_\alpha P_\rgt e)
        (\bar c \gamma^\alpha P_\rgt t)
    + \hc \Bigr)
    + \dotsb ~,
\end{multline}
where the ellipsis corresponds to terms where electrons and electron neutrinos are replaced with muons and muon neutrinos, or charm quarks are replaced with up quarks, as appropriate. The lepton-flavor--violating terms are given by  
\begin{multline}
    \label{eq:L_LFV}
    \Lambda^2 \mathcal L_\text{eff}^\text{LFV} =
    \Bigl(
        C_{e\mu tt}^{\LR} \bigl[ (\bar \nu_e \gamma_\alpha P_\lft \nu_\mu)
        + (\bar e \gamma_\alpha P_\lft \mu) \bigr] (\bar t \gamma^\alpha P_\rgt t)
        + C_{e\mu tt}^{\RR} (\bar e \gamma_\alpha P_\rgt \mu)
            (\bar t \gamma^\alpha P_\rgt t)
        + \hc
    \Bigr)  \\ 
    + \Bigl(
        C_{e\mu cc}^{\LR} \bigl[ (\bar \nu_e \gamma_\alpha P_\lft \nu_\mu)
        + (\bar e \gamma_\alpha P_\lft \mu) \bigr] (\bar c \gamma^\alpha P_\rgt c)
        + C_{e\mu cc}^{\RR} (\bar e \gamma_\alpha P_\rgt \mu)
            (\bar c \gamma^\alpha P_\rgt c)
        + \hc
    \Bigr) \\
    + \Bigl(
        C_{e\mu ct}^{\LR} \bigl[ (\bar \nu_e \gamma_\alpha P_\lft \nu_\mu)
        + (\bar e \gamma_\alpha P_\lft \mu) \bigr] (\bar c \gamma^\alpha P_\rgt t)
        + C_{e\mu ct}^{\RR} (\bar e \gamma_\alpha P_\rgt \mu)
            (\bar c \gamma^\alpha P_\rgt t)
        \\
        + C_{\mu ect}^{\LR} \bigl[ (\bar \nu_\mu \gamma_\alpha P_\lft \nu_e)
        + (\bar \mu \gamma_\alpha P_\lft e) \bigr] (\bar c \gamma^\alpha P_\rgt t)
        + C_{\mu ect}^{\RR} (\bar \mu \gamma_\alpha P_\rgt e)
            (\bar c \gamma^\alpha P_\rgt t)
        + \hc
    \Bigr)
    + \dotsb ~,
\end{multline}
\end{widetext}
with the ellipsis indicating terms in which charm quarks are replaced with up quarks.

\section{Theoretical Constraints on the Wilson Coefficients}
\label{sec:theory_bounds}

If the relations in \cref{eq:constraint_s} hold, one can derive constraints on the flavor-specific Wilson coefficients in \cref{eq:L_LFC,eq:L_LFV}. In particular, the flavor conserving Wilson coefficients have definite signs: if the UV is dominated by scalars, then
\begin{equation}
    \label{eq:sign_scalar}
    C^{\LR}_{\ell\ell qq} < 0
    ~~\textnormal{and}~~
    C^{\RR}_{\ell\ell qq} > 0, 
\end{equation}
while if the UV is dominated by vectors, then
\begin{equation} \label{eq:sign_vector}
    C^{\LR}_{\ell\ell qq} > 0
    ~~\textnormal{and}~~
    C^{\RR}_{\ell\ell qq} < 0.
\end{equation}
In either case, one finds that the flavor-violating coefficients are bounded from above by the flavor-conserving ones as \cite{Remmen:2020vts,Remmen:2020uze}
\begin{equation}
    \label{eq:simple-constraint}
    \def\arraystretch{1.5}
    \begin{array}{ll}
        \bigl|C_{eect}^{\LR}\bigr|^2
            < C_{eecc}^{\LR} C_{eett}^{\LR},
        ~~~~~
        &
        \bigl|C_{\mu\mu ct}^{\LR}\bigr|^2
            < C_{\mu\mu cc}^{\LR} C_{\mu\mu tt}^{\LR},
        \\
        \bigl|C_{eect}^{\RR}\bigr|^2
            < C_{eecc}^{\RR} C_{eett}^{\RR},
        &
        \bigl|C_{\mu\mu ct}^{\RR}\bigr|^2
            < C_{\mu\mu cc}^{\RR} C_{\mu\mu tt}^{\RR}.
    \end{array}
\end{equation}
If both scalar and vector contributions are comparable, there can be cancellations and no bounds can be established. Analogous bounds hold for lepton-flavor--violating and quark-flavor--conserving interactions. The case of interactions that violate both quark and lepton flavor is nontrivial, and we leave a detailed exploration for future work.

The relations in \cref{eq:simple-constraint} have very interesting implications. One can combine existing experimental constraints from flavor conserving processes on nonstandard interactions of $\ell\ell uu$, $\ell\ell cc$, and $\ell\ell tt$ to derive upper bounds on the rare top decays $t \to u \ell^+\ell^-$ and $t \to c \ell^+\ell^-$. These upper bounds then serve as targets for experimental searches. If rare top decays are observed above the bounds, one or more of the hypotheses of the theoretical bounds must be violated. Possible options are:
\begin{enumerate}
    \item The rare top decays are induced by effective interactions other than the ones spelled out in \cref{eq:L_LFC}---for example, by operators with scalar or tensor currents.
    \item The UV physics that generates the effective interactions is not dominated by either scalars or vectors. For example, both scalars and vectors could contribute at a comparable level such that nontrivial cancellations take place.
    \item The UV physics gives forward-scattering amplitudes that grow with $s^n$ for $n\geq 1$, where $s$ is the squared center-of-mass energy \cite{Remmen:2020uze, Davighi:2021osh}. For example, this can happen if the forward scattering is mediated by vectors in the $t$-channel.
\end{enumerate}
Therefore, the observation of a violation of the theoretical bounds allows one to exclude entire classes of possible UV models that lead to rare top decays.

\section{Experimental Constraints on the Wilson Coefficients} \label{sec:exp_constraints}

In this section, we discuss the most relevant experimental constraints on the Wilson coefficients.
\begin{itemize}
    \item In \cref{sec:flavorviolating}, we start with a discussion of processes that are induced by the top-flavor--violating operators in \cref{eq:L_LFC,eq:L_LFV}. We consider rare flavor-violating top decays at the LHC and single top production at $e^+ e^-$ colliders.
    \item In \cref{sec:lightflavorconserving}, we discuss the most relevant processes that constrain flavor-conserving operators with light quarks. This includes, in particular, high-mass tails in dilepton production at the LHC, as well as parity violation in low-energy electron-proton scattering.
    \item Finally, in \cref{sec:heavyflavorconserving}, we discuss constraints on flavor-conserving operators containing top quarks, including rare $B$ decays, decays of the $Z$ boson, and parity violation in M{\o}ller scattering.
    In principle, LHC measurements of the production cross section for $t \bar t$ and single top quarks are sensitive to the flavor conserving four-fermion operators with top quarks. The constraints from the CMS analysis of \refcite{Sirunyan:2020tqm} are weaker than the ones we find from $Z$ decays and $B$ decays, so we do not consider them here.
\end{itemize}
In all cases, we derive the expressions for the new physics contributions to the relevant observables, discuss the existing experimental bounds, and comment on the expected future sensitivities. We list constraints on all Wilson coefficients in \cref{tab:all-constraints}.

We note that the operators in \cref{eq:L_LFC,eq:L_LFV} also contain neutrinos. They can therefore modify the production of neutrinos at the LHC as well as the neutrino--nucleus scattering cross section measured at neutrino experiments. Nonstandard neutrino production at the LHC can be constrained by mono-$X$ searches. The most stringent bounds come from monojet searches \refcite{ATLAS:2021kxv,CMS:2021far}. However, we do not consider them in detail in this paper, since they are weaker than the bounds from LHC dilepton spectrum measurements that we discuss in \cref{sec:dilepton}. New physics contributions to neutrino--nucleus scattering are constrained by e.g. the results from the \textsc{coherent} experiment \cite{COHERENT:2017ipa}. Bounds on the operators with up quarks and neutrinos have been derived in \refcite{Altmannshofer:2018xyo}, and these bounds are indeed subdominant to the other bounds we consider.

\subsection{Flavor-violating processes}
\label{sec:flavorviolating}

\subsubsection{Rare top decays} \label{sec:raretop}

The effective couplings $C_{\ell \ell tq}^{\LR}$ and $C_{\ell \ell tq}^{\RR}$ enable the rare flavor-changing three-body top decays $t \to q \ell^+ \ell^-$, where $q\in\{u,c\}$ and $\ell\in\{e, \mu\}$. Normalizing to the dominant $t \to W b$ decay mode of the top quark, we find the leading-order branching ratios as
\begin{multline}
    \BR(t \to q \ell^+ \ell^-) \simeq
    \frac{1}{96 \pi^2} \frac{m_t^2 v^2}{\Lambda^4} \left(
        \bigl| C^{\LR}_{\ell\ell t q} \bigr|^2 +
        \bigl| C^{\RR}_{\ell\ell t q} \bigr|^2
    \right) \\
    \times \left(
    1 - \frac{m_W^2}{m_t^2}\right)^{-2}
        \left(1 + \frac{2m_W^2}{m_t^2}
    \right)^{-1}
    .
\end{multline}
Lepton flavor changing decays are also possible, with the leading-order branching ratios
\begin{multline}
    \BR(t \to q \ell \ell^\prime) \simeq
    \frac{1}{96 \pi^2} \frac{m_t^2 v^2}{\Lambda^4} \left(
        \bigl| C^{\LR}_{\ell\ell^\prime t q} \bigr|^2 +
        \bigl| C^{\RR}_{\ell\ell^\prime t q} \bigr|^2 
    \right. \\
    \left.  +
        \bigl| C^{\LR}_{\ell^\prime\ell t q} \bigr|^2 +
        \bigl| C^{\RR}_{\ell^\prime\ell t q} \bigr|^2
    \right) \\
    \times \left(1 - \frac{m_W^2}{m_t^2}\right)^{-2}
        \left(1 + \frac{2m_W^2}{m_t^2}\right)^{-1}
    ,
\end{multline}
where $\BR(t \to q \ell \ell^\prime)\equiv\BR(t \to q \ell^+ \ell^{\prime\,-}) + \BR(t \to q \ell^{\prime \, +} \ell^-)$. Our expressions for the branching ratios are consistent with the results in \refcite{Chala:2018agk}.

At the LHC, $t\bar t$ pairs are produced copiously, and searches have been conducted for the flavor-violating two-body decays $t \to q H$, $t \to q Z$, and $t \to q \gamma$. However, no direct searches for the three-body decays $t\to q \ell \ell$ exist so far. Bounds have been obtained in \refcite{Chala:2018agk} by recasting an ATLAS search for the decays $t\to Z q$ \cite{ATLAS:2018zsq} based on $\sim$\SI{36}{\per\femto\barn} of Run II data, taking into account the full set of dimension-six operators that can lead to the $t\to q \ell \ell$ decays. (See \refcite{CMS:2017wcz} for a related CMS search.) As we restrict our analysis to the vector operators $Q_{\ell\ell tq}^{\RR}$ and $Q_{\ell\ell tq}^{\LR}$, we can translate the results of \refcite{Chala:2018agk} into bounds on the $t\to q \ell^+ \ell^-$ branching ratios. We find, at 95\% C.L.,
\begin{equation}
\label{eq:raretop_BRs}
    \def\arraystretch{1.5}
    \begin{array}{l}
        \BR(t\to c   e^+   e^-) < \num{2.1e-4}~, \\
        \BR(t\to u   e^+   e^-) < \num{1.8e-4}~, \\
        \BR(t\to c \mu^+ \mu^-) < \num{1.5e-4}~, \\
        \BR(t\to u \mu^+ \mu^-) < \num{1.2e-4}~.
    \end{array}
\end{equation}
This translates into the following (rather weak) bounds on the new physics scale:
\begin{align}
    \label{eq:raretopbound1}
    \frac{|C^{\LR}_{eect}|}{\Lambda^2},
        \frac{|C^{\RR}_{eect}|}{\Lambda^2}
        <& \frac{1}{(\SI{0.32}{\tera\electronvolt})^2} ~, \\
    \label{eq:raretopbound2}
    \frac{|C^{\LR}_{eeut}|}{\Lambda^2},
        \frac{|C^{\RR}_{eeut}|}{\Lambda^2} 
        <& \frac{1}{(\SI{0.33}{\tera\electronvolt})^2} ~, \\
    \label{eq:raretopbound3}
    \frac{|C^{\LR}_{\mu\mu ct}|}{\Lambda^2},
        \frac{|C^{\RR}_{\mu\mu ct}|}{\Lambda^2}
        <& \frac{1}{(\SI{0.35}{\tera\electronvolt})^2} ~, \\
    \label{eq:raretopbound4}
    \frac{|C^{\LR}_{\mu\mu ut}|}{\Lambda^2},
        \frac{|C^{\RR}_{\mu\mu ut}|}{\Lambda^2}
        <& \frac{1}{(\SI{0.36}{\tera\electronvolt})^2} ~.
\end{align}
A dedicated search for $t \to q \ell^+ \ell^-$ at the High-Luminosity LHC might improve the above bounds on the branching ratios by almost two orders of magnitude \cite{Chala:2018agk}, corresponding to an improvement by a factor of $\sim$3 in sensitivity to the new physics scale. A future \SI{100}{\tera\electronvolt} collider might improve the sensitivity to rare top decay rates by another order of magnitude \cite{FCC:2018byv}. Improved sensitivity might also be obtained from searches for single top production in association with same-flavor dileptons \cite{Afik:2021jjh}.

LHC searches for rare lepton-flavor--violating top decays $t \to q \mu e$ have been proposed in \refcite{Davidson:2015zza}. It was estimated that with a center-of-mass energy of $\sqrt{s} = \SI{13}{\tera\electronvolt}$ and an integrated luminosity of $\SI{100}{\per\femto\barn}$, the LHC could probe the branching ratios $\BR(t \to q \mu e)$ to the level of $\sim$\num{e-5}. The first results from searches for lepton flavor violation in top decays in $\sim$\SI{80}{\per\femto\barn} of ATLAS data \cite{Gottardo:2018ptv} and $\sim$\SI{137}{\per\femto\barn} of CMS data \cite{CMS:2022ztx} have not observed any significant excess above expected SM backgrounds, placing bounds on the branching ratios. The CMS analysis of \refcite{CMS:2022ztx} considers not only $t \bar t$ production with a subsequent lepton-flavor--violating decay by one of the top quarks, but also single top production in association with $\mu e$. The results from searches for the flavor-changing decays and production are combined assuming the presence of a specific set of contact interactions and recast in terms of a bound on the branching ratio. For vector interactions, as in the case we study here, the bounds at 95\% C.L. are \cite{CMS:2022ztx,ATLAS:2023fcw}
\begin{equation}
    \def\arraystretch{1.5}
    \begin{array}{l}
        \BR(t \to c \mu e) < \num{1.3e-6}~, \\ 
        \BR(t \to u \mu e) < \num{1.3e-7}~.
    \end{array}
\end{equation}
Considering one lepton-flavor--violating top quark operator at a time, and assuming that the CMS bound on the vector interactions does not depend on the chirality of the quarks and leptons involved in the transition, these bounds correspond to
{\everymath={\displaystyle}
\begin{equation}
    \begin{array}{l}
        \frac{|C^{\LR}_{\mu e t c}|}{\Lambda^2},
        \frac{|C^{\RR}_{\mu e t c}|}{\Lambda^2},
        \frac{|C^{\LR}_{e \mu t c}|}{\Lambda^2},
        \frac{|C^{\RR}_{e \mu t c}|}{\Lambda^2} <
            \frac{1}{(\SI{1.1}{\tera\electronvolt})^2} ~, \\[0.5cm]
        \frac{|C^{\LR}_{\mu e t u}|}{\Lambda^2},
        \frac{|C^{\RR}_{\mu e t u}|}{\Lambda^2},
        \frac{|C^{\LR}_{e \mu t u}|}{\Lambda^2},
        \frac{|C^{\RR}_{e \mu t u}|}{\Lambda^2} <
            \frac{1}{(\SI{2.0}{\tera\electronvolt})^2} ~.
    \end{array}
\end{equation}
}
Assuming that sensitivity to the branching ratio scales with the square root of integrated luminosity, we expect improvement by a factor of $\sim$5 at the high-luminosity LHC. This corresponds to an improvement by a factor of $\sim$1.5 in reach to the new physics scale.

\subsubsection{Single top production}
\label{sec:singletop}
Single top production in lepton collisions, $\ell^+ \ell^- \to t q$, with $q\in\{u,c\}$, has been identified as an important probe of top-flavor--changing contact interactions \cite{Bar-Shalom:1999dtk,Durieux:2014xla,Bause:2020auq,Sun:2023cuf}. The tree-level production cross section from $e^+e^-$ collisions,
\begin{equation}
    \sigma(e^+ e^- \to t q) = \sigma(e^+ e^- \to t \bar q)
        + \sigma(e^+ e^- \to \bar t q)
    ,
\end{equation}
can be written in the following way:
\begin{multline}
    \label{eq:single-top}
    \sigma(e^+ e^- \to t q) =
    \frac{1}{6\pi} \frac{s}{\Lambda^4} \left(1-\frac{m_t^2}{s}\right)^2
        \left( 1 + \frac{m_t^2}{2s} \right) \\
    \times \left( \bigl| C^{\LR}_{eetq}\bigr|^2
        + \bigl| C^{\RR}_{eetq}\bigr|^2 \right)
    ,
\end{multline}
where $s$ is the squared center-of-mass energy. In our setup, single top production is sensitive to the exact same combination of Wilson coefficients as the $t \to q e^+ e^-$ decays.

Single top production has been searched for at LEP \cite{Aleph:2001dzz,L3:2002hbp,DELPHI:2011ab}. We use the combined results from all LEP experiments, reported in \refcite{Aleph:2001dzz}. We find that the strongest constraint on the flavor-changing effective interactions can be obtained from the quoted bound on the cross section at a center-of-mass energy of $\sqrt{s} = \SI{189}{\giga\electronvolt}$, resulting in the bound
\begin{equation}
    \label{eq:LEPbound}
    \sigma(e^+ e^- \to t q) < \SI{0.11}{\pico\barn}
    .
\end{equation}
We do not attempt to combine this bound with others that are given at different center-of-mass energies. As no flavor-tagging of the light quarks has been performed in \refcite{Aleph:2001dzz}, we interpret the bound in \cref{eq:LEPbound} as a bound on the combined cross section $\sigma(e^+ e^- \to t q) = \sigma(e^+ e^- \to t u) + \sigma(e^+ e^- \to t c)$. Switching on one effective operator at a time, and taking into account only the bound on the cross section quoted above, we find the following constraints on the Wilson coefficients:
\begin{equation} \label{eq:singletop_constraints}
    \frac{|C^{\LR}_{eect}|}{\Lambda^2},
    \frac{|C^{\RR}_{eect}|}{\Lambda^2},
    \frac{|C^{\LR}_{eeut}|}{\Lambda^2},
    \frac{|C^{\RR}_{eeut}|}{\Lambda^2} <
    \frac{1}{(\SI{0.7}{\tera\electronvolt})^2} ~.
\end{equation}
The bound on the new physics scale is stronger by a factor of $\sim$2 compared to the bounds from the rare top decays $t \to q e^+e^-$ discussed in the previous subsection.

Future $e^+ e^-$ colliders \cite{CEPCPhysicsStudyGroup:2022uwl, Bernardi:2022hny, ILCInternationalDevelopmentTeam:2022izu} can improve the sensitivity considerably. In \refcite{Shi:2019epw}, it is estimated that at CEPC, with a center-of-mass energy of $\sqrt{s} = \SI{240}{\giga\electronvolt}$ and an integrated luminosity of $\SI{5.6}{\per\atto\barn}$, new physics could be probed at a scale of several TeV. Similar sensitivities can be expected at FCC-ee and at the ILC.

At an $e^+e^-$ collider, the muonic operators also affect single top production, albeit only at the loop level. We have calculated the corresponding one-loop contribution to the cross section and found very weak constraints on the new physics scale, on the order of a $\textnormal{few}\times\SI{10}{\giga\electronvolt}$. This lies outside the regime of validity of the EFT, so no actual constraint can be obtained. We expect strong sensitivity to the muonic operators from single top production at a future high-energy muon collider \cite{AlAli:2021let, Sun:2023cuf}.

Finally, single top production in $ep$ collisions can also be used to constrain top-flavor--changing contact interactions. The process $ep \to et X$ has been searched for at HERA, and bounds on the corresponding cross sections have been obtained in \refscite{H1:2009yuy,ZEUS:2011mya}. These results can be used to constrain the Wilson coefficients $C^{\LR}_{eeut}$ and $C^{\RR}_{eeut}$. However, the constraints from $e^+e^- \to t q$ at LEP turn out to be stronger~\cite{Durieux:2014xla, Shi:2019epw}. We therefore do not consider single top production at $ep$ colliders for the remainder of this work.

\subsection{Light quark flavor-conserving processes}\label{sec:lightflavorconserving}

\subsubsection{Dilepton spectra at the LHC} \label{sec:dilepton}
Measurements of the dilepton invariant mass distribution at the LHC are well established as probes of new physics in the form of quark-lepton contact interactions \cite{Greljo:2017vvb, Aad:2020otl, CMS:2021ctt, Allwicher:2022gkm, Allwicher:2022mcg, Greljo:2022jac}. In the SM, for sufficiently large partonic center-of-mass energy $\sqrt{\hat s}$, the Drell-Yan parton-level cross sections $\hat \sigma_q = \sigma(q\bar q \to \ell^+ \ell^-)$ fall with $\hat s$, but they grow with $\hat s$ in the presence of the contact interactions. We find
\begin{equation}
    \hat \sigma_q(\hat s) = \frac{\hat s}{144\pi}
    \sum_{\mathrm X,\mathrm Y \in \{\mathrm L,\mathrm R\}}
        |A_q^{\mathrm{XY}}|^2
    ~,
\end{equation}
where in the limit $\hat s \gg m_Z^2$, the amplitudes can be approximated at leading order by
\begin{align}
    \label{eq:amplitudes}
    -i A_q^{\LL} &=
        \frac{1}{\hat s} e^2 Q_\ell Q_q +
        \frac{1}{\hat s}
            \frac{e^2}{s_W^2 c_W^2} Q_{\ell_L}^\text{w} Q_{q_L}^\text{w}
    ~, \\
   -i A_q^{\LR} &=
        \frac{1}{\hat s} e^2 Q_\ell Q_q +
        \frac{1}{\hat s}
            \frac{e^2}{s_W^2 c_W^2} Q_{\ell_L}^\text{w} Q_{q_R}^\text{w}
        + \frac{1}{\Lambda^2} C_{\ell\ell qq}^{\LR}
    ~, \\
   -i A_q^{\RL} &=
        \frac{1}{\hat s} e^2 Q_\ell Q_q +
        \frac{1}{\hat s}
            \frac{e^2}{s_W^2 c_W^2} Q_{\ell_R}^\text{w} Q_{q_L}^\text{w}
    ~, \\
   -i A_q^{\RR} &=
        \frac{1}{\hat s} e^2 Q_\ell Q_q +
        \frac{1}{\hat s}
            \frac{e^2}{s_W^2 c_W^2} Q_{\ell_R}^\text{w} Q_{q_R}^\text{w}
        + \frac{1}{\Lambda^2} C_{\ell\ell qq}^{\RR}
    ~.
\end{align}
Here $Q_f$ denotes the electric charges of the leptons and quarks, and $Q_f^\text{w} = T_3^f - s_W^2 Q_f$ denotes their weak charges. In the absence of the new-physics contact interactions, $-iA_q^{\LR}$ and $-iA_q^{\RR}$ are negative for up-type quarks. If the new physics effect is mediated by scalars, the theoretical bounds on the Wilson coefficients $C_{\ell\ell qq}^{\LR}$ and $C_{\ell\ell qq}^{\RR}$ from \cref{eq:sign_scalar} predict constructive interference in $A_{u,c}^{\LR}$ and destructive interference in $A_{u,c}^{\RR}$. Conversely, if the new physics effect is mediated by vectors, \cref{eq:sign_vector} implies destructive interference in $A_{u,c}^{\LR}$ and constructive interference in $A_{u,c}^{\RR}$.

Due to the growth with $\hat s$, the high invariant mass tails in $pp \to \ell^+ \ell^-$ are very sensitive to new-physics contact interactions. Consider a proton--proton scattering process with center-of-mass energy $\sqrt s$. The integrated cross section at large dilepton invariant mass $m_{\ell\ell}$ is given by
\begin{equation}
    \sigma(pp\to\ell^+\ell^-) =
    \sum_q \int_{\tau_\text{min}}^{\tau_\text{max}}\du\tau~
    2 \mathscr L_{q \bar q}(\tau) \hat \sigma_q(s\tau)
    ,
\end{equation}
where $\tau = m_{\ell\ell}^2/s$, with the parton luminosities given by 
\begin{equation}
    \mathscr L_{q\bar q}(\tau) =
    \int_\tau^1 \frac{\du x}{x} f_q(x) f_{\bar q}(\tau/x)
    .
\end{equation}
For our numerical analysis we include $q \in\{u,d,s,c\}$, and use the parton distribution functions $f_q$ and $f_{\bar q}$ from \refcite{Harland-Lang:2014zoa}, setting the factorization scale to the dynamical value $\mu_{\mathrm F} = m_{\ell\ell}$.

We use the ATLAS analysis of \refcite{Aad:2020otl} to constrain the Wilson coefficients $C_{\ell\ell qq}^{\LR}$ and $C_{\ell\ell qq}^{\RR}$, for $\ell\in\{e, \mu\}$ and $q\in\{u,c\}$. Similar constraints could be obtained from an analogous CMS analysis \cite{CMS:2021ctt}. At high $m_{\ell\ell}$, the main dilepton background to the new-physics signal is from SM Drell-Yan production, with only percent-level contributions from other background sources like $t\bar t$, single top, dibosons, or misidentified jets \cite{Aad:2020otl,CMS:2021ctt}. Using \verb|MadGraph5_aMC@NLO| \cite{Alwall:2014hca}, we explicitly checked that the product of acceptance and efficiency is very similar for SM Drell--Yan events and signal events in the presence of the contact interactions. We thus approximate the number of signal events $N_\text{sig}^\ell$ in a given signal region as
\begin{equation}
    N_\text{sig}^\ell = N_\text{bg}^\ell \left( \frac{\sigma(pp\to\ell^+\ell^-)}{\sigma(pp\to\ell^+\ell^-)_\SM}  - 1 \right)~,
\end{equation}
where $N_\text{bg}^\ell$ is the number of expected background events.

The ATLAS analysis uses signal regions that are optimized for either destructive or constructive interference. In our scenarios both destructive and constructive interference can occur, and in each case we use the signal regions that result in the stronger constraint. The signal regions optimized for constructive (destructive) interference are defined by $\tau_\text{min} = \SI{2.2}{\tera\electronvolt}$ (\SI{2.77}{\tera\electronvolt}) for electrons and $\tau_\text{min} = \SI{2.07}{\tera\electronvolt}$ (\SI{2.57}{\tera\electronvolt}) for muons. In all cases, $\tau_\text{max} = \SI{6.0}{\tera\electronvolt}$. The corresponding numbers of expected SM background events are $N_\text{bg}^e = 12.4$ (3.1) and $N_\text{bg}^\mu = 9.6$ (1.4) for electrons and for muons, respectively.

To obtain the bounds on the Wilson coefficients, we impose the model-independent upper limits at 95\% C.L. on the number of signal events: $N_\text{sig}^e < 16.0$ (4.4) for electrons, and $N_\text{sig}^\mu < 5.8$ (3.8) for muons \cite{Aad:2020otl}, in the signal regions optimized for constructive (destructive) interference. We have checked that this procedure reproduces the given bounds on the contact interactions studied in \refcite{Aad:2020otl}, finding agreement within 15\%.

\begin{table}
    \centering
    \begin{equation*}
        \arraycolsep=0.23cm
        \def\arraystretch{1.5}
        \begin{array}{lcc|lcc}
                \textnormal{Coeff.} &
                \Lambda_- &
                \Lambda_+ &
                \textnormal{Coeff.} &
                \Lambda_- &
                \Lambda_+
            \\\hline\hline
            C_{eeuu}^{\LR} & 8.0 & 6.5 & C_{\mu\mu uu}^{\LR} & 8.0 & 5.9 \\
            C_{eeuu}^{\RR} & 8.8 & 5.9 & C_{\mu\mu uu}^{\RR} & 9.2 & 5.1 \\
            C_{eecc}^{\LR} & 2.0 & 2.0 & C_{\mu\mu cc}^{\LR} & 2.1 & 2.0 \\
            C_{eecc}^{\RR} & 2.0 & 2.0 & C_{\mu\mu cc}^{\RR} & 2.1 & 2.0
        \end{array}
    \end{equation*}
    \caption{Bounds on the EFT scale in TeV units extracted from the LHC dilepton spectrum, measured in \refcite{Aad:2020otl}. Each row corresponds to a bound of the form $-1/\Lambda_{-}^2 < C/\Lambda^2 < 1/\Lambda_+^2$, with all other coefficients set to zero.}
    \label{tab:dilepton-bounds}
\end{table}

Switching on one operator at a time, we find the constraints in \cref{tab:dilepton-bounds}.
As expected, the strongest constraints are obtained for operators containing up quarks. We can estimate that at the high-luminosity LHC, which will increase the total luminosity by a factor of $\sim$20, the constraints on $\sigma$ will be improved by a factor of $\sim$4, and those on $\Lambda$ can therefore be improved by a factor of $\sim$2. This scaling represents only a rough estimate of the reach of the HL-LHC. A dedicated study along the lines of \refcite{CMS:2022gho} would be needed to assess the exact bound.

The electron-quark contact interactions can also be constrained from dijet production at LEP. The constraints on electron-charm interactions reported in \refcite{ALEPH:2013dgf} are comparable to the ones given above. In the case of electron-up interactions, the LHC constraints are considerably stronger.

Note that high-mass dilepton tails at the LHC can also be used to constrain lepton-flavor--violating and quark-flavor--conserving four-fermion contact interactions \cite{Angelescu:2020uug}. However, the theoretical relations among the Wilson coefficients that we consider in this work (\cref{eq:simple-constraint}) do not involve such contact interactions, so we will not consider these constraints.

\subsubsection{Atomic parity violation and electron nucleus scattering}
\label{sec:APV}

Experiments measuring atomic parity violation are sensitive probes of new physics at the TeV scale \cite{Safronova:2017xyt}. The most constraining result is currently a precision measurement of a parity-violating electric dipole transition in Cesium \cite{Wood:1997zq, Sahoo:2021thl}. Atomic parity violation is induced by the SM weak interaction, but can receive contributions from parity-violating new physics as well.  The Cesium measurement can be interpreted in terms of the nuclear weak charge, $Q_W$, which receives new physics contributions from the flavor-conserving operators in \cref{eq:L_LFC} that contain up quarks and electrons. The correction to $Q_W$ is given by
\begin{equation}
    \frac{\delta Q_W}{Q_W^\SM} =
    \frac{v^2}{\Lambda^2} \bigl( C_{eeuu}^{\RR} - C_{eeuu}^{\LR}  \bigr)
    \frac{2 \mathcal Z + \mathcal N}{\mathcal N - \mathcal Z (1-4 s_W^2)} ~,
\end{equation}
where $v = \SI{246}{\giga\electronvolt}$ is the vacuum expectation value (vev) of the Higgs, and $\mathcal Z$ and $\mathcal N$ are the numbers of protons and neutrons in the nucleus, respectively. Combining the Cesium measurement with the SM prediction for $Q_W$, one finds $Q_W^\text{exp} - Q_W^\SM = -0.48 \pm 0.35$ \cite{Sahoo:2021thl}, which at 95\% C.L. gives
\begin{equation}
  -0.3\% < \frac{\delta Q_W}{Q_W^\SM} < 1.6\%~.
\end{equation}
This translates into the following constraints on the Wilson coefficients:
\begin{equation}
    -\frac{1}{(\SI{7.2}{\tera\electronvolt})^2} <
    \frac{C_{eeuu}^{\RR}}{\Lambda^2} - \frac{C_{eeuu}^{\LR}}{\Lambda^2} <
    \frac{1}{(\SI{3.1}{\tera\electronvolt})^2} ~.
\end{equation}
The constraints from dilepton production at the LHC discussed in the previous section are more stringent (see \cref{tab:dilepton-bounds}).

A recent proposal has the potential to improve on this result. The measurements of elastic electron--proton or electron--\ce{^{12}C} scattering at the proposed P2 experiment \cite{Becker:2018ggl} at the Mainz Energy-recovering Superconducting Accelerator (MESA) facility can give a more stringent bound on these operators. The basic idea is to measure the parity-violating asymmetry,
\begin{equation}
    A_{\LR} = \frac{\sigma_L - \sigma_R}{\sigma_L + \sigma_R}
    ~,
 \end{equation}
where $\sigma_L$ ($\sigma_R$) is the cross section for the scattering of electrons with left (right) helicity. The asymmetry is directly related to the nuclear weak charge as
\begin{equation}
    \frac{\delta A_{\LR}}{A_{\LR}^\SM} = \frac{\delta Q_W}{Q_W^\SM}
    ~.
\end{equation}
The relative uncertainties $\delta A_{\LR}/A_{\LR}^\SM$ are expected to be
1.4\% and 0.3\% for electron--proton and electron--\ce{^{12}C} scattering, respectively~\cite{Becker:2018ggl}. This translates into the 95\% C.L. bounds
{\everymath={\displaystyle}
\begin{equation}
    \begin{array}{lcr}
        \left| \frac{C_{eeuu}^{\RR}}{\Lambda^2} - \frac{ C_{eeuu}^{\LR}}{\Lambda^2} \right| <
        \frac{1}{(\SI{7.5}{\tera\electronvolt})^2} &&
        \textnormal{($e^-$--$p^+$)}
        \\[0.5cm]
        \left| \frac{C_{eeuu}^{\RR}}{\Lambda^2} - \frac{ C_{eeuu}^{\LR}}{\Lambda^2} \right| <
        \frac{1}{(\SI{5.7}{\tera\electronvolt})^2} &&
        \textnormal{($e^-$--\ce{^{12}C})}
        .
    \end{array}
\end{equation}
}
These results are similar to those in \refcite{Dev:2021otb,Bischer:2021jqn}. For the numerical evaluation, we use $s_W^2 = 0.231$, corresponding to the $\overline{\text{MS}}$ value at the $Z$-pole. For the scattering with protons, one can expect sizable higher-order corrections due to the accidental suppression of the SM asymmetry $A_{\LR}^\SM \propto 1 - 4 s_W^2$. The resulting projected bounds are comparable to the current LHC bounds shown in \cref{tab:dilepton-bounds}.

\subsection{Heavy quark flavor conserving processes}\label{sec:heavyflavorconserving}

\subsubsection{Rare \texorpdfstring{$B$}{B} decays} \label{sec:rareB}

Rare decays of $b$ hadrons based on the $b \to s ee$ and $b \to s \mu\mu$ transitions are sensitive probes of new physics \cite{Bause:2021cna,Altmannshofer:2022hfs} and can be used to constrain both the flavor-conserving and the flavor-changing top quark operators. New physics contributions to rare $B$ decays are usually phrased in terms of an effective Hamiltonian
\begin{equation}
  \mathcal H_\text{eff}
  =   \mathcal H_\text{eff}^\SM
    - \frac{4G_F}{\sqrt{2}} V_{tb}V_{ts}^* \frac{e^2}{16\pi^2}
        \sum_{i} C_i O_i + \hc
  ~.
\end{equation}
In our case, the relevant operators are 
\begin{align}
O_9^{bs\ell\ell} &=
(\bar{s} \gamma_{\alpha} P_{L} b)(\bar{\ell} \gamma^\alpha \ell)\,, \\
O_{10}^{bs\ell\ell} &=
(\bar{s} \gamma_{\alpha} P_{L} b)(\bar{\ell} \gamma^\alpha \gamma_5 \ell)\,,
\end{align}
with $\ell \in \{e, \mu\}$. 

\begin{figure}[tb]
    \centering
    \includegraphics{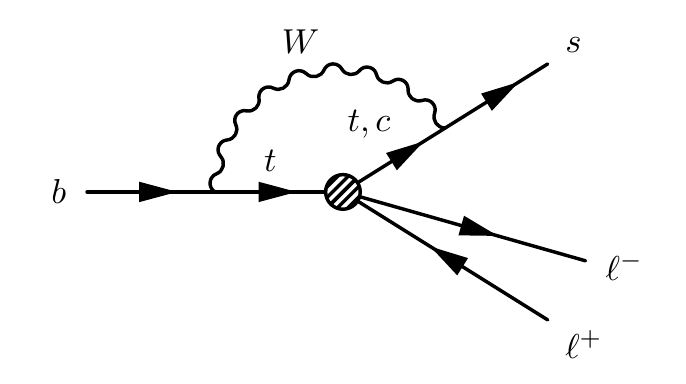}
    \caption{Example of a Feynman diagram contributing to the rare decay $b \to s \ell^{+} \ell^{-}$.}
    \label{fig:bsee}
\end{figure}

Starting from the top quark operators $Q_{\ell\ell tt}^{\LR}$, $Q_{\ell\ell ct}^{\LR}$, $Q_{\ell\ell tt}^{\RR}$, and $Q_{\ell\ell ct}^{\RR}$, one-loop corrections from the weak interactions induce the $O_{9}^{bs\ell\ell}$ and $O_{10}^{bs\ell\ell}$ operators through diagrams such as the one shown in \cref{fig:bsee}. We find for the corresponding Wilson coefficients
\begin{multline}
    \label{eq:C9}
    C_9^{bs\ell\ell} = \left(
        C^{\RR}_{\ell\ell tt} + C^{\LR}_{\ell\ell tt}
        + \frac{V_{cs}^*}{V_{ts}^*} \frac{m_c}{m_t} \bigl(C^{\RR}_{\ell\ell ct}
        + C^{\LR}_{\ell\ell ct} \bigr)
    \right) \\
   \times  \frac{1}{4 s_W^2} \frac{m_t^2}{4m_W^2} \frac{v^2}{\Lambda^2}
    \log\left(\frac{\Lambda^2}{m_W^2}\right)
    ~,
\end{multline}
\begin{multline}
    \label{eq:C10}
    C_{10}^{bs\ell\ell} = \left(
        C^{\RR}_{\ell\ell tt} - C^{\LR}_{\ell\ell tt}
        + \frac{V_{cs}^*}{V_{ts}^*}\frac{m_c}{m_t} \bigl(C^{\RR}_{\ell\ell ct}
        - C^{\LR}_{\ell\ell ct} \bigr)
    \right) \\
    \times \frac{1}{4 s_W^2} \frac{m_t^2}{4m_W^2} \frac{v^2}{\Lambda^2}
    \log\left(\frac{\Lambda^2}{m_W^2}\right)
    ~.
\end{multline}
The pieces proportional to $C^{\LR}_{\ell\ell tt}$, $C^{\RR}_{\ell\ell tt}$ have also been considered in~\refcite{Aebischer:2015fzz, Celis:2017doq, Camargo-Molina:2018cwu} (see also \refcite{Bissmann:2019gfc, Aoude:2020dwv, Bissmann:2020mfi,  Bruggisser:2021duo} for related studies of $B$ decay constraints on top operators). 
Note that we only take into account the logarithmically enhanced terms. Additional finite contributions from integrating out the top and $W$ are renormalization-scheme dependent and of the same order as unknown matching contributions at the scale $\Lambda$. Therefore, they are consistently neglected in our study.

The Wilson coefficients $C_9^{bs \ell\ell}$ and $C_{10}^{bs \ell\ell}$ are probed at the $\mathcal O(1)$ level by existing data on rare $B$ decays. (See e.g. the global fit in~\refcite{Altmannshofer:2021qrr}.)  For several years, global fits found a strong preference for nonstandard values of some of the Wilson coefficients due to a series of ``$B$-anomalies'', notably the hints for lepton flavor universality violation in rare $B$ decays through the observables $R_K$ and $R_{K^*}$ \cite{LHCb:2017avl,LHCb:2019efc,LHCb:2021lvy,LHCb:2021trn}, anomalous angular observables in the $B \to K^* \mu \mu$ decay \cite{LHCb:2020lmf}, and several anomalously low branching ratios of rare $B$ decays \cite{LHCb:2014cxe, LHCb:2016ykl, LHCb:2021zwz}. Top-philic new physics has been discussed e.g. in \refscite{Celis:2017doq, Becirevic:2017jtw, Kamenik:2017tnu, Fox:2018ldq, Camargo-Molina:2018cwu, Coy:2019rfr, Li:2021cty} as a possible explanation of the $B$ anomalies. 

However, the most recent updates on $R_K$ and $R_{K^*}$ \cite{LHCb:2022qnv, LHCb:2022zom} and the $B_s \to \mu^+ \mu^-$ branching ratio~\cite{CMS:2022dbz} are in good agreement with SM predictions. We use these results to constrain new physics in the Wilson coefficients $C_9^{bs \ell\ell}$ and $C_{10}^{bs \ell\ell}$, see also \cite{Ciuchini:2022wbq, Greljo:2022jac}.

We perform a fit including the following set of experimental results:
(i) the world average of the $B_s \to \mu^+ \mu^-$ branching ratio from HFLAV~\cite{HFLAV:2022pwe}; (ii) the new LHCb results on $R_K$ and $R_{K^*}$~\cite{LHCb:2022qnv, LHCb:2022zom}; (iii) the LHCb results on $R_{K_S}$ and $R_{K^{*+}}$~\cite{LHCb:2021lvy}; (iv) the Belle results on LFU tests in $b \to s \ell \ell$~\cite{Belle:2016fev, BELLE:2019xld, Belle:2019oag}.
The corresponding theory predictions are obtained with \texttt{flavio}~\cite[][version 2.3.3]{Straub:2018kue}. Instead of  \texttt{flavio}'s default values for the CKM matrix elements we use the current PDG values, $|V_{cb}| = (40.8 \pm 1.4)\times 10^{-3}$ and $|V_{ub}| = (3.82 \pm 0.20)\times 10^{-3}$ \cite{ParticleDataGroup:2022pth}. Considering either muon-specific or electron-specific Wilson coefficients, the fit gives approximately
\begin{equation}
    C_9^{bs\mu\mu} = -0.26\pm 0.33 ~,~~~ C_{10}^{bs\mu\mu} = -0.06 \pm 0.22 ~, 
\end{equation}
with an error correlation of $\rho = 87\%$, and
\begin{equation}
    C_9^{bsee} = 0.61 \pm 0.67 ~,~~~ C_{10}^{bsee} = 0.43 \pm 0.73 ~,
\end{equation}
with an error correlation of $\rho = 97.7\%$.
All new physics Wilson coefficients are compatible with zero at the $1\sigma$ level.

The above results can be translated into constraints on the
flavor-conserving top quark coefficients $C_{eett}^{\RR}$, $C_{eett}^{\LR}$,  $C_{\mu\mu tt}^{\RR}$, and $C_{\mu\mu tt}^{\LR}$. We find
{\everymath={\displaystyle}
\begin{equation}
    \begin{array}{rcl}
    \label{eq:b-bounds}
    \frac{-1}{(\SI{3.7}{\tera\electronvolt})^2} <&
        \frac{C_{eett}^{\LR}}{\Lambda^2}
        &< \frac{1}{(\SI{1.2}{\tera\electronvolt})^2}
    ~,\\
    \frac{-1}{(\SI{0.80}{\tera\electronvolt})^2} <&
        \frac{C_{eett}^{\RR}}{\Lambda^2}
        &< \frac{1}{(\SI{0.29}{\tera\electronvolt})^2}
    ~,\\
    \frac{-1}{(\SI{1.4}{\tera\electronvolt})^2} <&
        \frac{C_{\mu\mu tt}^{\LR}}{\Lambda^2}
        &< \frac{1}{(\SI{2.6}{\tera\electronvolt})^2}
    ~,\\
    \frac{-1}{(\SI{0.95}{\tera\electronvolt})^2} <&
        \frac{C_{\mu\mu tt}^{\RR}}{\Lambda^2}
        &< \frac{1}{(\SI{0.85}{\tera\electronvolt})^2}
    ~.
    \end{array}
\end{equation}
}

In contrast to the constraints discussed in previous sections, the new physics contributions to the rare $B$ decays are loop-induced and therefore contain a logarithmic dependence on the new physics scale $\Lambda$. The above best-fit ranges for the ratio of Wilson coefficients and new physics scale are derived setting the new physics Wilson coefficients $|C_{eett}^{\LR}| = 1$, $|C_{eett}^{\RR}| =1$, $|C_{\mu\mu tt}^{\LR}| = 1$, or $|C_{\mu\mu tt}^{\RR}| = 1$, one at a time. As the bounds are based on a leading logarithmic analysis, they hold to very good approximation also for other values of the Wilson coefficients, as long as the Wilson coefficients are not parametrically different from one.

Turning to the flavor changing coefficients $C_{eect}^{\LR}$, $C_{eect}^{\RR}$,  $C_{\mu\mu ct}^{\LR}$, and $C_{\mu\mu ct}^{\RR}$, we note that they can in principle be complex, and the respective bounds will depend on their complex phase. In most cases, the constraints on the imaginary parts of the Wilson coefficients are a factor of few weaker than the constraints on the real parts~\cite{Altmannshofer:2021qrr}. For simplicity, we will assume that the flavor-changing Wilson coefficients are real, and we will also neglect the tiny imaginary part of $V_{ts}$ that enters \cref{eq:C9,eq:C10}. 
Switching on only one of the flavor-changing coefficients $C_{eect}^{\LR}$, $C_{eect}^{\RR}$,  $C_{\mu\mu ct}^{\LR}$, $C_{\mu\mu ct}^{\RR}$ at a time, the corresponding best-fit ranges for the new physics scale are somewhat smaller than the ones corresponding to the flavor conserving coefficients, and they can be obtained from \cref{eq:b-bounds} by rescaling with a factor $\sqrt{|V_{ts}/V_{cs}|}\sqrt{m_t/m_c} \simeq 3$.

Rare $B$ decays could also be used to constrain the lepton-flavor--violating four-fermion operators \cite{Davidson:2018rqt}, but we will not consider such processes in this work.

\subsubsection{Decays of the \texorpdfstring{$Z$}{Z} boson}
\label{sec:Zdecay}

\begin{figure}[tb]
    \centering
    \includegraphics{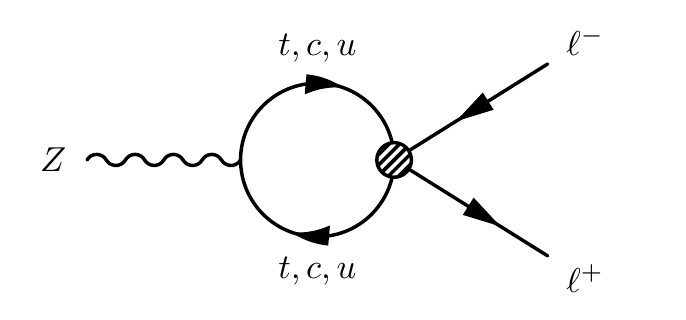}
    \caption{Feynman diagram for a new physics contribution to the decay $Z \rightarrow \ell^+\ell^-$. The corresponding Feynman diagrams for $Z \rightarrow \nu \bar\nu$ and $Z \rightarrow q \bar q$ can be obtained by an appropriate replacement of quark and lepton lines.}
    \label{fig:ZllLoop}
\end{figure}

The decays of the $Z$ boson to SM fermions have been measured with permille level precision at LEP. The good agreement with SM predictions gives stringent constraints on many new physics scenarios \cite{Carpentier:2010ue}. The new physics contact interactions that we consider in this work induce modifications of the $Z$ decays to charged leptons, neutrinos, and quarks, through diagrams such as the one shown in \cref{fig:ZllLoop}.

We parametrize the new-physics effects in $Z$ decays as modifications to the effective $Z$-fermion couplings. That is, we write
\begin{multline}
    \mathcal L\supset
    -\frac{g}{2 c_W}Z_\mu\Bigl[
        (g_\lft^f+\delta g_\lft^f)\bar f_\lft \gamma^\mu f_\lft \\
        + (g_\rgt^f+\delta g_\rgt^f)\bar f_\rgt \gamma^\mu f_\rgt
    \Bigr],
\end{multline}
where $g_{\lft,\rgt}^f$ are the SM $Z$ couplings and the $\delta g_{\lft,\rgt}^f$ are new-physics contributions. Using the above normalization, the SM couplings are given by
\begin{equation}
    \begin{array}{lll}
        g_\lft^u = 1 - \frac{4}{3} s_W^2,~~&
        g_\lft^\nu = 1,~~&
        g_\lft^\ell = -1 + 2 s_W^2,
        \\[0.25cm]
        g_\rgt^u = - \frac{4}{3} s_W^2,~~&
        g_\rgt^\nu = 0,~~&
        g_\rgt^\ell = 2 s_W^2
    \end{array}
\end{equation}
for up-type quarks, neutrinos, and charged leptons, respectively.
The down-type couplings remain SM-like in our new physics scenarios and are therefore not considered in the following.

For the new physics contributions to the charged lepton couplings, we find
\begin{multline}
    \label{eq:delta-g}
    \delta g_{\mathrm X}^\ell =
        -\left(\frac{3}{8 \pi^2} \frac{m_t^2}{\Lambda^2} -
        \frac{s_W^2}{6\pi^2} \frac{m_Z^2}{\Lambda^2} \right)
            C_{\ell\ell tt}^{\mathrm{XR}}
            \log\left(\frac{\Lambda^2}{m_t^2}\right) \\ +
        \frac{s_W^2}{6\pi^2} \frac{m_Z^2}{\Lambda^2}
        \bigl(C_{\ell\ell uu}^{\mathrm{XR}}+C_{\ell\ell cc}^{\mathrm{XR}}\bigr)
        \log\left(\frac{\Lambda^2}{m_Z^2}\right)
        ,
\end{multline}
where $\mathrm X\in\{\lft,\rgt\}$ and $\ell\in\{e,\mu\}$. As in the case of the rare $B$ decays discussed in the previous section, we consider the leading logarithmically enhanced new physics contributions and consistently neglect additional scheme-dependent terms. (The additional terms are, for example, considered in \cite{Dawson:2022bxd}.) The $Z$ couplings to left-handed neutrinos are shifted by
\begin{multline}
    \label{eq:delta-gnu}
    \delta g_\lft^{\nu_\ell} =
        -\left(\frac{3}{8 \pi^2} \frac{m_t^2}{\Lambda^2} -
        \frac{s_W^2}{6\pi^2} \frac{m_Z^2}{\Lambda^2} \right)
            C_{\ell\ell tt}^{\LR}  \log\left(\frac{\Lambda^2}{m_t^2}\right) \\ +
        \frac{s_W^2}{6\pi^2} \frac{m_Z^2}{\Lambda^2}
            \bigl(C_{\ell\ell uu}^{\LR}+C_{\ell\ell cc}^{\LR}\bigr)
          \log\left(\frac{\Lambda^2}{m_Z^2}\right)
          ,
\end{multline}
and the couplings to right-handed up-type quarks are shifted by
\begin{multline}
    \label{eq:delta-gq}
    \delta g_\rgt^q =
        \Bigl[ (1-2s_W^2) (C^{\LR}_{ee qq}+C^{\LR}_{\mu\mu qq}) -
            2s_W^2 (C^{\RR}_{ee qq}+C^{\RR}_{\mu\mu qq})\Bigr] \\ \times
        \frac{s_W^2}{24\pi^2} \frac{m_Z^2}{\Lambda^2}
           \log\left(\frac{\Lambda^2}{m_Z^2}\right)
        ,
\end{multline}
where $q\in\{u,c\}$. These modifications to the $Z$ couplings lead to deviations of the $Z$ partial widths from their SM predictions. For decays to charged leptons, we find
\begin{widetext}
\begin{multline}
    \label{eq:width-fraction-leptons}
    \frac{\Gamma(Z \to \ell \ell)}{\Gamma(Z\to \ell \ell)_\SM} =
    1 + \frac{(1-2s_W^2) C^{\LR}_{\ell\ell tt} -2s_W^2 C^{\RR}_{\ell\ell tt}}
             {1-4s_W^2 + 8 s_W^4}
    \left(
        \frac{3}{4 \pi^2} \frac{m_t^2}{\Lambda^2}
        - \frac{s_W^2}{3\pi^2} \frac{m_Z^2}{\Lambda^2}
    \right)\log\left(\frac{\Lambda^2}{m_t^2}\right)
    \\
    + \frac{
        (2s_W^2-1)\left(C^{\LR}_{\ell\ell uu}+C^{\LR}_{\ell\ell cc}\right)
        + 2s_W^2\left(C^{\RR}_{\ell\ell uu}+C^{\RR}_{\ell\ell cc}\right)
    }{1-4s_W^2 + 8 s_W^4}
    \frac{s_W^2}{3\pi^2} \frac{m_Z^2}{\Lambda^2}
    \log\left(\frac{\Lambda^2}{m_Z^2}\right)
    .
\end{multline}
Similarly, for the sum of the decay widths into neutrinos, we find
\begin{multline}
    \label{eq:Znunu}
    \frac{\Gamma(Z \to \nu \nu)}{\Gamma(Z\to \nu \nu)_\SM} =
    1 - \frac{1}{3}\bigl(C^{\LR}_{ee tt} + C^{\LR}_{\mu\mu tt} \bigr)
     \left(
        \frac{3}{4 \pi^2} \frac{m_t^2}{\Lambda^2}
        - \frac{s_W^2}{3\pi^2} \frac{m_Z^2}{\Lambda^2}
    \right) \log\left(\frac{\Lambda^2}{m_t^2}\right)
    \\
    + \frac{1}{3} \bigl( C^{\LR}_{ee uu} + C^{\LR}_{\mu\mu uu}
        + C^{\LR}_{ee cc} + C^{\LR}_{\mu\mu cc} \bigr)
     \frac{s_W^2}{3\pi^2} \frac{m_Z^2}{\Lambda^2}
    \log\left(\frac{\Lambda^2}{m_Z^2}\right)
    .
\end{multline}
For the decay into an up-type quark-antiquark pair, we find
\begin{equation}
    \label{eq:Zqq}
    \frac{\Gamma(Z \to qq)}{\Gamma(Z\to qq)_\SM} =
    1 + \frac{s_W^2}{\pi^2} \frac{m_Z^2}{\Lambda^2}
    \log\left(\frac{\Lambda^2}{m_Z^2}\right)
    \frac{
        (2s_W^2-1)(C^{\LR}_{ee qq}+C^{\LR}_{\mu\mu qq})
        + 2s_W^2(C^{\RR}_{ee qq}+C^{\RR}_{\mu\mu qq})
    }{9-24s_W^2 +32 s_W^4}~,
\end{equation}
\end{widetext}
where $q\in\{u,c\}$. The decay widths into down-type quarks remain SM-like.

To obtain bounds on the Wilson coefficients entering \cref{eq:width-fraction-leptons,eq:Znunu,eq:Zqq} above, we use the SM predictions and the LEP measurements of the $Z$ partial widths given in \refcite{ALEPH:2005ab} (Tables G.2 and 7.1, respectively). Considering one operator at a time, we find the bounds shown in \cref{tab:lep-width-bounds}. The bounds hold as long as the Wilson coefficients are of $\mathcal O(1)$.

\begin{table}[tb]
    \centering
    \begin{equation*}
        \arraycolsep=0.23cm
        \def\arraystretch{1.5}
        \begin{array}{llcc}
                \textnormal{Final state} &
                \textnormal{Coefficients} &
                \Lambda_- &
                \Lambda_+
            \\\hline\hline
                e^+e^-
                    & C^{\LR}_{ee tt} & 1.6 & 2.3 \\
                    & C^{\RR}_{ee tt} & 2.1 & 1.5 \\
                    & C^{\LR}_{ee uu}, \; C^{\LR}_{ee cc} & 0.2 & - \\
                    & C^{\RR}_{ee uu}, \; C^{\RR}_{ee cc} & - & 0.2
            \\\hline
                \mu^+\mu^-
                    & C^{\LR}_{\mu\mu tt} & 1.5 & 1.5 \\
                    & C^{\RR}_{\mu\mu tt} & 1.4 & 1.4
        \end{array}
    \end{equation*}
    \caption{Lower bounds on $\Lambda$ in \SI{}{\tera\electronvolt} units from $Z$ partial width measurements. Each row corresponds to a bound of the form $-1/\Lambda_{-}^2 < C/\Lambda^2 < 1/\Lambda_+^2$, with all other coefficients set to zero. In rows with multiple Wilson coefficients, the bound is the same for any of these coefficients set to 1 with the others set to zero. For the entries marked ``$-$'', and for Wilson coefficients not shown in this table, there is no bound in the regime of validity of the EFT, i.e., with $\Lambda\gg m_Z$.}
    \label{tab:lep-width-bounds}
\end{table}

\subsubsection{M{\o}ller scattering}
\label{sec:moller}
\begin{figure}
    \centering
    \includegraphics{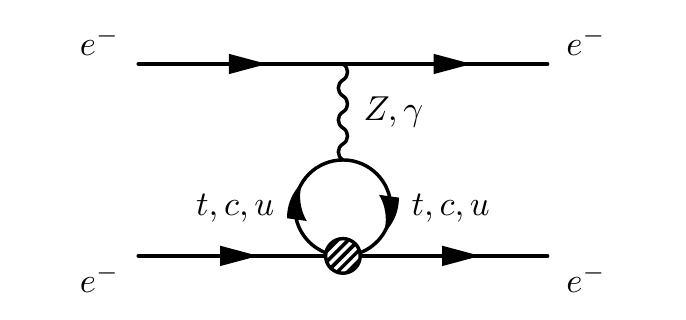}
    \caption{Example new physics contribution to the M\o ller scattering amplitude at one loop.}
    \label{fig:moller}
\end{figure}

The \textsc{moller} experiment~\cite{MOLLER:2014iki} will measure the parity-violating asymmetry $A_\text{PV}$ in polarized electron--electron scattering and determine the weak mixing angle at low energies with unprecedented precision. The asymmetry is defined by
\begin{equation}
    A_\text{PV} = \frac{\sigma_\lft - \sigma_\rgt}{\sigma_\lft + \sigma_\rgt}
    ~,
\end{equation}
where $\sigma_\lft$ ($\sigma_\rgt$) refers to the cross section of left-handed (right-handed) polarized electrons scattering on fixed target electrons. 
From the point of view of a low-energy EFT, the asymmetry is induced by the parity-violating four-electron operator
\begin{equation} \label{eq:moller}
    \mathcal{L} \supset \frac{C_\text{M\o ller}}{\Lambda^2}
        (\bar{e} \gamma^\mu \gamma_5 e) (\bar{e} \gamma_\mu e)
        .
\end{equation}
In the Standard Model, this operator can arise from tree-level $Z$ boson exchange, but it is suppressed by the accidentally small vector coupling of the $Z$ to electrons, proportional to $1 - 4 s_W^2 \simeq 0.0744$ \cite{Du:2019evk}. Precision measurements of $A_\text{PV}$ are thus sensitive to new physics contributions to the operator in \cref{eq:moller}.

The flavor-conserving $(ee)(qq)$ operators that we consider in this work contribute to $C_\text{M{\o}ller}$ at the one-loop level through diagrams as the one shown in \cref{fig:moller}. From a direct one-loop calculation, we find the following logarithmically enhanced terms:
\begin{multline}
    \label{eq:moller-one-loop}
    \frac{C_\text{M{\o}ller}}{\Lambda^2} =
    \frac{G_F}{\sqrt{2}} \Bigl[
        \frac{3}{8\pi^2} \frac{m_t^2}{\Lambda^2}
            C_{eett}^{\LR} \log\left(\frac{\Lambda^2}{m_t^2}\right)
       \\ + \Bigl(
            \frac{3 s_W^2 }{4\pi^2} \frac{m_t^2}{\Lambda^2}
            + \frac{s_W^2}{3\pi^2} \frac{m_W^2}{\Lambda^2}
        \Bigr)
        (C_{eett}^{\RR} - C_{eett}^{\LR})
        \log\left(\frac{\Lambda^2}{m_t^2}\right)
    \\
        + \frac{s_W^2}{3\pi^2} \frac{m_W^2}{\Lambda^2}
            \sum_{q=u,c} (C_{eeqq}^{\RR} - C_{eeqq}^{\LR})
            \log\left(\frac{\Lambda^2}{m_q^2}\right)
    \Bigr]
    .
\end{multline}
As in the other loop-induced processes discussed above, we consistently neglect additional nonlogarithmic terms. Note that in the up quark contribution ($q=u$), the quark mass in the logarithm should be replaced with an appropriate hadronic scale, e.g. $m_u^2 \to m_\pi^2 \simeq (\SI{140}{\mega\electronvolt})^2$. We cross-checked the above result using the known anomalous dimensions of dimension-six operators in the Standard Model EFT and the low energy EFT below the electroweak scale \cite{Jenkins:2013wua, Alonso:2013hga, Jenkins:2017jig, Jenkins:2017dyc}.

The correction to the parity-violating asymmetry in M{\o}ller scattering that corresponds to the above Wilson coefficient can be written in the following way:
\begin{equation}
    \label{eq:moller-APV}
    \frac{\delta A_\text{PV}}{A_\text{PV}^\SM} =
    \frac{2}{1-4s_W^2+\Delta Q_W^e} \frac{\sqrt{2}}{G_F}
    \frac{C_\text{M{\o}ller}}{\Lambda^2}
    ~,
\end{equation}
where we include higher-order electroweak corrections to the SM prediction, $\Delta Q_W^e \simeq -0.0249$ \cite{Czarnecki:1995fw,Du:2019evk}. 

The \textsc{moller} experiment aims at a percent-level uncertainty, $|\delta A_\text{PV}|/A_\text{PV}^\SM < 2.4\%$ \cite{MOLLER:2014iki}, resulting in sensitivity to one-loop induced new physics at the TeV scale. More precisely, if one of the new physics operators is considered at a time, we find the following projected bounds:
{
\everymath={\displaystyle}
\begin{equation}
    \begin{array}{ll}
        \frac{|C_{eett}^{\LR}|}{\Lambda^2} <
            \frac{1}{(\SI{1.32}{\tera\electronvolt})^2} ~,~~&
        \frac{|C_{eett}^{\RR}|}{\Lambda^2} <
            \frac{1}{(\SI{1.36}{\tera\electronvolt})^2} ~, \\
        \frac{|C_{eecc}^{\LR}|}{\Lambda^2} <
            \frac{1}{(\SI{0.78}{\tera\electronvolt})^2} ~,~~&
        \frac{|C_{eecc}^{\RR}|}{\Lambda^2} <
            \frac{1}{(\SI{0.78}{\tera\electronvolt})^2} ~, \\
        \frac{|C_{eeuu}^{\LR}|}{\Lambda^2} <
            \frac{1}{(\SI{0.86}{\tera\electronvolt})^2} ~,~~&
        \frac{|C_{eeuu}^{\RR}|}{\Lambda^2} <
            \frac{1}{(\SI{0.86}{\tera\electronvolt})^2} ~.
    \end{array}
\end{equation}
}
Because the bounds are based on a leading logarithmic approximation of the one-loop contributions, they hold as long as the Wilson coefficients are of $\mathcal O(1)$. We note that in our setup these expected sensitivities are weaker (stronger) than the existing constraints from $Z$ decays in the case of operators involving top quarks (light quarks), see \cref{tab:lep-width-bounds}.

\begin{figure*}
    \centering
    \includegraphics[width=\textwidth]{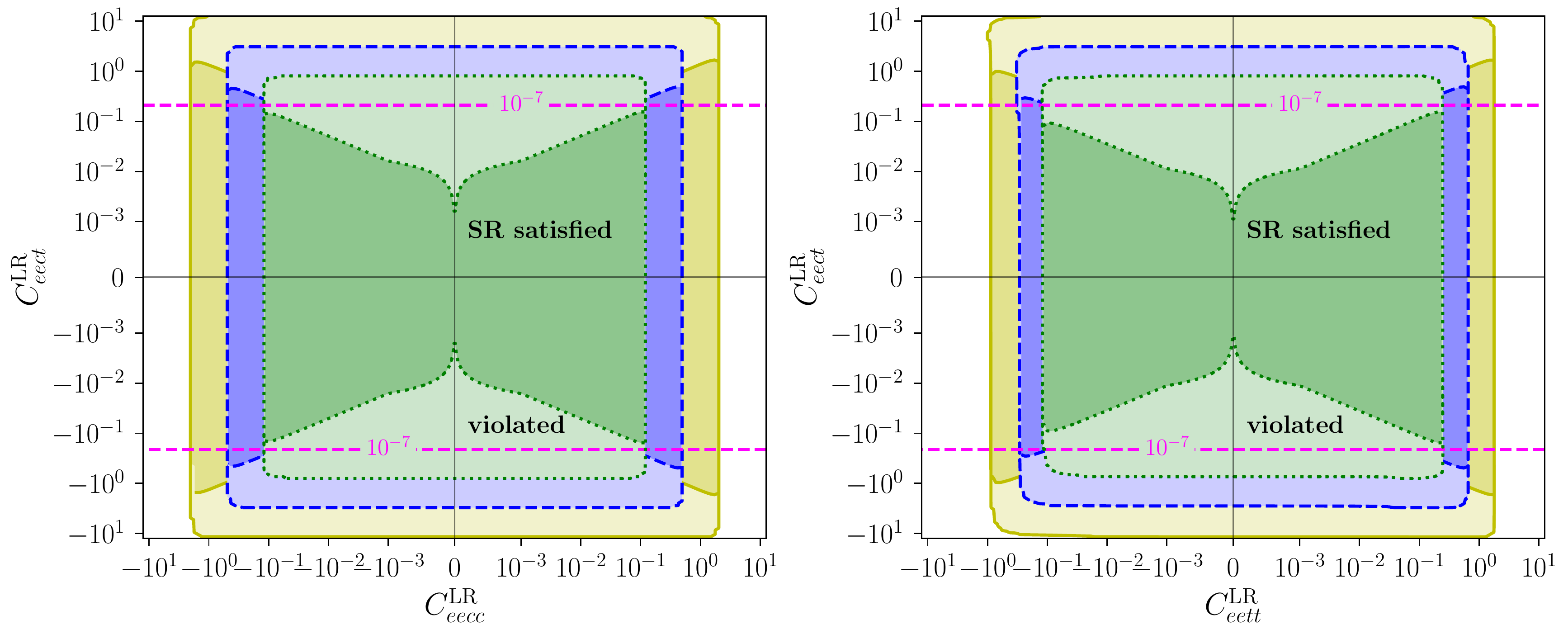}
    \caption{Implications of the sum rules for the Wilson coefficients. In the left (right) panel, $C^\LR_{\ell\ell tt}$ ($C^\LR_{\ell\ell cc}$) is fixed to its maximum value consistent with experimental bounds and perturbativity. The green, blue, and yellow regions correspond to $\Lambda = \SI{500}{\giga\electronvolt}, \SI{1}{\tera\electronvolt}, \SI{2}{\tera\electronvolt}$. For each color, the darker shaded region is consistent with sum rules for some allowed value of the suppressed coefficient. The light shaded regions are inconsistent with experimental data if sum rules are imposed. Note that the regions corresponding to higher $\Lambda$ are partially obscured by those corresponding to lower $\Lambda$. The dashed magenta line shows the contour $\BR(t\to c e^+e^-)=\num{e-7}$ for $\Lambda = \SI{1}{\tera\electronvolt}$, corresponding to the blue region.}
    \label{fig:wc-constraints}
\end{figure*}

\section{Numerical Analysis} \label{sec:numerics}
We now combine the theoretical constraints of \cref{sec:theory_bounds} with the experimental constraints of \cref{sec:exp_constraints} to highlight constrained parameter space for both the Wilson coefficients and the observables themselves. In particular, if the sum rules in \cref{eq:simple-constraint} hold, then experimental bounds on the flavor-conserving Wilson coefficients translate into restrictions on the flavor-violating coefficients. If observable data are eventually fitted to a region in the space of Wilson coefficients that violates the sum rules, this will imply one of the exceptional scenarios described in \cref{sec:theory_bounds}, providing a wealth of information about the underlying UV physics.

We start by discussing a simple example for illustration. Consider the decays $t\to c ee$ and $t \to u ee$, assuming that they are induced by the Wilson coefficients $C_{eect}^{\RR}$ and $C_{eeut}^{\RR}$, respectively. According to \cref{eq:simple-constraint}, these flavor-violating Wilson coefficients are bounded by the allowed sizes of the flavor-conserving $C_{eett}^{\RR}$, $C_{eecc}^{\RR}$, and $C_{eeuu}^{\RR}$. In \cref{sec:exp_constraints}, we derived constraints on these flavor-conserving coefficients that in many cases remain strong even for values of $\Lambda$ exceeding the TeV scale. The most stringent constraints on $C_{eecc}^{\RR}$ and $C_{eeuu}^{\RR}$ come from dilepton production at the LHC (see \cref{tab:dilepton-bounds}) and read as follows:
\begin{align}
    -\frac{1}{(\SI{8.8}{\tera\electronvolt})^2}
        &< \frac{C_{eeuu}^{\RR}}{\Lambda^2} <
    \frac{1}{(\SI{5.9}{\tera\electronvolt})^2}
    ~,\\
    -\frac{1}{(\SI{2.0}{\tera\electronvolt})^2}
        &< \frac{C_{eecc}^{\RR}}{\Lambda^2} <
    \frac{1}{(\SI{2.0}{\tera\electronvolt})^2}
    ~.
\end{align}
The strongest constraint on $C_{eett}^{\RR}$ is from the LEP precision measurements of $Z$ decays (see \cref{tab:lep-width-bounds}), leading to
\begin{equation}
    -\frac{1}{(\SI{2.1}{\tera\electronvolt})^2}
        < \frac{C_{eett}^{\RR}}{\Lambda^2} <
    \frac{1}{(\SI{1.5}{\tera\electronvolt})^2}
    ~.
\end{equation}
The current experimental constraints on the corresponding flavor-violating Wilson coefficients, however, correspond to $\Lambda < \SI{1}{\tera\electronvolt}$, per \cref{eq:raretopbound1,eq:raretopbound2}. Thus, the theoretical constraints provide a new, stronger bound. In this example, the sum rules imply
\begin{equation}
    \frac{|C_{eeut}^\RR|}{\Lambda^2} <
        \frac{1}{(\SI{3.0}{\tera\electronvolt})^2} ~,~~~
    \frac{|C_{eect}^\RR|}{\Lambda^2} <
        \frac{1}{(\SI{1.7}{\tera\electronvolt})^2} ~,
\end{equation}
which are indeed much more stringent than the direct experimental bounds in \cref{eq:raretopbound1,eq:raretopbound2}. This translates into the following upper bounds on rare top decay branching ratios:
\begin{align}
    \label{eq:target1}
    \text{BR}(t\to c e^+e^-) &\lesssim 3 \times 10^{-7}, \\
    \label{eq:target2}
    \text{BR}(t\to u e^+e^-) &\lesssim 3 \times 10^{-8}.
\end{align}
Future observation of the rare top decays above these target values would exclude the full class of new physics models that can be described by the Wilson coefficients $C_{eect}^{\RR}$, $C_{eeut}^{\RR}$, $C_{eett}^{\RR}$, $C_{eecc}^{\RR}$, and $C_{eeuu}^{\RR}$ and that obey the sum rules in \cref{eq:simple-constraint}. Note that the targets in \cref{eq:target1,eq:target2} are 3--4 orders of magnitude below the current direct bounds on the branching ratios given in \cref{eq:raretop_BRs}, leaving ample parameter space to be probed.

\begin{figure*}
    \centering
    \includegraphics[width=\textwidth]{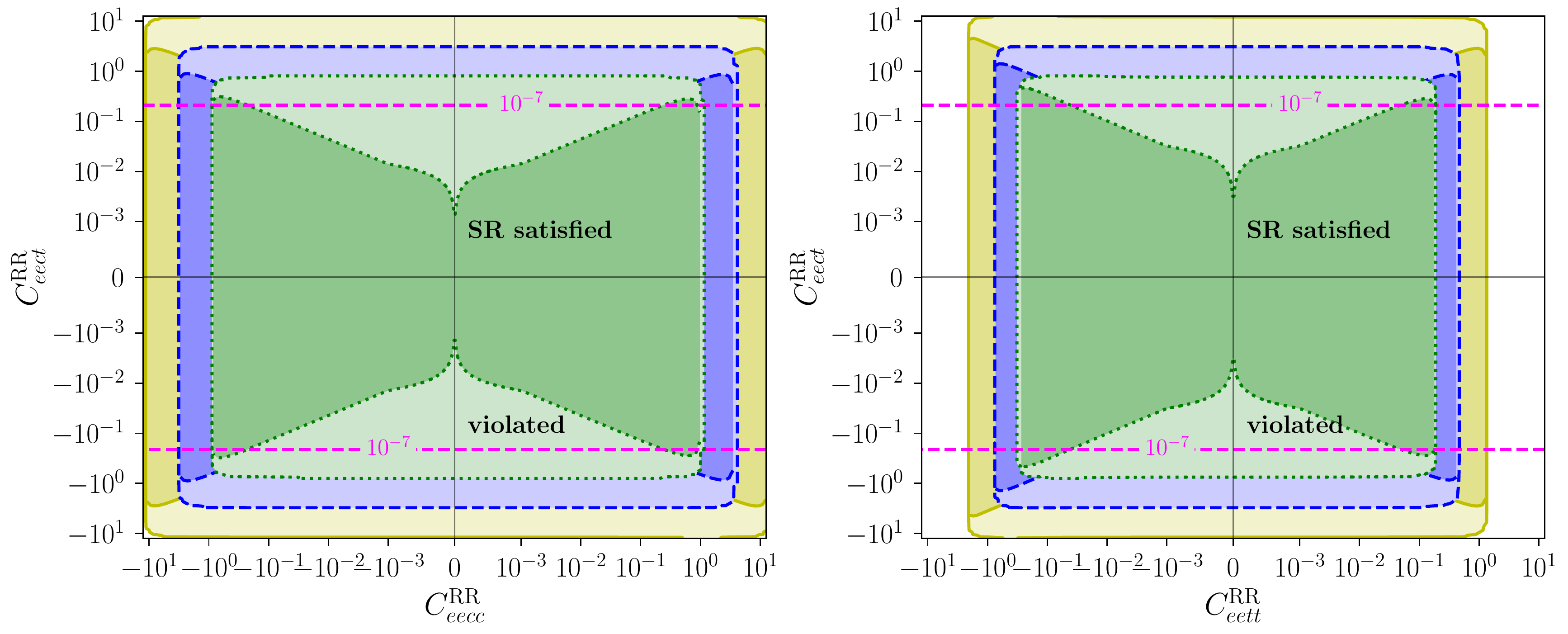}
    \\
    \includegraphics[width=\textwidth]{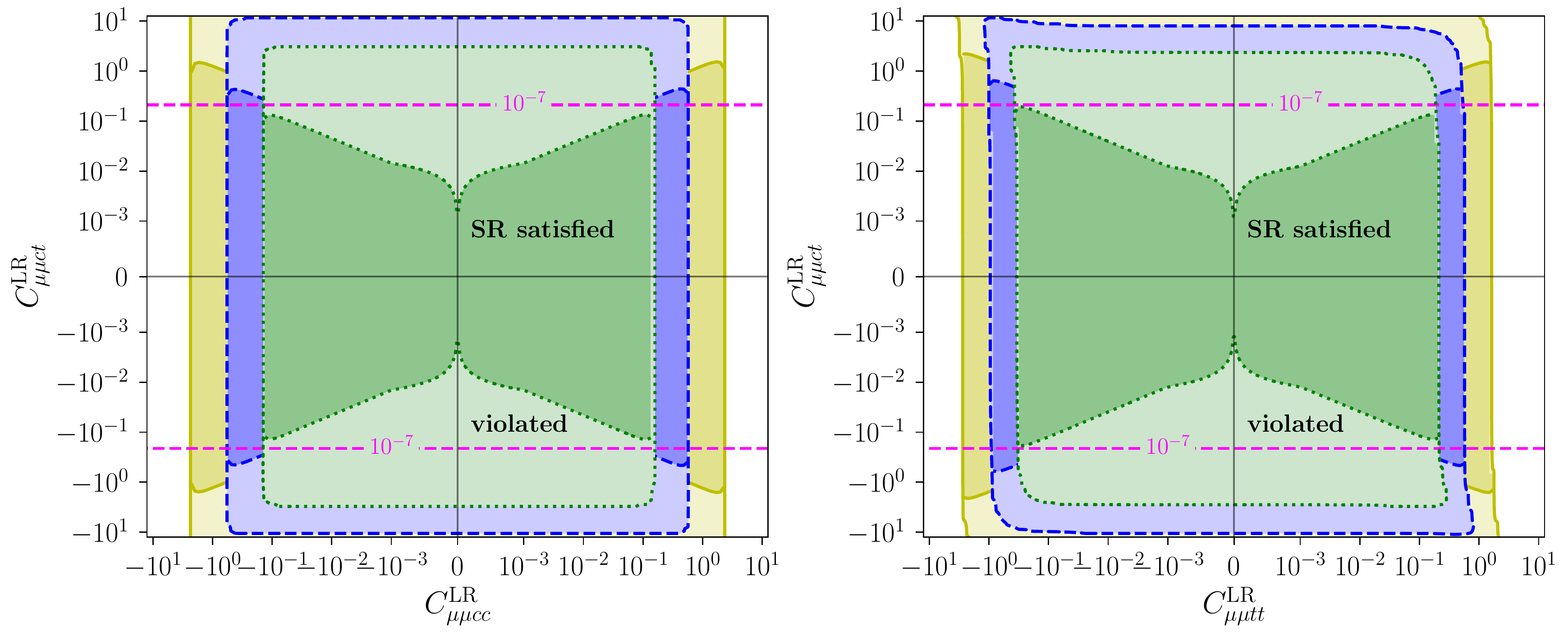}
    \\
    \includegraphics[width=\textwidth]{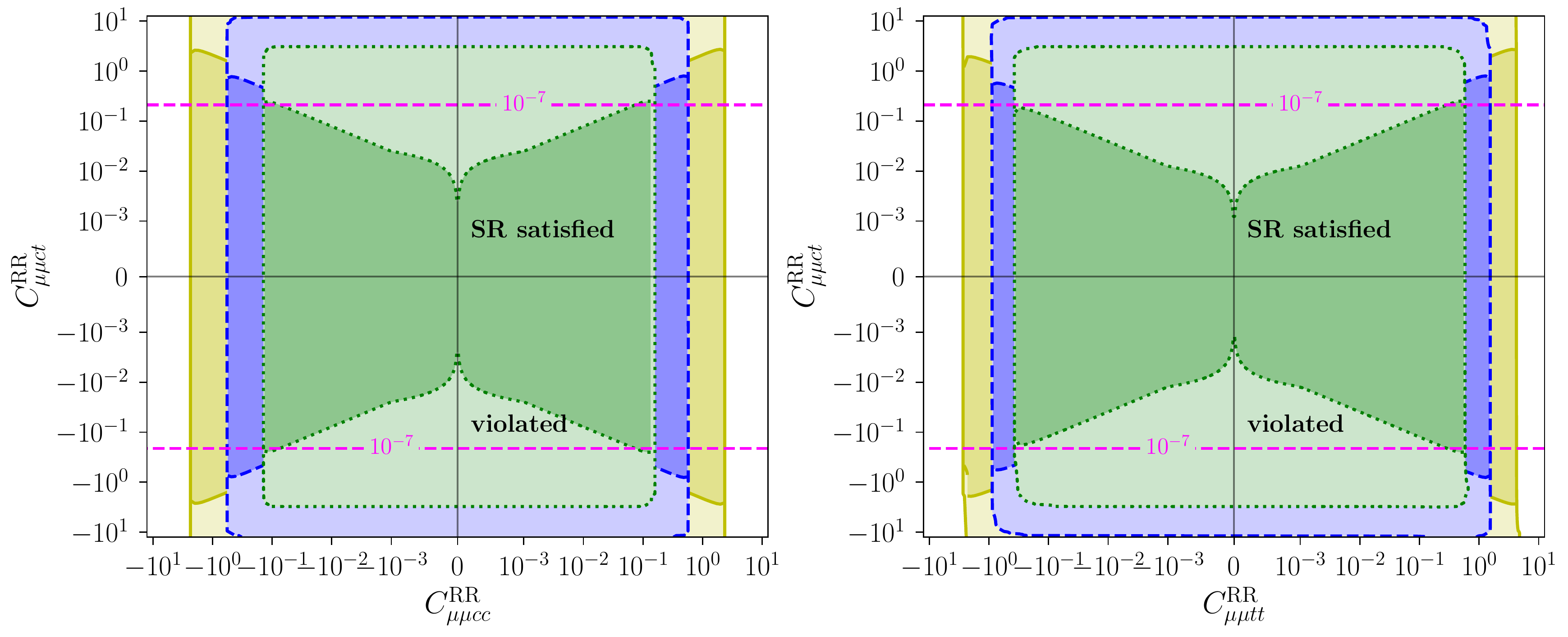}
    \caption{Implications of the sum rules, as in \cref{fig:wc-constraints}, for additional pairs of Wilson coefficients.
    }
    \label{fig:wcs}
\end{figure*}

Following the pattern of this example, we now explore the implications of the sum rules in a systematic way. First, we consider the implications of the sum rules for the Wilson coefficients themselves. In general, each sum rule determines a region in a parameter space with five real degrees of freedom: the EFT scale $\Lambda$, two real flavor-conserving Wilson coefficients, and one complex flavor-violating Wilson coefficient. We show two-dimensional slices through this parameter space determined by the following conditions:
\begin{enumerate}
    \item The EFT scale $\Lambda$ is fixed to a single value.
    \item We vary one flavor-conserving and one flavor-violating Wilson coefficient, restricting the flavor-violating coefficient to be real and positive.
    \item The second flavor-conserving coefficient is set to its largest value compatible with experimental bounds with all other coefficients set to zero. This leads to the most conservative form of the sum rules (i.e., the least stringent theoretical bound on the flavor violating Wilson coefficients).
    \item We impose perturbativity, requiring $\left|C_{abcd}\right| < 4\pi$ for all Wilson coefficients. This is required in order to self-consistently relate the Wilson coefficients to observables.
\end{enumerate}
The constrained regions are shown in \cref{fig:wc-constraints} for the case of the LR operator with $c$--$t$ flavor violation coupled to left-handed electrons. For completeness, other cases are qualitatively very similar and shown in \cref{fig:wcs}. The sum rules are always violated for sufficiently large values of the flavor-violating coefficient, indicating that the measurement of a single flavor-violating observable can diagnose the failure of the sum rules to apply.

The sum rules also impose restrictions in the space of observables: that is, there exist points in the space of observables that cannot be produced by any combination of perturbative Wilson coefficients satisfying the sum rules. Thus, if future experimental data were to prefer such a point, this would imply either one of the exceptional scenarios discussed in \cref{sec:theory_bounds} or a nonperturbative theory.

To locate such points, we take a purely numerical approach: we sample the space of Wilson coefficients with a set of points $\{\bb C_i\}$ and discard all points that are inconsistent with experimental bounds at 95\% C.L. We repeat the process including only samples that satisfy the sum rules, producing a set of points $\{\bb C_i^{\mathrm{SR}}\}$. Each point $\bb C_i$ in the space of Wilson coefficients maps to a point $\bb O_i$ in the space of observables, so we obtain two corresponding sets of points $\{\bb O_i\}$ and $\{\bb O_i^{\mathrm{SR}}\}$ in observable space. From these points, we determine regions $R$ and $R^{\mathrm{SR}} \subset R$ in observable space that are compatible with observables and additionally compatible with the sum rules. In the limit of a large number of sample points, the region $R \setminus R^{\mathrm{SR}}$ consists of points in observable space that are consistent with data but inconsistent with the sum rules in the space of Wilson coefficients.

We carry out the sampling using the Metropolis--Hastings algorithm, taking the target probability density to be proportional to $\left\|\bb O\right\|\exp(-\chi^2(\Lambda, \bb C)/\chi^2_{\mathrm{cut}})$, where $\bb C$ denotes the Wilson coefficients, $\chi^2(\Lambda, \bb C)$ denotes the associated chi-square test statistic, $\bb O$ denotes the point in the plane of observables, and $\chi^2_{\mathrm{cut}}$ is a constant cutoff. This functional form is chosen to ensure an enhanced sampling probability for points that are in tension with experiments, in order to fully sample the boundaries of $R$ and $R^{\mathrm{SR}}$. Note that after imposing perturbativity, the space of Wilson coefficients is bounded, so it is also possible to sample the space exhaustively. While this approach is inefficient, we have performed log-uniform sampling to check that the results are qualitatively similar to those obtained with the Metropolis--Hastings approach.

In the numerical sampling, we include complex phases for flavor-violating coefficients, and we vary both $C^\LR_{abcd}$ and $C^{\RR}_{abcd}$. We show results for electronic observables, and neglect any coupling to muons. In addition to the Wilson coefficients, we simultaneously vary the new physics scale $\Lambda$ over a bounded domain: to accord with experimental constraints and capabilities, we impose $\SI{100}{\giga\electronvolt}<\Lambda<\SI{100}{\tera\electronvolt}$.

\begin{figure*}
    \centering
    \strut\hfill
    \includegraphics[width=\columnwidth]{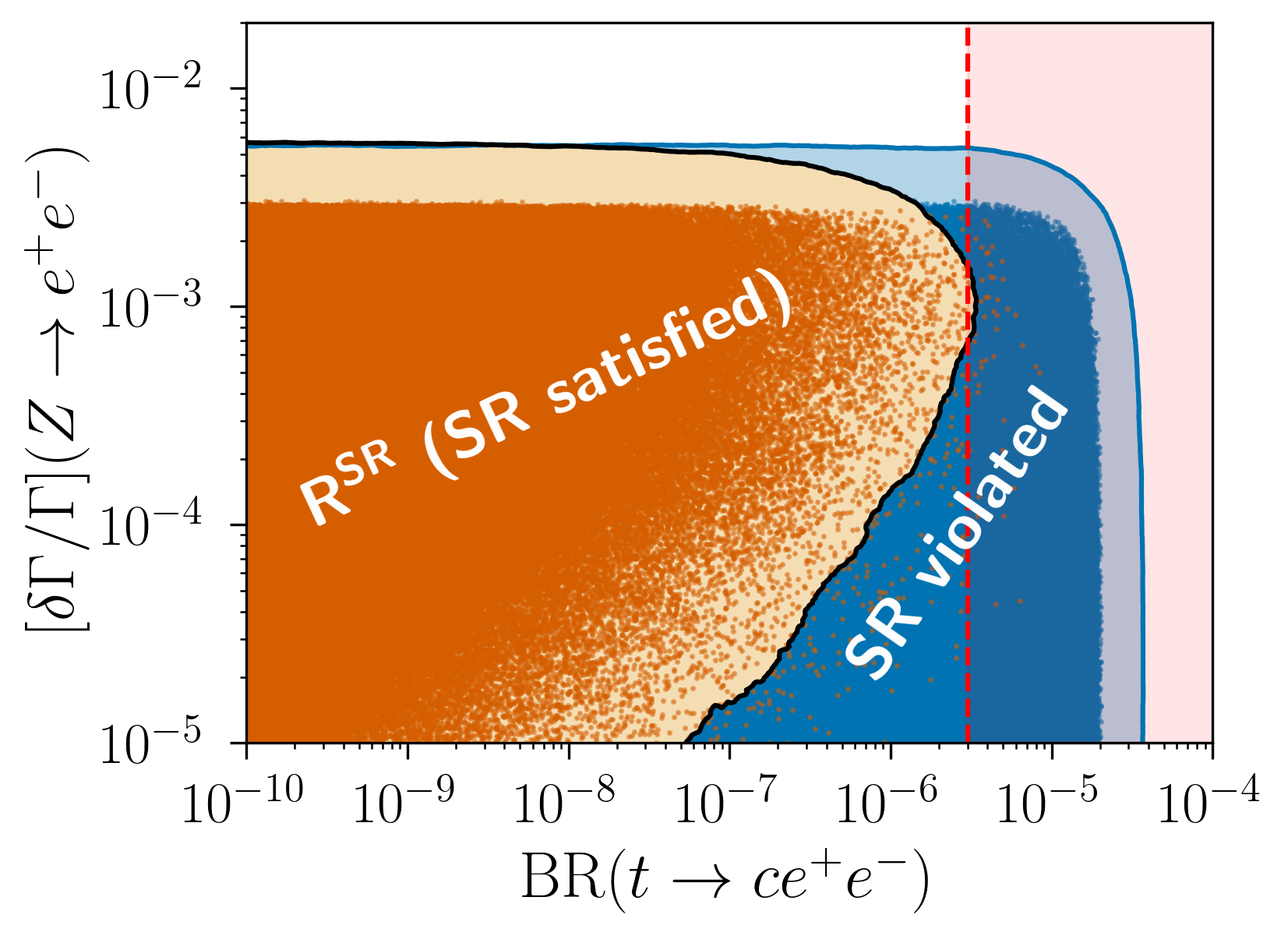}
    \hfill
    \includegraphics[width=\columnwidth]{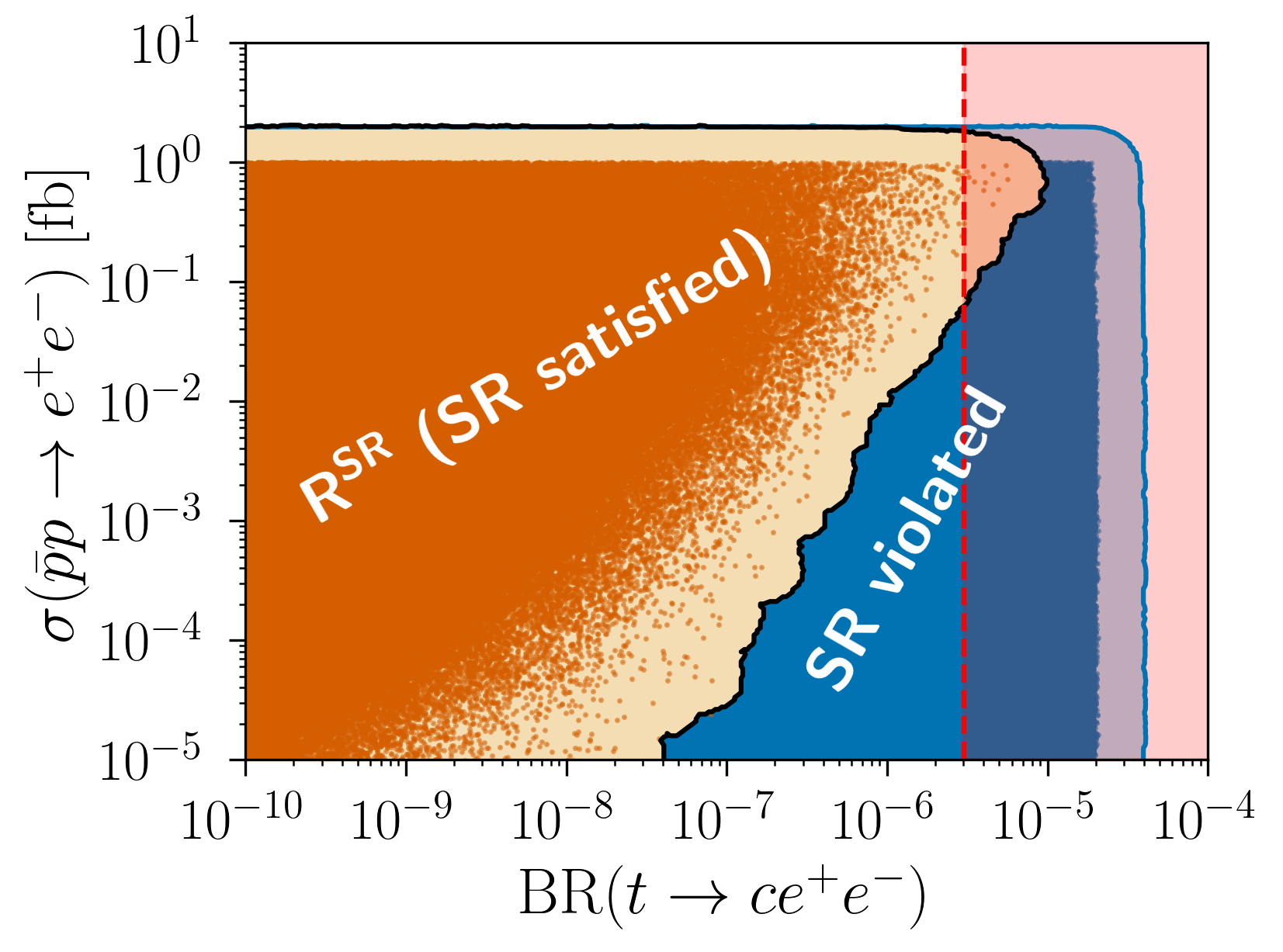}
    \hfill\strut
    \\
    \strut\hfill
    \includegraphics[width=\columnwidth]{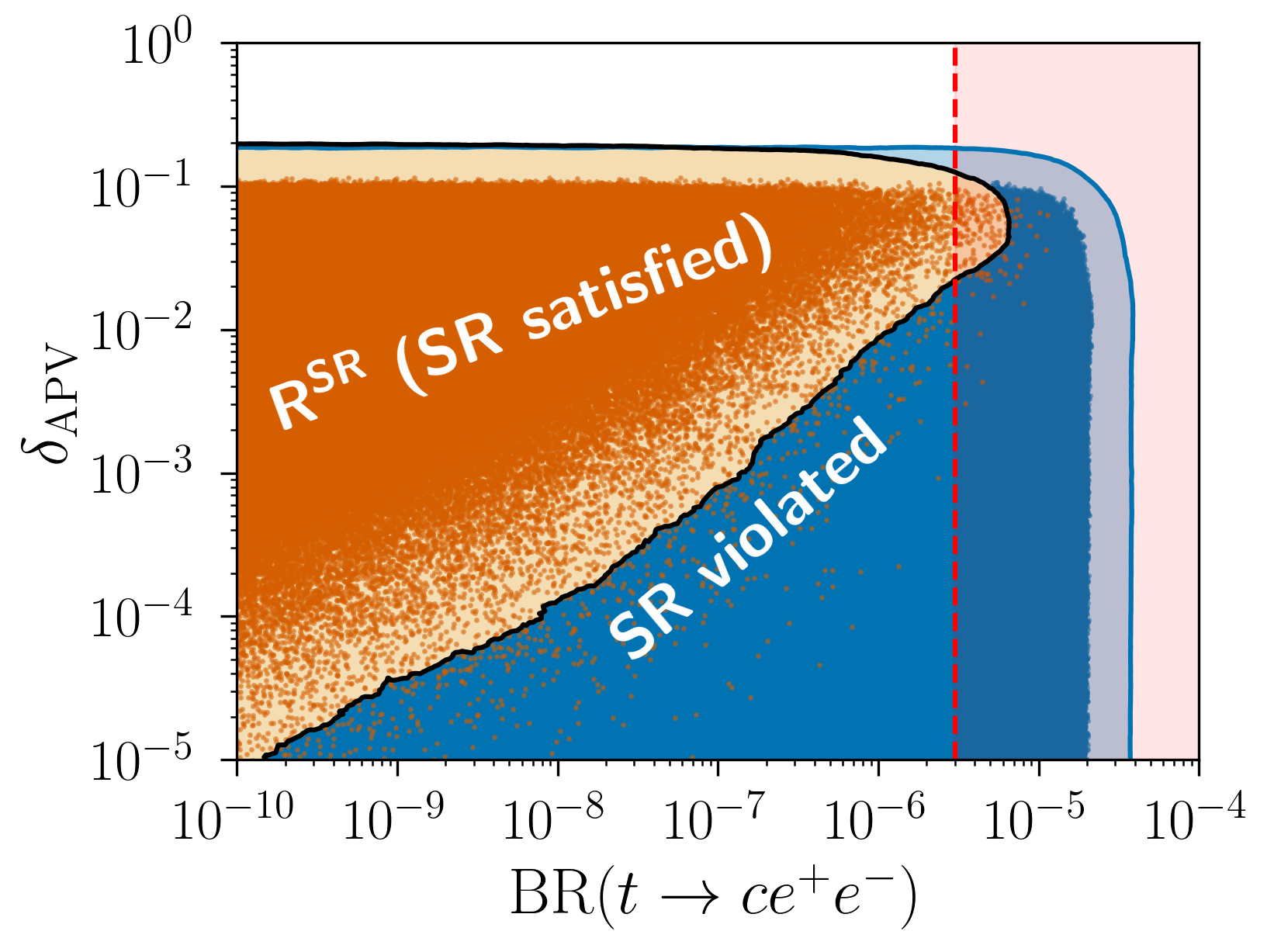}
    \hfill
    \includegraphics[width=\columnwidth]{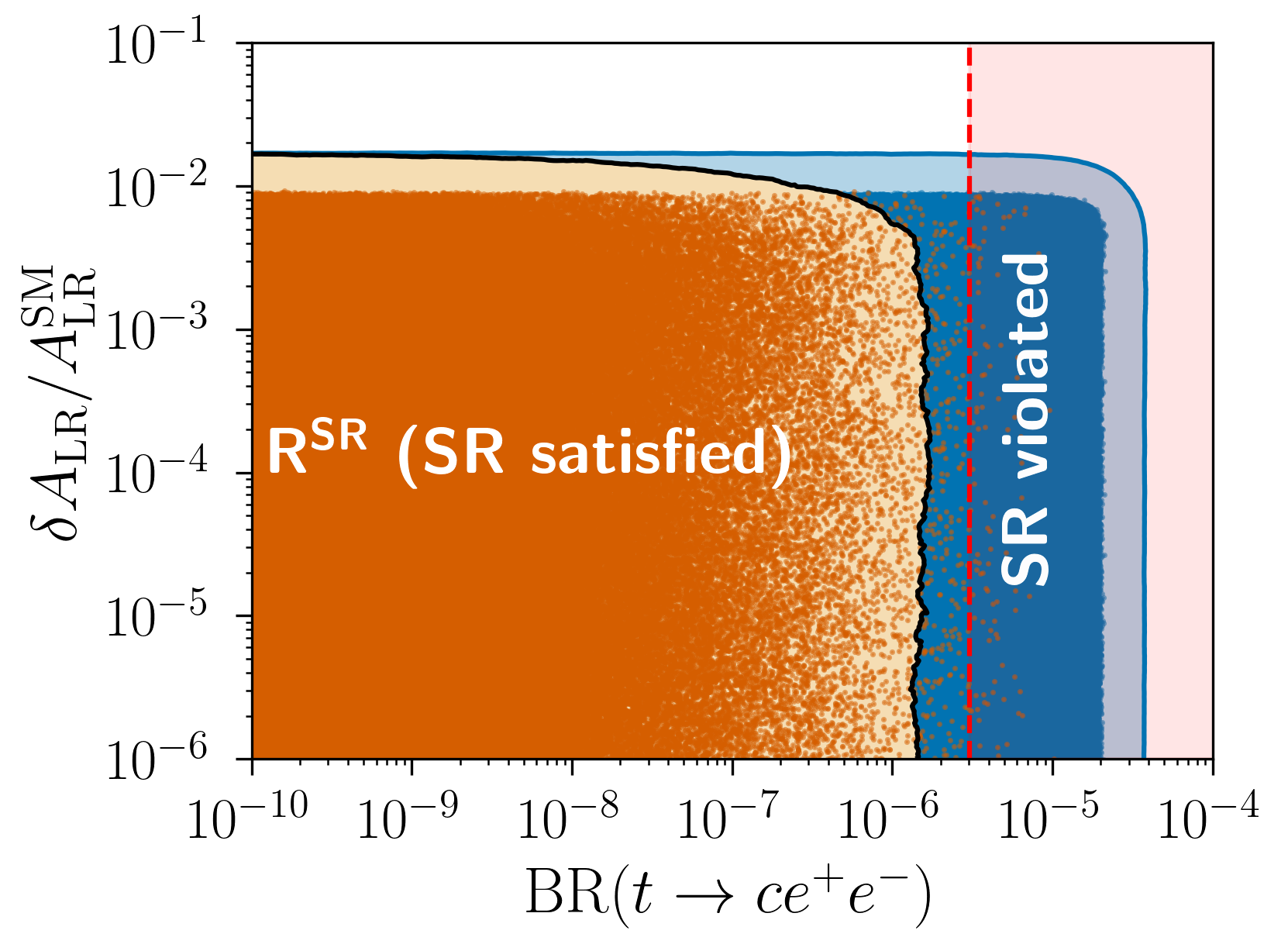}
    \hfill\strut
    \caption{Allowed regions with sample points in planes of complementary observables with $\SI{100}{\giga\electronvolt} < \Lambda < \SI{100}{\tera\electronvolt}$. All points are compatible with all the observables considered in this work. The orange points correspond to combinations of Wilson coefficients that satisfy the sum rules. The lighter shaded regions show the regions $R$ (orange) and $R^{\mathrm{SR}}$ (blue), determined as described in the text. The red regions indicate parameter space that will be constrained by expected experimental searches of $\BR(t\to ce^+e^-)$~\cite{Chala:2018agk}. In particular, almost any nonzero measurement of $\BR(t\to ce^+e^-)$ in the upcoming generation of experiments (see dashed red line in the figure) would likely be incompatible with the sum rules under the assumptions made here.}
    \label{fig:observables}
\end{figure*}

We show the results of this computation in two-dimensional slices through the space of observables in \cref{fig:observables}. To demonstrate the most nontrivial structures, we pair the flavor-violating observable $\BR(t\to c\bar e^+e^-)$ with flavor-conserving observables. For example, the top-left panel shows the regions $R$ (light blue) and $R^{\mathrm{SR}}$ (light orange) in the plane of the branching ratio $\BR(t\to c e^+e^-)$ and the decay width of the $Z$ boson to electrons $\Gamma(Z\to\bar ee)$, with the latter shown as the fractional shift from the SM value. The actual points sampled are shown in dark blue and dark orange. The difference between $R$ and $R^{\mathrm{SR}}$ consists of points that are inconsistent with either the sum rules or the bounds imposed on the Wilson coefficients and new physics scale.

Here, a sufficiently large value of the flavor-violating observable $\BR(t\to c\bar ee)$ always implies violation of the sum rules due to the experimental bounds on the corresponding flavor-conserving Wilson coefficients. For smaller values of $\BR(t\to c\bar ee)$, consistency with the sum rules imposes a restriction on $\Gamma(Z\to\bar ee)$. The red region bounded by the red dashed line in each plot indicates the expected future sensitivity to the flavor-violating top branching ratio estimated in \refcite{Chala:2018agk}, which arises from a phenomenological recasting of a ATLAS search for $t\to Zq$. All in all, the sum rules together with the experimental bounds on flavor conserving operators imply an upper bound on the branching ratio of $t\to c e^+e^-$ roughly an order of magnitude more stringent than the present experimental bound on the branching ratio.

Translating a discrete set of sample points to continuous regions $R$ and $R^{\mathrm{SR}}$ requires a prescription for determining the inclusion of arbitrary points in the plane. In \cref{fig:observables}, we include a test point $p$ in a region if 100 sample points lie within a distance corresponding to a factor of $0.3$ in each direction (i.e., within a ball of radius $0.3$ in log space). This means that some small number of points satisfy the sum rules but nonetheless lie outside the region $R^{\mathrm{SR}}$ as shown. However, these points are extremely sparse, and reflect the fact that it is possible to evade our constraints with extreme tuning.

\section{Conclusions}
\label{sec:conclusions}

The existence of the sum rules of \cref{eq:simple-constraint} at dimension six relating flavor-conserving and flavor-violating quark--lepton operators has nontrivial consequences for low-energy BSM phenomenology, both for the viability of particular classes of models and for the interpretation of a BSM signal in future experiments. We now summarize the implications of these theoretical bounds for the experimental outlook.

Typically, theoretical bounds based on positivity and unitarity are invoked to rule out regions of parameter space, motivating searches in other regimes. With a few key assumptions, it is possible to set similar constraints with the sum rules considered in this work. In this case, \cref{fig:observables} can be interpreted as theoretical constraints on observationally accessible new physics, given the following conditions:
\begin{enumerate}
    \item The sum rules hold, i.e., the assumptions in \cref{sec:theory_bounds} are satisfied.
    \item Any new-physics couplings to muons can be neglected, a simplifying assumption made to reduce the dimensionality of the parameter space.
    \item The scale of new physics lies between \SI{100}{\giga\electronvolt} and \SI{100}{\tera\electronvolt}, such that BSM effects are observable but not already in frank conflict with observations.
    \item The new physics is perturbative, such that the contributions to observables can be reliably computed.
\end{enumerate}
Given these assumptions, it is possible to make direct statements about observables. For example, in the absence of severe fine tuning, \cref{fig:observables} implies that $\BR(t\to c\bar ee) \lesssim \num{e-5}$, an order of magnitude below current experimental bounds.

However, the first assumption above is interestingly fragile: the sum rules fail to hold for some UV structures. In particular, the sum rules can be violated if the effective interactions arise from a combination of scalar and vector interactions in the UV, or if the forward scattering amplitude grows faster than $s^1$. Thus, rather than ruling out parameter space, the sum rules highlight parameter space where low-energy observables encode nontrivial features of the UV physics. The measurement of a combination of observables incompatible with the sum rules would immediately imply one of the alternative scenarios in \cref{sec:theory_bounds}. Thus, rather than excluding parameter space, these constraints \textit{motivate} experimental searches in sum rule--violating regimes.

The present work establishes the first such step toward establishing UV properties from low-energy observables: in particular, we have found that rare top decays provide the best observational prospects to probe the violation of the sum rules. If the sum rules hold, then the relationship between flavor-conserving and flavor-violating observables is substantially constrained. Indeed, any positive detection of flavor-violating top decays in the upcoming generation of experiments must signal a violation of the sum rules under the assumptions of this work. As indicated by the red bands in \cref{fig:observables}, upcoming experiments will not be sensitive to flavor-violating top decays in most of the parameter space consistent with the sum rules in the absence of an extreme fine-tuning. Thus, ongoing searches for flavor-violating top decays nontrivially probe the structure of UV physics. This is one of the main results of the present work.

In our numerical analysis we have focused on the flavor-violating process $t\to c\ell\ell$, which is poorly constrained by direct searches. However, similar results apply for e.g. $t\to u\ell\ell$. Extensions of the theoretical constraints used in this paper may give rise to additional targets elsewhere in the parameter space. Beyond the opportunities to exploit other sum rules in a similar context, we note that we have considered only quark flavor violation. Lepton flavor violation would provide an alternative set of observables that may allow for complementary tests of the sum rules. The use of theoretical constraints may prove especially important for the program of global fits of large numbers of SMEFT coefficients to all available data (see, e.g., \refscite{Durieux:2014xla, Brivio:2019ius, Ellis:2020unq, Ethier:2021bye}). In particular, such global fits may give substantially different results with or without these sum rules as a prior constraint, which further motivates experimental searches along these lines.

\section*{Note Added}
In \cref{sec:singletop}, we have considered the constraints to the top-flavor--changing operators from single top production at LEP. The cross section bound we quote in \cref{eq:LEPbound}, $\sigma(e^+ e^- \to t q) < \SI{0.11}{\pico\barn}$ for $\sqrt{s} = \SI{189}{\giga\electronvolt}$~\cite{Aleph:2001dzz}, is in fact not the most stringent. As pointed out in \refcite{Durieux:2014xla}, a slightly better constraint can be derived from $\sigma(e^+ e^- \to t q) < \SI{0.17}{\pico\barn}$ for $\sqrt{s} = \SI{207}{\giga\electronvolt}$~\cite{Aleph:2001dzz}. We find
\begin{equation}
    \frac{|C^{\LR}_{eect}|}{\Lambda^2},
    \frac{|C^{\RR}_{eect}|}{\Lambda^2},
    \frac{|C^{\LR}_{eeut}|}{\Lambda^2},
    \frac{|C^{\RR}_{eeut}|}{\Lambda^2} <
    \frac{1}{(\SI{0.9}{\tera\electronvolt})^2} ~,
\end{equation}
which updates \cref{eq:singletop_constraints} above. We thank Gauthier Durieux for pointing this out. The new constraint does not change any of our conclusions.

\section*{Acknowledgements}
The research of W.A. and B.V.L. is supported in part by the U.S. Department of Energy grant No. DE-SC0010107. The research of S.G. is supported in part by the NSF CAREER grant No. PHY-1915852 and in part by the U.S. Department of Energy grant No. DESC0023093. The research of B.V.L. is supported by the Josephine de Karman Fellowship Trust and by the MIT Pappalardo Fellowship. The research of J.Z. is supported by the Foundational Questions Institute (FQXi.org), and the Faggin Presidential Chair Fund. B.V.L. is grateful to SLAC and the Kavli Institute for Particle Astrophysics and Cosmology for hospitality while portions of this work were completed. We thank Grant Remmen and Nick Rodd for useful clarifications and for making version 2 of \refcite{Remmen:2020uze} available to us prior to publication. We especially thank Hiren Patel for key contributions in earlier stages of this work. While circumstances compelled Hiren to pursue a career in a different field, we are deeply grateful for his involvement and insights.

\begin{table*}[h]
    \centering
    \begin{equation*}
        \arraycolsep=0.23cm
        \def\arraystretch{1.5}
        \begin{array}{lcc|lcc}
                \textnormal{Coefficient} &
                \Lambda_- &
                \Lambda_+ &
                \textnormal{Coefficient} &
                \Lambda_- &
                \Lambda_+
            \\\hline\hline
                \textnormal{\textbf{Rare $t$ decays}}
                &&&
                \textnormal{\textbf{Rare $t$ decays}}
            \\
                C^{\LR}_{eect}            & 0.32 & 0.32
                    & C^{\RR}_{eect}      & 0.32 & 0.32 \\
                C^{\LR}_{eeut}            & 0.33 & 0.33
                    & C^{\RR}_{eeut}      & 0.33 & 0.33 \\
                C^{\LR}_{\mu\mu ct}       & 0.35 & 0.35
                    & C^{\RR}_{\mu\mu ct} & 0.35 & 0.35 \\
                C^{\LR}_{\mu\mu ut}       & 0.36 & 0.36
                    & C^{\RR}_{\mu\mu ut} & 0.36 & 0.36 \\
                C^{\LR}_{\mu e t c}       & 1.1  & 1.1 
                    & C^{\RR}_{\mu e t c} & 1.1  & 1.1  \\
                C^{\LR}_{e \mu t c}       & 1.1  & 1.1
                    & C^{\RR}_{e \mu t c} & 1.1  & 1.1  \\
                C^{\LR}_{\mu e t u}       & 2.0  & 2.0
                    & C^{\RR}_{\mu e t u} & 2.0  & 2.0  \\
                C^{\LR}_{e \mu t u}       & 2.0  & 2.0
                    & C^{\RR}_{e \mu t u} & 2.0  & 2.0
            \\\hline
                \textnormal{\textbf{Single $t$ production}}
                &&&
                \textnormal{\textbf{Single $t$ production}}
            \\
                C^{\LR}_{eect}            & 0.70  & 0.70
                    & C^{\RR}_{eect}      & 0.70  & 0.70 \\
                C^{\LR}_{eeut}            & 0.70  & 0.70
                    & C^{\RR}_{eeut}      & 0.70  & 0.70
            \\\hline
                \textnormal{\textbf{Dilepton spectra}}
                &&&
                \textnormal{\textbf{Dilepton spectra}}
            \\
                C_{eeuu}^{\LR}            & 8.0  & 6.5
                    & C_{\mu\mu uu}^{\LR} & 8.0  & 5.9  \\
                C_{eeuu}^{\RR}            & 8.8  & 5.9
                    & C_{\mu\mu uu}^{\RR} & 9.2  & 5.1  \\
                C_{eecc}^{\LR}            & 2.0  & 2.0
                    & C_{\mu\mu cc}^{\LR} & 2.1  & 2.0  \\
                C_{eecc}^{\RR}            & 2.0  & 2.0
                    & C_{\mu\mu cc}^{\RR} & 2.1  & 2.0
            \\\hline
                \textnormal{\textbf{Rare $B$ decays}}
                &&&
                \textnormal{\textbf{Rare $B$ decays}}
            \\
                C_{eett}^{\LR}            & 3.7  & 1.2
                    & C_{eett}^{\RR}      & 0.80 & 0.29 \\
                C_{\mu\mu tt}^{\LR}       & 1.4  & 2.6
                    & C_{\mu\mu tt}^{\RR} & 0.95 & 0.85
            \\\hline
                \textnormal{\textbf{$Z$ decays}}
                &&&
                \textnormal{\textbf{$Z$ decays}}
            \\
                C^{\LR}_{ee tt}           & 1.6  & 2.3
                    & C^{\RR}_{ee tt}     & 2.1  & 1.5  \\
                C^{\LR}_{ee uu}           & 0.20 & -
                    & C^{\LR}_{ee cc}     & 0.20 & -    \\
                C^{\RR}_{ee uu}           & -    & 0.20
                    &  C^{\RR}_{ee cc}    & -    & 0.20 \\
                C^{\LR}_{\mu\mu tt}       & 1.5  & 1.5
                    & C^{\RR}_{\mu\mu tt} & 1.4  & 1.4
            \\\hline
                \textnormal{\textbf{M{\o}ller scattering} (expected)}
                &&&
                \textnormal{\textbf{M{\o}ller scattering} (expected)}
            \\
                C_{eett}^{\LR}            &1.32  & 1.32
                    & C_{eett}^{\RR}      &1.36  & 1.36 \\
                C_{eecc}^{\LR}            &0.78  & 0.78
                    & C_{eecc}^{\RR}      &0.78  & 0.78 \\
                C_{ee uu}^{\LR}           &0.86  & 0.86
                    & C_{ee uu}^{\RR}     &0.86  & 0.86
            \\\hline
                \textnormal{\textbf{\ce{^{12}C}--$e^-$ scattering} (expected)}
                &&&
                \textnormal{\textbf{$p^+$--$e^-$ scattering} (expected)}
            \\
                C_{eeuu}^{\RR}-C_{eeuu}^{\LR}    & 5.7 & 5.7
                    & C_{eeuu}^{\RR}-C_{eeuu}^{\LR}    & 7.5 & 7.5
            \\\hline
                \textnormal{\textbf{Atomic parity violation}}
                &&&
            \\
                C_{eeuu}^{\RR}-C_{eeuu}^{\LR}    & 7.2 & 3.1
        \end{array}
    \end{equation*}
    \caption{Summary of all bounds on the EFT scale $\Lambda$ in TeV units. Each row corresponds to a bound of the form $-1/\Lambda_{-}^2 < C/\Lambda^2 < 1/\Lambda_+^2$, where $C$ denotes the Wilson coefficient in question.}
    \label{tab:all-constraints}
\end{table*}

\clearpage

\bibliography{references}

\begin{thebibliography}{134}%
\makeatletter
\providecommand \@ifxundefined [1]{%
 \@ifx{#1\undefined}
}%
\providecommand \@ifnum [1]{%
 \ifnum #1\expandafter \@firstoftwo
 \else \expandafter \@secondoftwo
 \fi
}%
\providecommand \@ifx [1]{%
 \ifx #1\expandafter \@firstoftwo
 \else \expandafter \@secondoftwo
 \fi
}%
\providecommand \natexlab [1]{#1}%
\providecommand \enquote  [1]{``#1''}%
\providecommand \bibnamefont  [1]{#1}%
\providecommand \bibfnamefont [1]{#1}%
\providecommand \citenamefont [1]{#1}%
\providecommand \href@noop [0]{\@secondoftwo}%
\providecommand \href [0]{\begingroup \@sanitize@url \@href}%
\providecommand \@href[1]{\@@startlink{#1}\@@href}%
\providecommand \@@href[1]{\endgroup#1\@@endlink}%
\providecommand \@sanitize@url [0]{\catcode `\\12\catcode `\$12\catcode
  `\&12\catcode `\#12\catcode `\^12\catcode `\_12\catcode `\%12\relax}%
\providecommand \@@startlink[1]{}%
\providecommand \@@endlink[0]{}%
\providecommand \url  [0]{\begingroup\@sanitize@url \@url }%
\providecommand \@url [1]{\endgroup\@href {#1}{\urlprefix }}%
\providecommand \urlprefix  [0]{URL }%
\providecommand \Eprint [0]{\href }%
\providecommand \doibase [0]{https://doi.org/}%
\providecommand \selectlanguage [0]{\@gobble}%
\providecommand \bibinfo  [0]{\@secondoftwo}%
\providecommand \bibfield  [0]{\@secondoftwo}%
\providecommand \translation [1]{[#1]}%
\providecommand \BibitemOpen [0]{}%
\providecommand \bibitemStop [0]{}%
\providecommand \bibitemNoStop [0]{.\EOS\space}%
\providecommand \EOS [0]{\spacefactor3000\relax}%
\providecommand \BibitemShut  [1]{\csname bibitem#1\endcsname}%
\let\auto@bib@innerbib\@empty
\bibitem [{\citenamefont {Aguilar-Saavedra}(2004)}]{Aguilar-Saavedra:2004mfd}%
  \BibitemOpen
  \bibfield  {author} {\bibinfo {author} {\bibfnamefont {J.~A.}\ \bibnamefont
  {Aguilar-Saavedra}},\ }\bibfield  {title} {\bibinfo {title} {{Top
  flavor-changing neutral interactions: Theoretical expectations and
  experimental detection}},\ }\href@noop {} {\bibfield  {journal} {\bibinfo
  {journal} {Acta Phys. Polon. B}\ }\textbf {\bibinfo {volume} {35}},\ \bibinfo
  {pages} {2695} (\bibinfo {year} {2004})},\ \Eprint
  {https://arxiv.org/abs/hep-ph/0409342} {arXiv:hep-ph/0409342} \BibitemShut
  {NoStop}%
\bibitem [{\citenamefont {Altmannshofer}\ \emph
  {et~al.}(2019{\natexlab{a}})\citenamefont {Altmannshofer}, \citenamefont
  {Maddock},\ and\ \citenamefont {Tuckler}}]{Altmannshofer:2019ogm}%
  \BibitemOpen
  \bibfield  {author} {\bibinfo {author} {\bibfnamefont {W.}~\bibnamefont
  {Altmannshofer}}, \bibinfo {author} {\bibfnamefont {B.}~\bibnamefont
  {Maddock}},\ and\ \bibinfo {author} {\bibfnamefont {D.}~\bibnamefont
  {Tuckler}},\ }\bibfield  {title} {\bibinfo {title} {{Rare Top Decays as
  Probes of Flavorful Higgs Bosons}},\ }\href
  {https://doi.org/10.1103/PhysRevD.100.015003} {\bibfield  {journal} {\bibinfo
   {journal} {Phys. Rev. D}\ }\textbf {\bibinfo {volume} {100}},\ \bibinfo
  {pages} {015003} (\bibinfo {year} {2019}{\natexlab{a}})},\ \Eprint
  {https://arxiv.org/abs/1904.10956} {arXiv:1904.10956 [hep-ph]} \BibitemShut
  {NoStop}%
\bibitem [{\citenamefont {Grzadkowski}\ \emph {et~al.}(2010)\citenamefont
  {Grzadkowski}, \citenamefont {Iskrzynski}, \citenamefont {Misiak},\ and\
  \citenamefont {Rosiek}}]{Grzadkowski:2010es}%
  \BibitemOpen
  \bibfield  {author} {\bibinfo {author} {\bibfnamefont {B.}~\bibnamefont
  {Grzadkowski}}, \bibinfo {author} {\bibfnamefont {M.}~\bibnamefont
  {Iskrzynski}}, \bibinfo {author} {\bibfnamefont {M.}~\bibnamefont {Misiak}},\
  and\ \bibinfo {author} {\bibfnamefont {J.}~\bibnamefont {Rosiek}},\
  }\bibfield  {title} {\bibinfo {title} {{Dimension-Six Terms in the Standard
  Model Lagrangian}},\ }\href {https://doi.org/10.1007/JHEP10(2010)085}
  {\bibfield  {journal} {\bibinfo  {journal} {JHEP}\ }\textbf {\bibinfo
  {volume} {10}},\ \bibinfo {pages} {085}},\ \Eprint
  {https://arxiv.org/abs/1008.4884} {arXiv:1008.4884 [hep-ph]} \BibitemShut
  {NoStop}%
\bibitem [{\citenamefont {Drobnak}\ \emph {et~al.}(2009)\citenamefont
  {Drobnak}, \citenamefont {Fajfer},\ and\ \citenamefont
  {Kamenik}}]{Drobnak:2008br}%
  \BibitemOpen
  \bibfield  {author} {\bibinfo {author} {\bibfnamefont {J.}~\bibnamefont
  {Drobnak}}, \bibinfo {author} {\bibfnamefont {S.}~\bibnamefont {Fajfer}},\
  and\ \bibinfo {author} {\bibfnamefont {J.~F.}\ \bibnamefont {Kamenik}},\
  }\bibfield  {title} {\bibinfo {title} {{Signatures of NP models in top FCNC
  decay t ---\ensuremath{>} c(u) l+ l-}},\ }\href
  {https://doi.org/10.1088/1126-6708/2009/03/077} {\bibfield  {journal}
  {\bibinfo  {journal} {JHEP}\ }\textbf {\bibinfo {volume} {03}},\ \bibinfo
  {pages} {077}},\ \Eprint {https://arxiv.org/abs/0812.0294} {arXiv:0812.0294
  [hep-ph]} \BibitemShut {NoStop}%
\bibitem [{\citenamefont {Durieux}\ \emph {et~al.}(2015)\citenamefont
  {Durieux}, \citenamefont {Maltoni},\ and\ \citenamefont
  {Zhang}}]{Durieux:2014xla}%
  \BibitemOpen
  \bibfield  {author} {\bibinfo {author} {\bibfnamefont {G.}~\bibnamefont
  {Durieux}}, \bibinfo {author} {\bibfnamefont {F.}~\bibnamefont {Maltoni}},\
  and\ \bibinfo {author} {\bibfnamefont {C.}~\bibnamefont {Zhang}},\ }\bibfield
   {title} {\bibinfo {title} {{Global approach to top-quark flavor-changing
  interactions}},\ }\href {https://doi.org/10.1103/PhysRevD.91.074017}
  {\bibfield  {journal} {\bibinfo  {journal} {Phys. Rev. D}\ }\textbf {\bibinfo
  {volume} {91}},\ \bibinfo {pages} {074017} (\bibinfo {year} {2015})},\
  \Eprint {https://arxiv.org/abs/1412.7166} {arXiv:1412.7166 [hep-ph]}
  \BibitemShut {NoStop}%
\bibitem [{\citenamefont {Forslund}\ and\ \citenamefont
  {Kidonakis}(2018)}]{Forslund:2018qcp}%
  \BibitemOpen
  \bibfield  {author} {\bibinfo {author} {\bibfnamefont {M.}~\bibnamefont
  {Forslund}}\ and\ \bibinfo {author} {\bibfnamefont {N.}~\bibnamefont
  {Kidonakis}},\ }\bibfield  {title} {\bibinfo {title} {{Associated production
  of a top quark with a photon via anomalous couplings}},\ }\href
  {https://doi.org/10.1103/PhysRevD.98.074017} {\bibfield  {journal} {\bibinfo
  {journal} {Phys. Rev. D}\ }\textbf {\bibinfo {volume} {98}},\ \bibinfo
  {pages} {074017} (\bibinfo {year} {2018})},\ \Eprint
  {https://arxiv.org/abs/1808.09014} {arXiv:1808.09014 [hep-ph]} \BibitemShut
  {NoStop}%
\bibitem [{\citenamefont {Chala}\ \emph {et~al.}(2019)\citenamefont {Chala},
  \citenamefont {Santiago},\ and\ \citenamefont {Spannowsky}}]{Chala:2018agk}%
  \BibitemOpen
  \bibfield  {author} {\bibinfo {author} {\bibfnamefont {M.}~\bibnamefont
  {Chala}}, \bibinfo {author} {\bibfnamefont {J.}~\bibnamefont {Santiago}},\
  and\ \bibinfo {author} {\bibfnamefont {M.}~\bibnamefont {Spannowsky}},\
  }\bibfield  {title} {\bibinfo {title} {{Constraining four-fermion operators
  using rare top decays}},\ }\href {https://doi.org/10.1007/JHEP04(2019)014}
  {\bibfield  {journal} {\bibinfo  {journal} {JHEP}\ }\textbf {\bibinfo
  {volume} {04}},\ \bibinfo {pages} {014}},\ \Eprint
  {https://arxiv.org/abs/1809.09624} {arXiv:1809.09624 [hep-ph]} \BibitemShut
  {NoStop}%
\bibitem [{\citenamefont {Shi}\ and\ \citenamefont
  {Zhang}(2019)}]{Shi:2019epw}%
  \BibitemOpen
  \bibfield  {author} {\bibinfo {author} {\bibfnamefont {L.}~\bibnamefont
  {Shi}}\ and\ \bibinfo {author} {\bibfnamefont {C.}~\bibnamefont {Zhang}},\
  }\bibfield  {title} {\bibinfo {title} {{Probing the top quark flavor-changing
  couplings at CEPC}},\ }\href {https://doi.org/10.1088/1674-1137/43/11/113104}
  {\bibfield  {journal} {\bibinfo  {journal} {Chin. Phys. C}\ }\textbf
  {\bibinfo {volume} {43}},\ \bibinfo {pages} {113104} (\bibinfo {year}
  {2019})},\ \Eprint {https://arxiv.org/abs/1906.04573} {arXiv:1906.04573
  [hep-ph]} \BibitemShut {NoStop}%
\bibitem [{\citenamefont {Afik}\ \emph {et~al.}(2021)\citenamefont {Afik},
  \citenamefont {Bar-Shalom}, \citenamefont {Soni},\ and\ \citenamefont
  {Wudka}}]{Afik:2021jjh}%
  \BibitemOpen
  \bibfield  {author} {\bibinfo {author} {\bibfnamefont {Y.}~\bibnamefont
  {Afik}}, \bibinfo {author} {\bibfnamefont {S.}~\bibnamefont {Bar-Shalom}},
  \bibinfo {author} {\bibfnamefont {A.}~\bibnamefont {Soni}},\ and\ \bibinfo
  {author} {\bibfnamefont {J.}~\bibnamefont {Wudka}},\ }\bibfield  {title}
  {\bibinfo {title} {{New flavor physics in di- and trilepton events from
  single-top production at the LHC and beyond}},\ }\href
  {https://doi.org/10.1103/PhysRevD.103.075031} {\bibfield  {journal} {\bibinfo
   {journal} {Phys. Rev. D}\ }\textbf {\bibinfo {volume} {103}},\ \bibinfo
  {pages} {075031} (\bibinfo {year} {2021})},\ \Eprint
  {https://arxiv.org/abs/2101.05286} {arXiv:2101.05286 [hep-ph]} \BibitemShut
  {NoStop}%
\bibitem [{\citenamefont {Degrande}\ \emph {et~al.}(2011)\citenamefont
  {Degrande}, \citenamefont {Gerard}, \citenamefont {Grojean}, \citenamefont
  {Maltoni},\ and\ \citenamefont {Servant}}]{Degrande:2010kt}%
  \BibitemOpen
  \bibfield  {author} {\bibinfo {author} {\bibfnamefont {C.}~\bibnamefont
  {Degrande}}, \bibinfo {author} {\bibfnamefont {J.-M.}\ \bibnamefont
  {Gerard}}, \bibinfo {author} {\bibfnamefont {C.}~\bibnamefont {Grojean}},
  \bibinfo {author} {\bibfnamefont {F.}~\bibnamefont {Maltoni}},\ and\ \bibinfo
  {author} {\bibfnamefont {G.}~\bibnamefont {Servant}},\ }\bibfield  {title}
  {\bibinfo {title} {{Non-resonant New Physics in Top Pair Production at Hadron
  Colliders}},\ }\href {https://doi.org/10.1007/JHEP03(2011)125} {\bibfield
  {journal} {\bibinfo  {journal} {JHEP}\ }\textbf {\bibinfo {volume} {03}},\
  \bibinfo {pages} {125}},\ \Eprint {https://arxiv.org/abs/1010.6304}
  {arXiv:1010.6304 [hep-ph]} \BibitemShut {NoStop}%
\bibitem [{\citenamefont {Zhang}\ and\ \citenamefont
  {Willenbrock}(2011)}]{Zhang:2010dr}%
  \BibitemOpen
  \bibfield  {author} {\bibinfo {author} {\bibfnamefont {C.}~\bibnamefont
  {Zhang}}\ and\ \bibinfo {author} {\bibfnamefont {S.}~\bibnamefont
  {Willenbrock}},\ }\bibfield  {title} {\bibinfo {title}
  {{Effective-Field-Theory Approach to Top-Quark Production and Decay}},\
  }\href {https://doi.org/10.1103/PhysRevD.83.034006} {\bibfield  {journal}
  {\bibinfo  {journal} {Phys. Rev. D}\ }\textbf {\bibinfo {volume} {83}},\
  \bibinfo {pages} {034006} (\bibinfo {year} {2011})},\ \Eprint
  {https://arxiv.org/abs/1008.3869} {arXiv:1008.3869 [hep-ph]} \BibitemShut
  {NoStop}%
\bibitem [{\citenamefont {Buckley}\ \emph {et~al.}(2016)\citenamefont
  {Buckley}, \citenamefont {Englert}, \citenamefont {Ferrando}, \citenamefont
  {Miller}, \citenamefont {Moore}, \citenamefont {Russell},\ and\ \citenamefont
  {White}}]{Buckley:2015lku}%
  \BibitemOpen
  \bibfield  {author} {\bibinfo {author} {\bibfnamefont {A.}~\bibnamefont
  {Buckley}}, \bibinfo {author} {\bibfnamefont {C.}~\bibnamefont {Englert}},
  \bibinfo {author} {\bibfnamefont {J.}~\bibnamefont {Ferrando}}, \bibinfo
  {author} {\bibfnamefont {D.~J.}\ \bibnamefont {Miller}}, \bibinfo {author}
  {\bibfnamefont {L.}~\bibnamefont {Moore}}, \bibinfo {author} {\bibfnamefont
  {M.}~\bibnamefont {Russell}},\ and\ \bibinfo {author} {\bibfnamefont {C.~D.}\
  \bibnamefont {White}},\ }\bibfield  {title} {\bibinfo {title} {{Constraining
  top quark effective theory in the LHC Run II era}},\ }\href
  {https://doi.org/10.1007/JHEP04(2016)015} {\bibfield  {journal} {\bibinfo
  {journal} {JHEP}\ }\textbf {\bibinfo {volume} {04}},\ \bibinfo {pages}
  {015}},\ \Eprint {https://arxiv.org/abs/1512.03360} {arXiv:1512.03360
  [hep-ph]} \BibitemShut {NoStop}%
\bibitem [{\citenamefont {Schulze}\ and\ \citenamefont
  {Soreq}(2016)}]{Schulze:2016qas}%
  \BibitemOpen
  \bibfield  {author} {\bibinfo {author} {\bibfnamefont {M.}~\bibnamefont
  {Schulze}}\ and\ \bibinfo {author} {\bibfnamefont {Y.}~\bibnamefont
  {Soreq}},\ }\bibfield  {title} {\bibinfo {title} {{Pinning down electroweak
  dipole operators of the top quark}},\ }\href
  {https://doi.org/10.1140/epjc/s10052-016-4263-x} {\bibfield  {journal}
  {\bibinfo  {journal} {Eur. Phys. J. C}\ }\textbf {\bibinfo {volume} {76}},\
  \bibinfo {pages} {466} (\bibinfo {year} {2016})},\ \Eprint
  {https://arxiv.org/abs/1603.08911} {arXiv:1603.08911 [hep-ph]} \BibitemShut
  {NoStop}%
\bibitem [{\citenamefont {Maltoni}\ \emph {et~al.}(2016)\citenamefont
  {Maltoni}, \citenamefont {Vryonidou},\ and\ \citenamefont
  {Zhang}}]{Maltoni:2016yxb}%
  \BibitemOpen
  \bibfield  {author} {\bibinfo {author} {\bibfnamefont {F.}~\bibnamefont
  {Maltoni}}, \bibinfo {author} {\bibfnamefont {E.}~\bibnamefont {Vryonidou}},\
  and\ \bibinfo {author} {\bibfnamefont {C.}~\bibnamefont {Zhang}},\ }\bibfield
   {title} {\bibinfo {title} {{Higgs production in association with a
  top-antitop pair in the Standard Model Effective Field Theory at NLO in
  QCD}},\ }\href {https://doi.org/10.1007/JHEP10(2016)123} {\bibfield
  {journal} {\bibinfo  {journal} {JHEP}\ }\textbf {\bibinfo {volume} {10}},\
  \bibinfo {pages} {123}},\ \Eprint {https://arxiv.org/abs/1607.05330}
  {arXiv:1607.05330 [hep-ph]} \BibitemShut {NoStop}%
\bibitem [{\citenamefont {Degrande}\ \emph {et~al.}(2018)\citenamefont
  {Degrande}, \citenamefont {Maltoni}, \citenamefont {Mimasu}, \citenamefont
  {Vryonidou},\ and\ \citenamefont {Zhang}}]{Degrande:2018fog}%
  \BibitemOpen
  \bibfield  {author} {\bibinfo {author} {\bibfnamefont {C.}~\bibnamefont
  {Degrande}}, \bibinfo {author} {\bibfnamefont {F.}~\bibnamefont {Maltoni}},
  \bibinfo {author} {\bibfnamefont {K.}~\bibnamefont {Mimasu}}, \bibinfo
  {author} {\bibfnamefont {E.}~\bibnamefont {Vryonidou}},\ and\ \bibinfo
  {author} {\bibfnamefont {C.}~\bibnamefont {Zhang}},\ }\bibfield  {title}
  {\bibinfo {title} {{Single-top associated production with a $Z$ or $H$ boson
  at the LHC: the SMEFT interpretation}},\ }\href
  {https://doi.org/10.1007/JHEP10(2018)005} {\bibfield  {journal} {\bibinfo
  {journal} {JHEP}\ }\textbf {\bibinfo {volume} {10}},\ \bibinfo {pages}
  {005}},\ \Eprint {https://arxiv.org/abs/1804.07773} {arXiv:1804.07773
  [hep-ph]} \BibitemShut {NoStop}%
\bibitem [{\citenamefont {de~Beurs}\ \emph {et~al.}(2018)\citenamefont
  {de~Beurs}, \citenamefont {Laenen}, \citenamefont {Vreeswijk},\ and\
  \citenamefont {Vryonidou}}]{deBeurs:2018pvs}%
  \BibitemOpen
  \bibfield  {author} {\bibinfo {author} {\bibfnamefont {M.}~\bibnamefont
  {de~Beurs}}, \bibinfo {author} {\bibfnamefont {E.}~\bibnamefont {Laenen}},
  \bibinfo {author} {\bibfnamefont {M.}~\bibnamefont {Vreeswijk}},\ and\
  \bibinfo {author} {\bibfnamefont {E.}~\bibnamefont {Vryonidou}},\ }\bibfield
  {title} {\bibinfo {title} {{Effective operators in $t$-channel single top
  production and decay}},\ }\href
  {https://doi.org/10.1140/epjc/s10052-018-6399-3} {\bibfield  {journal}
  {\bibinfo  {journal} {Eur. Phys. J. C}\ }\textbf {\bibinfo {volume} {78}},\
  \bibinfo {pages} {919} (\bibinfo {year} {2018})},\ \Eprint
  {https://arxiv.org/abs/1807.03576} {arXiv:1807.03576 [hep-ph]} \BibitemShut
  {NoStop}%
\bibitem [{\citenamefont {Farina}\ \emph {et~al.}(2019)\citenamefont {Farina},
  \citenamefont {Mondino}, \citenamefont {Pappadopulo},\ and\ \citenamefont
  {Ruderman}}]{Farina:2018lqo}%
  \BibitemOpen
  \bibfield  {author} {\bibinfo {author} {\bibfnamefont {M.}~\bibnamefont
  {Farina}}, \bibinfo {author} {\bibfnamefont {C.}~\bibnamefont {Mondino}},
  \bibinfo {author} {\bibfnamefont {D.}~\bibnamefont {Pappadopulo}},\ and\
  \bibinfo {author} {\bibfnamefont {J.~T.}\ \bibnamefont {Ruderman}},\
  }\bibfield  {title} {\bibinfo {title} {{New Physics from High Energy Tops}},\
  }\href {https://doi.org/10.1007/JHEP01(2019)231} {\bibfield  {journal}
  {\bibinfo  {journal} {JHEP}\ }\textbf {\bibinfo {volume} {01}},\ \bibinfo
  {pages} {231}},\ \Eprint {https://arxiv.org/abs/1811.04084} {arXiv:1811.04084
  [hep-ph]} \BibitemShut {NoStop}%
\bibitem [{\citenamefont {Hartland}\ \emph {et~al.}(2019)\citenamefont
  {Hartland}, \citenamefont {Maltoni}, \citenamefont {Nocera}, \citenamefont
  {Rojo}, \citenamefont {Slade}, \citenamefont {Vryonidou},\ and\ \citenamefont
  {Zhang}}]{Hartland:2019bjb}%
  \BibitemOpen
  \bibfield  {author} {\bibinfo {author} {\bibfnamefont {N.~P.}\ \bibnamefont
  {Hartland}}, \bibinfo {author} {\bibfnamefont {F.}~\bibnamefont {Maltoni}},
  \bibinfo {author} {\bibfnamefont {E.~R.}\ \bibnamefont {Nocera}}, \bibinfo
  {author} {\bibfnamefont {J.}~\bibnamefont {Rojo}}, \bibinfo {author}
  {\bibfnamefont {E.}~\bibnamefont {Slade}}, \bibinfo {author} {\bibfnamefont
  {E.}~\bibnamefont {Vryonidou}},\ and\ \bibinfo {author} {\bibfnamefont
  {C.}~\bibnamefont {Zhang}},\ }\bibfield  {title} {\bibinfo {title} {{A Monte
  Carlo global analysis of the Standard Model Effective Field Theory: the top
  quark sector}},\ }\href {https://doi.org/10.1007/JHEP04(2019)100} {\bibfield
  {journal} {\bibinfo  {journal} {JHEP}\ }\textbf {\bibinfo {volume} {04}},\
  \bibinfo {pages} {100}},\ \Eprint {https://arxiv.org/abs/1901.05965}
  {arXiv:1901.05965 [hep-ph]} \BibitemShut {NoStop}%
\bibitem [{\citenamefont {Durieux}\ \emph {et~al.}(2019)\citenamefont
  {Durieux}, \citenamefont {Irles}, \citenamefont {Miralles}, \citenamefont
  {Pe\~nuelas}, \citenamefont {P\"oschl}, \citenamefont {Perell\'o},\ and\
  \citenamefont {Vos}}]{Durieux:2019rbz}%
  \BibitemOpen
  \bibfield  {author} {\bibinfo {author} {\bibfnamefont {G.}~\bibnamefont
  {Durieux}}, \bibinfo {author} {\bibfnamefont {A.}~\bibnamefont {Irles}},
  \bibinfo {author} {\bibfnamefont {V.}~\bibnamefont {Miralles}}, \bibinfo
  {author} {\bibfnamefont {A.}~\bibnamefont {Pe\~nuelas}}, \bibinfo {author}
  {\bibfnamefont {R.}~\bibnamefont {P\"oschl}}, \bibinfo {author}
  {\bibfnamefont {M.}~\bibnamefont {Perell\'o}},\ and\ \bibinfo {author}
  {\bibfnamefont {M.}~\bibnamefont {Vos}},\ }\bibfield  {title} {\bibinfo
  {title} {{The electro-weak couplings of the top and bottom quarks - Global
  fit and future prospects}},\ }\href {https://doi.org/10.1007/JHEP12(2019)098}
  {\bibfield  {journal} {\bibinfo  {journal} {JHEP}\ }\textbf {\bibinfo
  {volume} {12}},\ \bibinfo {pages} {98}},\ \bibinfo {note} {[Erratum: JHEP 01,
  195 (2021)]},\ \Eprint {https://arxiv.org/abs/1907.10619} {arXiv:1907.10619
  [hep-ph]} \BibitemShut {NoStop}%
\bibitem [{\citenamefont {Brivio}\ \emph {et~al.}(2020)\citenamefont {Brivio},
  \citenamefont {Bruggisser}, \citenamefont {Maltoni}, \citenamefont
  {Moutafis}, \citenamefont {Plehn}, \citenamefont {Vryonidou}, \citenamefont
  {Westhoff},\ and\ \citenamefont {Zhang}}]{Brivio:2019ius}%
  \BibitemOpen
  \bibfield  {author} {\bibinfo {author} {\bibfnamefont {I.}~\bibnamefont
  {Brivio}}, \bibinfo {author} {\bibfnamefont {S.}~\bibnamefont {Bruggisser}},
  \bibinfo {author} {\bibfnamefont {F.}~\bibnamefont {Maltoni}}, \bibinfo
  {author} {\bibfnamefont {R.}~\bibnamefont {Moutafis}}, \bibinfo {author}
  {\bibfnamefont {T.}~\bibnamefont {Plehn}}, \bibinfo {author} {\bibfnamefont
  {E.}~\bibnamefont {Vryonidou}}, \bibinfo {author} {\bibfnamefont
  {S.}~\bibnamefont {Westhoff}},\ and\ \bibinfo {author} {\bibfnamefont
  {C.}~\bibnamefont {Zhang}},\ }\bibfield  {title} {\bibinfo {title} {{O new
  physics, where art thou? A global search in the top sector}},\ }\href
  {https://doi.org/10.1007/JHEP02(2020)131} {\bibfield  {journal} {\bibinfo
  {journal} {JHEP}\ }\textbf {\bibinfo {volume} {02}},\ \bibinfo {pages}
  {131}},\ \Eprint {https://arxiv.org/abs/1910.03606} {arXiv:1910.03606
  [hep-ph]} \BibitemShut {NoStop}%
\bibitem [{\citenamefont {Stolarski}\ and\ \citenamefont
  {Tonero}(2020)}]{Stolarski:2020cvf}%
  \BibitemOpen
  \bibfield  {author} {\bibinfo {author} {\bibfnamefont {D.}~\bibnamefont
  {Stolarski}}\ and\ \bibinfo {author} {\bibfnamefont {A.}~\bibnamefont
  {Tonero}},\ }\bibfield  {title} {\bibinfo {title} {{Constraining new physics
  with single top productionat LHC}},\ }\href
  {https://doi.org/10.1007/JHEP08(2020)036} {\bibfield  {journal} {\bibinfo
  {journal} {JHEP}\ }\textbf {\bibinfo {volume} {08}}\bibfield  {number}
  {\bibinfo  {number} { (08)},\ \bibinfo {pages} {036}},\ }\Eprint
  {https://arxiv.org/abs/2004.07856} {arXiv:2004.07856 [hep-ph]} \BibitemShut
  {NoStop}%
\bibitem [{\citenamefont {Ellis}\ \emph
  {et~al.}(2021{\natexlab{a}})\citenamefont {Ellis}, \citenamefont {Madigan},
  \citenamefont {Mimasu}, \citenamefont {Sanz},\ and\ \citenamefont
  {You}}]{Ellis:2020unq}%
  \BibitemOpen
  \bibfield  {author} {\bibinfo {author} {\bibfnamefont {J.}~\bibnamefont
  {Ellis}}, \bibinfo {author} {\bibfnamefont {M.}~\bibnamefont {Madigan}},
  \bibinfo {author} {\bibfnamefont {K.}~\bibnamefont {Mimasu}}, \bibinfo
  {author} {\bibfnamefont {V.}~\bibnamefont {Sanz}},\ and\ \bibinfo {author}
  {\bibfnamefont {T.}~\bibnamefont {You}},\ }\bibfield  {title} {\bibinfo
  {title} {{Top, Higgs, Diboson and Electroweak Fit to the Standard Model
  Effective Field Theory}},\ }\href {https://doi.org/10.1007/JHEP04(2021)279}
  {\bibfield  {journal} {\bibinfo  {journal} {JHEP}\ }\textbf {\bibinfo
  {volume} {04}},\ \bibinfo {pages} {279}},\ \Eprint
  {https://arxiv.org/abs/2012.02779} {arXiv:2012.02779 [hep-ph]} \BibitemShut
  {NoStop}%
\bibitem [{\citenamefont {Afik}\ \emph {et~al.}(2022)\citenamefont {Afik},
  \citenamefont {Bar-Shalom}, \citenamefont {Pal}, \citenamefont {Soni},\ and\
  \citenamefont {Wudka}}]{Afik:2021xmi}%
  \BibitemOpen
  \bibfield  {author} {\bibinfo {author} {\bibfnamefont {Y.}~\bibnamefont
  {Afik}}, \bibinfo {author} {\bibfnamefont {S.}~\bibnamefont {Bar-Shalom}},
  \bibinfo {author} {\bibfnamefont {K.}~\bibnamefont {Pal}}, \bibinfo {author}
  {\bibfnamefont {A.}~\bibnamefont {Soni}},\ and\ \bibinfo {author}
  {\bibfnamefont {J.}~\bibnamefont {Wudka}},\ }\bibfield  {title} {\bibinfo
  {title} {{Multi-lepton probes of new physics and lepton-universality in
  top-quark interactions}},\ }\href
  {https://doi.org/10.1016/j.nuclphysb.2022.115849} {\bibfield  {journal}
  {\bibinfo  {journal} {Nucl. Phys. B}\ }\textbf {\bibinfo {volume} {980}},\
  \bibinfo {pages} {115849} (\bibinfo {year} {2022})},\ \Eprint
  {https://arxiv.org/abs/2111.13711} {arXiv:2111.13711 [hep-ph]} \BibitemShut
  {NoStop}%
\bibitem [{\citenamefont {Miralles}\ \emph {et~al.}(2022)\citenamefont
  {Miralles}, \citenamefont {L\'opez}, \citenamefont {Ll\'acer}, \citenamefont
  {Pe\~nuelas}, \citenamefont {Perell\'o},\ and\ \citenamefont
  {Vos}}]{Miralles:2021dyw}%
  \BibitemOpen
  \bibfield  {author} {\bibinfo {author} {\bibfnamefont {V.}~\bibnamefont
  {Miralles}}, \bibinfo {author} {\bibfnamefont {M.~M.}\ \bibnamefont
  {L\'opez}}, \bibinfo {author} {\bibfnamefont {M.~M.}\ \bibnamefont
  {Ll\'acer}}, \bibinfo {author} {\bibfnamefont {A.}~\bibnamefont
  {Pe\~nuelas}}, \bibinfo {author} {\bibfnamefont {M.}~\bibnamefont
  {Perell\'o}},\ and\ \bibinfo {author} {\bibfnamefont {M.}~\bibnamefont
  {Vos}},\ }\bibfield  {title} {\bibinfo {title} {{The top quark electro-weak
  couplings after LHC Run 2}},\ }\href
  {https://doi.org/10.1007/JHEP02(2022)032} {\bibfield  {journal} {\bibinfo
  {journal} {JHEP}\ }\textbf {\bibinfo {volume} {02}},\ \bibinfo {pages}
  {032}},\ \Eprint {https://arxiv.org/abs/2107.13917} {arXiv:2107.13917
  [hep-ph]} \BibitemShut {NoStop}%
\bibitem [{\citenamefont {Ethier}\ \emph {et~al.}(2021)\citenamefont {Ethier},
  \citenamefont {Magni}, \citenamefont {Maltoni}, \citenamefont {Mantani},
  \citenamefont {Nocera}, \citenamefont {Rojo}, \citenamefont {Slade},
  \citenamefont {Vryonidou},\ and\ \citenamefont {Zhang}}]{Ethier:2021bye}%
  \BibitemOpen
  \bibfield  {author} {\bibinfo {author} {\bibfnamefont {J.~J.}\ \bibnamefont
  {Ethier}}, \bibinfo {author} {\bibfnamefont {G.}~\bibnamefont {Magni}},
  \bibinfo {author} {\bibfnamefont {F.}~\bibnamefont {Maltoni}}, \bibinfo
  {author} {\bibfnamefont {L.}~\bibnamefont {Mantani}}, \bibinfo {author}
  {\bibfnamefont {E.~R.}\ \bibnamefont {Nocera}}, \bibinfo {author}
  {\bibfnamefont {J.}~\bibnamefont {Rojo}}, \bibinfo {author} {\bibfnamefont
  {E.}~\bibnamefont {Slade}}, \bibinfo {author} {\bibfnamefont
  {E.}~\bibnamefont {Vryonidou}},\ and\ \bibinfo {author} {\bibfnamefont
  {C.}~\bibnamefont {Zhang}} (\bibinfo {collaboration} {SMEFiT}),\ }\bibfield
  {title} {\bibinfo {title} {{Combined SMEFT interpretation of Higgs, diboson,
  and top quark data from the LHC}},\ }\href
  {https://doi.org/10.1007/JHEP11(2021)089} {\bibfield  {journal} {\bibinfo
  {journal} {JHEP}\ }\textbf {\bibinfo {volume} {11}},\ \bibinfo {pages}
  {089}},\ \Eprint {https://arxiv.org/abs/2105.00006} {arXiv:2105.00006
  [hep-ph]} \BibitemShut {NoStop}%
\bibitem [{\citenamefont {Liu}\ \emph {et~al.}(2022)\citenamefont {Liu},
  \citenamefont {Wang}, \citenamefont {Zhang}, \citenamefont {Zhang},\ and\
  \citenamefont {Gu}}]{Liu:2022vgo}%
  \BibitemOpen
  \bibfield  {author} {\bibinfo {author} {\bibfnamefont {Y.}~\bibnamefont
  {Liu}}, \bibinfo {author} {\bibfnamefont {Y.}~\bibnamefont {Wang}}, \bibinfo
  {author} {\bibfnamefont {C.}~\bibnamefont {Zhang}}, \bibinfo {author}
  {\bibfnamefont {L.}~\bibnamefont {Zhang}},\ and\ \bibinfo {author}
  {\bibfnamefont {J.}~\bibnamefont {Gu}},\ }\bibfield  {title} {\bibinfo
  {title} {{Probing Top-quark Operators with Precision Electroweak
  Measurements}},\ }\href@noop {} {\  (\bibinfo {year} {2022})},\ \Eprint
  {https://arxiv.org/abs/2205.05655} {arXiv:2205.05655 [hep-ph]} \BibitemShut
  {NoStop}%
\bibitem [{\citenamefont {Barman}\ and\ \citenamefont
  {Ismail}(2022)}]{Barman:2022vjd}%
  \BibitemOpen
  \bibfield  {author} {\bibinfo {author} {\bibfnamefont {R.~K.}\ \bibnamefont
  {Barman}}\ and\ \bibinfo {author} {\bibfnamefont {A.}~\bibnamefont
  {Ismail}},\ }\bibfield  {title} {\bibinfo {title} {{Constraining the top
  electroweak sector of the SMEFT through $Z$ associated top pair and single
  top production at the HL-LHC}},\ }\href@noop {} {\  (\bibinfo {year}
  {2022})},\ \Eprint {https://arxiv.org/abs/2205.07912} {arXiv:2205.07912
  [hep-ph]} \BibitemShut {NoStop}%
\bibitem [{\citenamefont {Adams}\ \emph {et~al.}(2006)\citenamefont {Adams},
  \citenamefont {Arkani-Hamed}, \citenamefont {Dubovsky}, \citenamefont
  {Nicolis},\ and\ \citenamefont {Rattazzi}}]{Adams:2006sv}%
  \BibitemOpen
  \bibfield  {author} {\bibinfo {author} {\bibfnamefont {A.}~\bibnamefont
  {Adams}}, \bibinfo {author} {\bibfnamefont {N.}~\bibnamefont {Arkani-Hamed}},
  \bibinfo {author} {\bibfnamefont {S.}~\bibnamefont {Dubovsky}}, \bibinfo
  {author} {\bibfnamefont {A.}~\bibnamefont {Nicolis}},\ and\ \bibinfo {author}
  {\bibfnamefont {R.}~\bibnamefont {Rattazzi}},\ }\bibfield  {title} {\bibinfo
  {title} {{Causality, analyticity and an IR obstruction to UV completion}},\
  }\href {https://doi.org/10.1088/1126-6708/2006/10/014} {\bibfield  {journal}
  {\bibinfo  {journal} {JHEP}\ }\textbf {\bibinfo {volume} {10}},\ \bibinfo
  {pages} {014}},\ \Eprint {https://arxiv.org/abs/hep-th/0602178}
  {arXiv:hep-th/0602178} \BibitemShut {NoStop}%
\bibitem [{\citenamefont {Distler}\ \emph {et~al.}(2007)\citenamefont
  {Distler}, \citenamefont {Grinstein}, \citenamefont {Porto},\ and\
  \citenamefont {Rothstein}}]{Distler:2006if}%
  \BibitemOpen
  \bibfield  {author} {\bibinfo {author} {\bibfnamefont {J.}~\bibnamefont
  {Distler}}, \bibinfo {author} {\bibfnamefont {B.}~\bibnamefont {Grinstein}},
  \bibinfo {author} {\bibfnamefont {R.~A.}\ \bibnamefont {Porto}},\ and\
  \bibinfo {author} {\bibfnamefont {I.~Z.}\ \bibnamefont {Rothstein}},\
  }\bibfield  {title} {\bibinfo {title} {{Falsifying Models of New Physics via
  WW Scattering}},\ }\href {https://doi.org/10.1103/PhysRevLett.98.041601}
  {\bibfield  {journal} {\bibinfo  {journal} {Phys. Rev. Lett.}\ }\textbf
  {\bibinfo {volume} {98}},\ \bibinfo {pages} {041601} (\bibinfo {year}
  {2007})},\ \Eprint {https://arxiv.org/abs/hep-ph/0604255}
  {arXiv:hep-ph/0604255} \BibitemShut {NoStop}%
\bibitem [{\citenamefont {Vecchi}(2007)}]{Vecchi:2007na}%
  \BibitemOpen
  \bibfield  {author} {\bibinfo {author} {\bibfnamefont {L.}~\bibnamefont
  {Vecchi}},\ }\bibfield  {title} {\bibinfo {title} {{Causal versus analytic
  constraints on anomalous quartic gauge couplings}},\ }\href
  {https://doi.org/10.1088/1126-6708/2007/11/054} {\bibfield  {journal}
  {\bibinfo  {journal} {JHEP}\ }\textbf {\bibinfo {volume} {11}},\ \bibinfo
  {pages} {054}},\ \Eprint {https://arxiv.org/abs/0704.1900} {arXiv:0704.1900
  [hep-ph]} \BibitemShut {NoStop}%
\bibitem [{\citenamefont {Zhang}\ and\ \citenamefont
  {Zhou}(2019)}]{Zhang:2018shp}%
  \BibitemOpen
  \bibfield  {author} {\bibinfo {author} {\bibfnamefont {C.}~\bibnamefont
  {Zhang}}\ and\ \bibinfo {author} {\bibfnamefont {S.-Y.}\ \bibnamefont
  {Zhou}},\ }\bibfield  {title} {\bibinfo {title} {{Positivity bounds on vector
  boson scattering at the LHC}},\ }\href
  {https://doi.org/10.1103/PhysRevD.100.095003} {\bibfield  {journal} {\bibinfo
   {journal} {Phys. Rev. D}\ }\textbf {\bibinfo {volume} {100}},\ \bibinfo
  {pages} {095003} (\bibinfo {year} {2019})},\ \Eprint
  {https://arxiv.org/abs/1808.00010} {arXiv:1808.00010 [hep-ph]} \BibitemShut
  {NoStop}%
\bibitem [{\citenamefont {Bi}\ \emph {et~al.}(2019)\citenamefont {Bi},
  \citenamefont {Zhang},\ and\ \citenamefont {Zhou}}]{Bi:2019phv}%
  \BibitemOpen
  \bibfield  {author} {\bibinfo {author} {\bibfnamefont {Q.}~\bibnamefont
  {Bi}}, \bibinfo {author} {\bibfnamefont {C.}~\bibnamefont {Zhang}},\ and\
  \bibinfo {author} {\bibfnamefont {S.-Y.}\ \bibnamefont {Zhou}},\ }\bibfield
  {title} {\bibinfo {title} {{Positivity constraints on aQGC: carving out the
  physical parameter space}},\ }\href {https://doi.org/10.1007/JHEP06(2019)137}
  {\bibfield  {journal} {\bibinfo  {journal} {JHEP}\ }\textbf {\bibinfo
  {volume} {06}},\ \bibinfo {pages} {137}},\ \Eprint
  {https://arxiv.org/abs/1902.08977} {arXiv:1902.08977 [hep-ph]} \BibitemShut
  {NoStop}%
\bibitem [{\citenamefont {Remmen}\ and\ \citenamefont
  {Rodd}(2019)}]{Remmen:2019cyz}%
  \BibitemOpen
  \bibfield  {author} {\bibinfo {author} {\bibfnamefont {G.~N.}\ \bibnamefont
  {Remmen}}\ and\ \bibinfo {author} {\bibfnamefont {N.~L.}\ \bibnamefont
  {Rodd}},\ }\bibfield  {title} {\bibinfo {title} {{Consistency of the Standard
  Model Effective Field Theory}},\ }\href
  {https://doi.org/10.1007/JHEP12(2019)032} {\bibfield  {journal} {\bibinfo
  {journal} {JHEP}\ }\textbf {\bibinfo {volume} {12}},\ \bibinfo {pages}
  {032}},\ \Eprint {https://arxiv.org/abs/1908.09845} {arXiv:1908.09845
  [hep-ph]} \BibitemShut {NoStop}%
\bibitem [{\citenamefont {Ellis}\ \emph {et~al.}(2020)\citenamefont {Ellis},
  \citenamefont {Ge}, \citenamefont {He},\ and\ \citenamefont
  {Xiao}}]{Ellis:2019zex}%
  \BibitemOpen
  \bibfield  {author} {\bibinfo {author} {\bibfnamefont {J.}~\bibnamefont
  {Ellis}}, \bibinfo {author} {\bibfnamefont {S.-F.}\ \bibnamefont {Ge}},
  \bibinfo {author} {\bibfnamefont {H.-J.}\ \bibnamefont {He}},\ and\ \bibinfo
  {author} {\bibfnamefont {R.-Q.}\ \bibnamefont {Xiao}},\ }\bibfield  {title}
  {\bibinfo {title} {{Probing the scale of new physics in the $ZZ\gamma$
  coupling at $e^+e^-$ colliders}},\ }\href
  {https://doi.org/10.1088/1674-1137/44/6/063106} {\bibfield  {journal}
  {\bibinfo  {journal} {Chin. Phys. C}\ }\textbf {\bibinfo {volume} {44}},\
  \bibinfo {pages} {063106} (\bibinfo {year} {2020})},\ \Eprint
  {https://arxiv.org/abs/1902.06631} {arXiv:1902.06631 [hep-ph]} \BibitemShut
  {NoStop}%
\bibitem [{\citenamefont {Ellis}\ \emph
  {et~al.}(2021{\natexlab{b}})\citenamefont {Ellis}, \citenamefont {He},\ and\
  \citenamefont {Xiao}}]{Ellis:2020ljj}%
  \BibitemOpen
  \bibfield  {author} {\bibinfo {author} {\bibfnamefont {J.}~\bibnamefont
  {Ellis}}, \bibinfo {author} {\bibfnamefont {H.-J.}\ \bibnamefont {He}},\ and\
  \bibinfo {author} {\bibfnamefont {R.-Q.}\ \bibnamefont {Xiao}},\ }\bibfield
  {title} {\bibinfo {title} {{Probing new physics in dimension-8 neutral gauge
  couplings at $e^{+} e^{-}$ colliders}},\ }\href
  {https://doi.org/10.1007/s11433-020-1617-3} {\bibfield  {journal} {\bibinfo
  {journal} {Sci. China Phys. Mech. Astron.}\ }\textbf {\bibinfo {volume}
  {64}},\ \bibinfo {pages} {221062} (\bibinfo {year} {2021}{\natexlab{b}})},\
  \Eprint {https://arxiv.org/abs/2008.04298} {arXiv:2008.04298 [hep-ph]}
  \BibitemShut {NoStop}%
\bibitem [{\citenamefont {Yamashita}\ \emph {et~al.}(2021)\citenamefont
  {Yamashita}, \citenamefont {Zhang},\ and\ \citenamefont
  {Zhou}}]{Yamashita:2020gtt}%
  \BibitemOpen
  \bibfield  {author} {\bibinfo {author} {\bibfnamefont {K.}~\bibnamefont
  {Yamashita}}, \bibinfo {author} {\bibfnamefont {C.}~\bibnamefont {Zhang}},\
  and\ \bibinfo {author} {\bibfnamefont {S.-Y.}\ \bibnamefont {Zhou}},\
  }\bibfield  {title} {\bibinfo {title} {{Elastic positivity vs extremal
  positivity bounds in SMEFT: a case study in transversal electroweak
  gauge-boson scatterings}},\ }\href {https://doi.org/10.1007/JHEP01(2021)095}
  {\bibfield  {journal} {\bibinfo  {journal} {JHEP}\ }\textbf {\bibinfo
  {volume} {01}},\ \bibinfo {pages} {095}},\ \Eprint
  {https://arxiv.org/abs/2009.04490} {arXiv:2009.04490 [hep-ph]} \BibitemShut
  {NoStop}%
\bibitem [{\citenamefont {Ellis}\ \emph {et~al.}(2023)\citenamefont {Ellis},
  \citenamefont {He},\ and\ \citenamefont {Xiao}}]{Ellis:2022zdw}%
  \BibitemOpen
  \bibfield  {author} {\bibinfo {author} {\bibfnamefont {J.}~\bibnamefont
  {Ellis}}, \bibinfo {author} {\bibfnamefont {H.-J.}\ \bibnamefont {He}},\ and\
  \bibinfo {author} {\bibfnamefont {R.-Q.}\ \bibnamefont {Xiao}},\ }\bibfield
  {title} {\bibinfo {title} {{Probing neutral triple gauge couplings at the LHC
  and future hadron colliders}},\ }\href
  {https://doi.org/10.1103/PhysRevD.107.035005} {\bibfield  {journal} {\bibinfo
   {journal} {Phys. Rev. D}\ }\textbf {\bibinfo {volume} {107}},\ \bibinfo
  {pages} {035005} (\bibinfo {year} {2023})},\ \Eprint
  {https://arxiv.org/abs/2206.11676} {arXiv:2206.11676 [hep-ph]} \BibitemShut
  {NoStop}%
\bibitem [{\citenamefont {Bellazzini}\ and\ \citenamefont
  {Riva}(2018)}]{Bellazzini:2018paj}%
  \BibitemOpen
  \bibfield  {author} {\bibinfo {author} {\bibfnamefont {B.}~\bibnamefont
  {Bellazzini}}\ and\ \bibinfo {author} {\bibfnamefont {F.}~\bibnamefont
  {Riva}},\ }\bibfield  {title} {\bibinfo {title} {{New phenomenological and
  theoretical perspective on anomalous $ZZ$ and $Z\gamma$ processes}},\ }\href
  {https://doi.org/10.1103/PhysRevD.98.095021} {\bibfield  {journal} {\bibinfo
  {journal} {Phys. Rev. D}\ }\textbf {\bibinfo {volume} {98}},\ \bibinfo
  {pages} {095021} (\bibinfo {year} {2018})},\ \Eprint
  {https://arxiv.org/abs/1806.09640} {arXiv:1806.09640 [hep-ph]} \BibitemShut
  {NoStop}%
\bibitem [{\citenamefont {Gu}\ \emph {et~al.}(2022)\citenamefont {Gu},
  \citenamefont {Wang},\ and\ \citenamefont {Zhang}}]{Gu:2020ldn}%
  \BibitemOpen
  \bibfield  {author} {\bibinfo {author} {\bibfnamefont {J.}~\bibnamefont
  {Gu}}, \bibinfo {author} {\bibfnamefont {L.-T.}\ \bibnamefont {Wang}},\ and\
  \bibinfo {author} {\bibfnamefont {C.}~\bibnamefont {Zhang}},\ }\bibfield
  {title} {\bibinfo {title} {{Unambiguously Testing Positivity at Lepton
  Colliders}},\ }\href {https://doi.org/10.1103/PhysRevLett.129.011805}
  {\bibfield  {journal} {\bibinfo  {journal} {Phys. Rev. Lett.}\ }\textbf
  {\bibinfo {volume} {129}},\ \bibinfo {pages} {011805} (\bibinfo {year}
  {2022})},\ \Eprint {https://arxiv.org/abs/2011.03055} {arXiv:2011.03055
  [hep-ph]} \BibitemShut {NoStop}%
\bibitem [{\citenamefont {Fuks}\ \emph {et~al.}(2021)\citenamefont {Fuks},
  \citenamefont {Liu}, \citenamefont {Zhang},\ and\ \citenamefont
  {Zhou}}]{Fuks:2020ujk}%
  \BibitemOpen
  \bibfield  {author} {\bibinfo {author} {\bibfnamefont {B.}~\bibnamefont
  {Fuks}}, \bibinfo {author} {\bibfnamefont {Y.}~\bibnamefont {Liu}}, \bibinfo
  {author} {\bibfnamefont {C.}~\bibnamefont {Zhang}},\ and\ \bibinfo {author}
  {\bibfnamefont {S.-Y.}\ \bibnamefont {Zhou}},\ }\bibfield  {title} {\bibinfo
  {title} {{Positivity in electron-positron scattering: testing the axiomatic
  quantum field theory principles and probing the existence of UV states}},\
  }\href {https://doi.org/10.1088/1674-1137/abcd8c} {\bibfield  {journal}
  {\bibinfo  {journal} {Chin. Phys. C}\ }\textbf {\bibinfo {volume} {45}},\
  \bibinfo {pages} {023108} (\bibinfo {year} {2021})},\ \Eprint
  {https://arxiv.org/abs/2009.02212} {arXiv:2009.02212 [hep-ph]} \BibitemShut
  {NoStop}%
\bibitem [{\citenamefont {Li}\ \emph {et~al.}(2022{\natexlab{a}})\citenamefont
  {Li}, \citenamefont {Mimasu}, \citenamefont {Yamashita}, \citenamefont
  {Yang}, \citenamefont {Zhang},\ and\ \citenamefont {Zhou}}]{Li:2022rag}%
  \BibitemOpen
  \bibfield  {author} {\bibinfo {author} {\bibfnamefont {X.}~\bibnamefont
  {Li}}, \bibinfo {author} {\bibfnamefont {K.}~\bibnamefont {Mimasu}}, \bibinfo
  {author} {\bibfnamefont {K.}~\bibnamefont {Yamashita}}, \bibinfo {author}
  {\bibfnamefont {C.}~\bibnamefont {Yang}}, \bibinfo {author} {\bibfnamefont
  {C.}~\bibnamefont {Zhang}},\ and\ \bibinfo {author} {\bibfnamefont {S.-Y.}\
  \bibnamefont {Zhou}},\ }\bibfield  {title} {\bibinfo {title} {{Moments for
  positivity: using Drell-Yan data to test positivity bounds and
  reverse-engineer new physics}},\ }\href@noop {} {\  (\bibinfo {year}
  {2022}{\natexlab{a}})},\ \Eprint {https://arxiv.org/abs/2204.13121}
  {arXiv:2204.13121 [hep-ph]} \BibitemShut {NoStop}%
\bibitem [{\citenamefont {Remmen}\ and\ \citenamefont
  {Rodd}(2020)}]{Remmen:2020vts}%
  \BibitemOpen
  \bibfield  {author} {\bibinfo {author} {\bibfnamefont {G.~N.}\ \bibnamefont
  {Remmen}}\ and\ \bibinfo {author} {\bibfnamefont {N.~L.}\ \bibnamefont
  {Rodd}},\ }\bibfield  {title} {\bibinfo {title} {{Flavor Constraints from
  Unitarity and Analyticity}},\ }\href
  {https://doi.org/10.1103/PhysRevLett.125.081601} {\bibfield  {journal}
  {\bibinfo  {journal} {Phys. Rev. Lett.}\ }\textbf {\bibinfo {volume} {125}},\
  \bibinfo {pages} {081601} (\bibinfo {year} {2020})},\ \Eprint
  {https://arxiv.org/abs/2004.02885} {arXiv:2004.02885 [hep-ph]} \BibitemShut
  {NoStop}%
\bibitem [{\citenamefont {Bonnefoy}\ \emph {et~al.}(2021)\citenamefont
  {Bonnefoy}, \citenamefont {Gendy},\ and\ \citenamefont
  {Grojean}}]{Bonnefoy:2020yee}%
  \BibitemOpen
  \bibfield  {author} {\bibinfo {author} {\bibfnamefont {Q.}~\bibnamefont
  {Bonnefoy}}, \bibinfo {author} {\bibfnamefont {E.}~\bibnamefont {Gendy}},\
  and\ \bibinfo {author} {\bibfnamefont {C.}~\bibnamefont {Grojean}},\
  }\bibfield  {title} {\bibinfo {title} {{Positivity bounds on Minimal Flavor
  Violation}},\ }\href {https://doi.org/10.1007/JHEP04(2021)115} {\bibfield
  {journal} {\bibinfo  {journal} {JHEP}\ }\textbf {\bibinfo {volume} {04}},\
  \bibinfo {pages} {115}},\ \Eprint {https://arxiv.org/abs/2011.12855}
  {arXiv:2011.12855 [hep-ph]} \BibitemShut {NoStop}%
\bibitem [{\citenamefont {Remmen}\ and\ \citenamefont
  {Rodd}(2022{\natexlab{a}})}]{Remmen:2020uze}%
  \BibitemOpen
  \bibfield  {author} {\bibinfo {author} {\bibfnamefont {G.~N.}\ \bibnamefont
  {Remmen}}\ and\ \bibinfo {author} {\bibfnamefont {N.~L.}\ \bibnamefont
  {Rodd}},\ }\bibfield  {title} {\bibinfo {title} {{Signs, spin, SMEFT: Sum
  rules at dimension six}},\ }\href
  {https://doi.org/10.1103/PhysRevD.105.036006} {\bibfield  {journal} {\bibinfo
   {journal} {Phys. Rev. D}\ }\textbf {\bibinfo {volume} {105}},\ \bibinfo
  {pages} {036006} (\bibinfo {year} {2022}{\natexlab{a}})},\ \Eprint
  {https://arxiv.org/abs/2010.04723} {arXiv:2010.04723 [hep-ph]} \BibitemShut
  {NoStop}%
\bibitem [{\citenamefont {Azatov}\ \emph {et~al.}(2022)\citenamefont {Azatov},
  \citenamefont {Ghosh},\ and\ \citenamefont {Singh}}]{Azatov:2021ygj}%
  \BibitemOpen
  \bibfield  {author} {\bibinfo {author} {\bibfnamefont {A.}~\bibnamefont
  {Azatov}}, \bibinfo {author} {\bibfnamefont {D.}~\bibnamefont {Ghosh}},\ and\
  \bibinfo {author} {\bibfnamefont {A.~H.}\ \bibnamefont {Singh}},\ }\bibfield
  {title} {\bibinfo {title} {{Four-fermion operators at dimension 6: Dispersion
  relations and UV completions}},\ }\href
  {https://doi.org/10.1103/PhysRevD.105.115019} {\bibfield  {journal} {\bibinfo
   {journal} {Phys. Rev. D}\ }\textbf {\bibinfo {volume} {105}},\ \bibinfo
  {pages} {115019} (\bibinfo {year} {2022})},\ \Eprint
  {https://arxiv.org/abs/2112.02302} {arXiv:2112.02302 [hep-ph]} \BibitemShut
  {NoStop}%
\bibitem [{\citenamefont {Zhang}(2021)}]{Zhang:2021eeo}%
  \BibitemOpen
  \bibfield  {author} {\bibinfo {author} {\bibfnamefont {C.}~\bibnamefont
  {Zhang}},\ }\bibfield  {title} {\bibinfo {title} {{SMEFTs living on the edge:
  determining the UV theories from positivity and extremality}},\ }\href@noop
  {} {\  (\bibinfo {year} {2021})},\ \Eprint {https://arxiv.org/abs/2112.11665}
  {arXiv:2112.11665 [hep-ph]} \BibitemShut {NoStop}%
\bibitem [{\citenamefont {Remmen}\ and\ \citenamefont
  {Rodd}(2022{\natexlab{b}})}]{Remmen:2022orj}%
  \BibitemOpen
  \bibfield  {author} {\bibinfo {author} {\bibfnamefont {G.~N.}\ \bibnamefont
  {Remmen}}\ and\ \bibinfo {author} {\bibfnamefont {N.~L.}\ \bibnamefont
  {Rodd}},\ }\bibfield  {title} {\bibinfo {title} {{Spinning Sum Rules for the
  Dimension-Six SMEFT}},\ }\href@noop {} {\  (\bibinfo {year}
  {2022}{\natexlab{b}})},\ \Eprint {https://arxiv.org/abs/2206.13524}
  {arXiv:2206.13524 [hep-ph]} \BibitemShut {NoStop}%
\bibitem [{\citenamefont {Fox}\ \emph {et~al.}(2008)\citenamefont {Fox},
  \citenamefont {Ligeti}, \citenamefont {Papucci}, \citenamefont {Perez},\ and\
  \citenamefont {Schwartz}}]{Fox:2007in}%
  \BibitemOpen
  \bibfield  {author} {\bibinfo {author} {\bibfnamefont {P.~J.}\ \bibnamefont
  {Fox}}, \bibinfo {author} {\bibfnamefont {Z.}~\bibnamefont {Ligeti}},
  \bibinfo {author} {\bibfnamefont {M.}~\bibnamefont {Papucci}}, \bibinfo
  {author} {\bibfnamefont {G.}~\bibnamefont {Perez}},\ and\ \bibinfo {author}
  {\bibfnamefont {M.~D.}\ \bibnamefont {Schwartz}},\ }\bibfield  {title}
  {\bibinfo {title} {{Deciphering top flavor violation at the LHC with $B$
  factories}},\ }\href {https://doi.org/10.1103/PhysRevD.78.054008} {\bibfield
  {journal} {\bibinfo  {journal} {Phys. Rev. D}\ }\textbf {\bibinfo {volume}
  {78}},\ \bibinfo {pages} {054008} (\bibinfo {year} {2008})},\ \Eprint
  {https://arxiv.org/abs/0704.1482} {arXiv:0704.1482 [hep-ph]} \BibitemShut
  {NoStop}%
\bibitem [{\citenamefont {Davighi}\ \emph {et~al.}(2022)\citenamefont
  {Davighi}, \citenamefont {Melville},\ and\ \citenamefont
  {You}}]{Davighi:2021osh}%
  \BibitemOpen
  \bibfield  {author} {\bibinfo {author} {\bibfnamefont {J.}~\bibnamefont
  {Davighi}}, \bibinfo {author} {\bibfnamefont {S.}~\bibnamefont {Melville}},\
  and\ \bibinfo {author} {\bibfnamefont {T.}~\bibnamefont {You}},\ }\bibfield
  {title} {\bibinfo {title} {{Natural selection rules: new positivity bounds
  for massive spinning particles}},\ }\href
  {https://doi.org/10.1007/JHEP02(2022)167} {\bibfield  {journal} {\bibinfo
  {journal} {JHEP}\ }\textbf {\bibinfo {volume} {02}},\ \bibinfo {pages}
  {167}},\ \Eprint {https://arxiv.org/abs/2108.06334} {arXiv:2108.06334
  [hep-th]} \BibitemShut {NoStop}%
\bibitem [{\citenamefont {Sirunyan}\ \emph
  {et~al.}(2021{\natexlab{a}})\citenamefont {Sirunyan} \emph
  {et~al.}}]{Sirunyan:2020tqm}%
  \BibitemOpen
  \bibfield  {author} {\bibinfo {author} {\bibfnamefont {A.~M.}\ \bibnamefont
  {Sirunyan}} \emph {et~al.} (\bibinfo {collaboration} {CMS}),\ }\bibfield
  {title} {\bibinfo {title} {{Search for new physics in top quark production
  with additional leptons in proton-proton collisions at $\sqrt{s} = $ 13 TeV
  using effective field theory}},\ }\href
  {https://doi.org/10.1007/JHEP03(2021)095} {\bibfield  {journal} {\bibinfo
  {journal} {JHEP}\ }\textbf {\bibinfo {volume} {03}},\ \bibinfo {pages}
  {095}},\ \Eprint {https://arxiv.org/abs/2012.04120} {arXiv:2012.04120
  [hep-ex]} \BibitemShut {NoStop}%
\bibitem [{\citenamefont {Aad}\ \emph {et~al.}(2021)\citenamefont {Aad} \emph
  {et~al.}}]{ATLAS:2021kxv}%
  \BibitemOpen
  \bibfield  {author} {\bibinfo {author} {\bibfnamefont {G.}~\bibnamefont
  {Aad}} \emph {et~al.} (\bibinfo {collaboration} {ATLAS}),\ }\bibfield
  {title} {\bibinfo {title} {{Search for new phenomena in events with an
  energetic jet and missing transverse momentum in $pp$ collisions at $\sqrt
  {s}$ =13 TeV with the ATLAS detector}},\ }\href
  {https://doi.org/10.1103/PhysRevD.103.112006} {\bibfield  {journal} {\bibinfo
   {journal} {Phys. Rev. D}\ }\textbf {\bibinfo {volume} {103}},\ \bibinfo
  {pages} {112006} (\bibinfo {year} {2021})},\ \Eprint
  {https://arxiv.org/abs/2102.10874} {arXiv:2102.10874 [hep-ex]} \BibitemShut
  {NoStop}%
\bibitem [{\citenamefont {Tumasyan}\ \emph {et~al.}(2021)\citenamefont
  {Tumasyan} \emph {et~al.}}]{CMS:2021far}%
  \BibitemOpen
  \bibfield  {author} {\bibinfo {author} {\bibfnamefont {A.}~\bibnamefont
  {Tumasyan}} \emph {et~al.} (\bibinfo {collaboration} {CMS}),\ }\bibfield
  {title} {\bibinfo {title} {{Search for new particles in events with energetic
  jets and large missing transverse momentum in proton-proton collisions at $
  \sqrt{s} $ = 13 TeV}},\ }\href {https://doi.org/10.1007/JHEP11(2021)153}
  {\bibfield  {journal} {\bibinfo  {journal} {JHEP}\ }\textbf {\bibinfo
  {volume} {11}},\ \bibinfo {pages} {153}},\ \Eprint
  {https://arxiv.org/abs/2107.13021} {arXiv:2107.13021 [hep-ex]} \BibitemShut
  {NoStop}%
\bibitem [{\citenamefont {Akimov}\ \emph {et~al.}(2017)\citenamefont {Akimov}
  \emph {et~al.}}]{COHERENT:2017ipa}%
  \BibitemOpen
  \bibfield  {author} {\bibinfo {author} {\bibfnamefont {D.}~\bibnamefont
  {Akimov}} \emph {et~al.} (\bibinfo {collaboration} {COHERENT}),\ }\bibfield
  {title} {\bibinfo {title} {{Observation of Coherent Elastic Neutrino-Nucleus
  Scattering}},\ }\href {https://doi.org/10.1126/science.aao0990} {\bibfield
  {journal} {\bibinfo  {journal} {Science}\ }\textbf {\bibinfo {volume}
  {357}},\ \bibinfo {pages} {1123} (\bibinfo {year} {2017})},\ \Eprint
  {https://arxiv.org/abs/1708.01294} {arXiv:1708.01294 [nucl-ex]} \BibitemShut
  {NoStop}%
\bibitem [{\citenamefont {Altmannshofer}\ \emph
  {et~al.}(2019{\natexlab{b}})\citenamefont {Altmannshofer}, \citenamefont
  {Tammaro},\ and\ \citenamefont {Zupan}}]{Altmannshofer:2018xyo}%
  \BibitemOpen
  \bibfield  {author} {\bibinfo {author} {\bibfnamefont {W.}~\bibnamefont
  {Altmannshofer}}, \bibinfo {author} {\bibfnamefont {M.}~\bibnamefont
  {Tammaro}},\ and\ \bibinfo {author} {\bibfnamefont {J.}~\bibnamefont
  {Zupan}},\ }\bibfield  {title} {\bibinfo {title} {{Non-standard neutrino
  interactions and low energy experiments}},\ }\href
  {https://doi.org/10.1007/JHEP11(2021)113} {\bibfield  {journal} {\bibinfo
  {journal} {JHEP}\ }\textbf {\bibinfo {volume} {09}},\ \bibinfo {pages}
  {083}},\ \bibinfo {note} {[Erratum: JHEP 11, 113 (2021)]},\ \Eprint
  {https://arxiv.org/abs/1812.02778} {arXiv:1812.02778 [hep-ph]} \BibitemShut
  {NoStop}%
\bibitem [{\citenamefont {Aaboud}\ \emph {et~al.}(2018)\citenamefont {Aaboud}
  \emph {et~al.}}]{ATLAS:2018zsq}%
  \BibitemOpen
  \bibfield  {author} {\bibinfo {author} {\bibfnamefont {M.}~\bibnamefont
  {Aaboud}} \emph {et~al.} (\bibinfo {collaboration} {ATLAS}),\ }\bibfield
  {title} {\bibinfo {title} {{Search for flavour-changing neutral current
  top-quark decays $t\to qZ$ in proton-proton collisions at $\sqrt{s}=13$ TeV
  with the ATLAS detector}},\ }\href {https://doi.org/10.1007/JHEP07(2018)176}
  {\bibfield  {journal} {\bibinfo  {journal} {JHEP}\ }\textbf {\bibinfo
  {volume} {07}},\ \bibinfo {pages} {176}},\ \Eprint
  {https://arxiv.org/abs/1803.09923} {arXiv:1803.09923 [hep-ex]} \BibitemShut
  {NoStop}%
\bibitem [{\citenamefont {Sirunyan}\ \emph {et~al.}(2017)\citenamefont
  {Sirunyan} \emph {et~al.}}]{CMS:2017wcz}%
  \BibitemOpen
  \bibfield  {author} {\bibinfo {author} {\bibfnamefont {A.~M.}\ \bibnamefont
  {Sirunyan}} \emph {et~al.} (\bibinfo {collaboration} {CMS}),\ }\bibfield
  {title} {\bibinfo {title} {{Search for associated production of a Z boson
  with a single top quark and for tZ flavour-changing interactions in pp
  collisions at $ \sqrt{s}=8 $ TeV}},\ }\href
  {https://doi.org/10.1007/JHEP07(2017)003} {\bibfield  {journal} {\bibinfo
  {journal} {JHEP}\ }\textbf {\bibinfo {volume} {07}},\ \bibinfo {pages}
  {003}},\ \Eprint {https://arxiv.org/abs/1702.01404} {arXiv:1702.01404
  [hep-ex]} \BibitemShut {NoStop}%
\bibitem [{\citenamefont {Abada}\ \emph {et~al.}(2019)\citenamefont {Abada}
  \emph {et~al.}}]{FCC:2018byv}%
  \BibitemOpen
  \bibfield  {author} {\bibinfo {author} {\bibfnamefont {A.}~\bibnamefont
  {Abada}} \emph {et~al.} (\bibinfo {collaboration} {FCC}),\ }\bibfield
  {title} {\bibinfo {title} {{FCC Physics Opportunities}: {Future Circular
  Collider Conceptual Design Report Volume 1}},\ }\href
  {https://doi.org/10.1140/epjc/s10052-019-6904-3} {\bibfield  {journal}
  {\bibinfo  {journal} {Eur. Phys. J. C}\ }\textbf {\bibinfo {volume} {79}},\
  \bibinfo {pages} {474} (\bibinfo {year} {2019})}\BibitemShut {NoStop}%
\bibitem [{\citenamefont {Davidson}\ \emph {et~al.}(2015)\citenamefont
  {Davidson}, \citenamefont {Mangano}, \citenamefont {Perries},\ and\
  \citenamefont {Sordini}}]{Davidson:2015zza}%
  \BibitemOpen
  \bibfield  {author} {\bibinfo {author} {\bibfnamefont {S.}~\bibnamefont
  {Davidson}}, \bibinfo {author} {\bibfnamefont {M.~L.}\ \bibnamefont
  {Mangano}}, \bibinfo {author} {\bibfnamefont {S.}~\bibnamefont {Perries}},\
  and\ \bibinfo {author} {\bibfnamefont {V.}~\bibnamefont {Sordini}},\
  }\bibfield  {title} {\bibinfo {title} {{Lepton Flavour Violating top decays
  at the LHC}},\ }\href {https://doi.org/10.1140/epjc/s10052-015-3649-5}
  {\bibfield  {journal} {\bibinfo  {journal} {Eur. Phys. J. C}\ }\textbf
  {\bibinfo {volume} {75}},\ \bibinfo {pages} {450} (\bibinfo {year} {2015})},\
  \Eprint {https://arxiv.org/abs/1507.07163} {arXiv:1507.07163 [hep-ph]}
  \BibitemShut {NoStop}%
\bibitem [{\citenamefont {Gottardo}(2018)}]{Gottardo:2018ptv}%
  \BibitemOpen
  \bibfield  {author} {\bibinfo {author} {\bibfnamefont {C.~A.}\ \bibnamefont
  {Gottardo}} (\bibinfo {collaboration} {ATLAS}),\ }\bibfield  {title}
  {\bibinfo {title} {{Search for charged lepton-flavour violation in top-quark
  decays at the LHC with the ATLAS detector}},\ }in\ \href@noop {} {\emph
  {\bibinfo {booktitle} {{11th International Workshop on Top Quark Physics}}}}\
  (\bibinfo {year} {2018})\ \Eprint {https://arxiv.org/abs/1809.09048}
  {arXiv:1809.09048 [hep-ex]} \BibitemShut {NoStop}%
\bibitem [{\citenamefont {Tumasyan}\ \emph {et~al.}(2022)\citenamefont
  {Tumasyan} \emph {et~al.}}]{CMS:2022ztx}%
  \BibitemOpen
  \bibfield  {author} {\bibinfo {author} {\bibfnamefont {A.}~\bibnamefont
  {Tumasyan}} \emph {et~al.} (\bibinfo {collaboration} {CMS}),\ }\bibfield
  {title} {\bibinfo {title} {{Search for charged-lepton flavor violation in top
  quark production and decay in pp collisions at $ \sqrt{s} $ = 13 TeV}},\
  }\href {https://doi.org/10.1007/JHEP06(2022)082} {\bibfield  {journal}
  {\bibinfo  {journal} {JHEP}\ }\textbf {\bibinfo {volume} {06}},\ \bibinfo
  {pages} {082}},\ \Eprint {https://arxiv.org/abs/2201.07859} {arXiv:2201.07859
  [hep-ex]} \BibitemShut {NoStop}%
\bibitem [{ATL(2023)}]{ATLAS:2023fcw}%
  \BibitemOpen
  \bibfield  {title} {\bibinfo {title} {{Search for charged-lepton-flavour
  violating $\mu\tau qt$ interactions in top-quark production and decay with
  the ATLAS detector at the LHC}},\ }\href@noop {} {\  (\bibinfo {year}
  {2023})}\BibitemShut {NoStop}%
\bibitem [{\citenamefont {Bar-Shalom}\ and\ \citenamefont
  {Wudka}(1999)}]{Bar-Shalom:1999dtk}%
  \BibitemOpen
  \bibfield  {author} {\bibinfo {author} {\bibfnamefont {S.}~\bibnamefont
  {Bar-Shalom}}\ and\ \bibinfo {author} {\bibfnamefont {J.}~\bibnamefont
  {Wudka}},\ }\bibfield  {title} {\bibinfo {title} {{Flavor changing single top
  quark production channels at e+ e- colliders in the effective Lagrangian
  description}},\ }\href {https://doi.org/10.1103/PhysRevD.60.094016}
  {\bibfield  {journal} {\bibinfo  {journal} {Phys. Rev. D}\ }\textbf {\bibinfo
  {volume} {60}},\ \bibinfo {pages} {094016} (\bibinfo {year} {1999})},\
  \Eprint {https://arxiv.org/abs/hep-ph/9905407} {arXiv:hep-ph/9905407}
  \BibitemShut {NoStop}%
\bibitem [{\citenamefont {Bause}\ \emph {et~al.}(2022)\citenamefont {Bause},
  \citenamefont {Gisbert}, \citenamefont {Golz},\ and\ \citenamefont
  {Hiller}}]{Bause:2020auq}%
  \BibitemOpen
  \bibfield  {author} {\bibinfo {author} {\bibfnamefont {R.}~\bibnamefont
  {Bause}}, \bibinfo {author} {\bibfnamefont {H.}~\bibnamefont {Gisbert}},
  \bibinfo {author} {\bibfnamefont {M.}~\bibnamefont {Golz}},\ and\ \bibinfo
  {author} {\bibfnamefont {G.}~\bibnamefont {Hiller}},\ }\bibfield  {title}
  {\bibinfo {title} {{Lepton universality and lepton flavor conservation tests
  with dineutrino modes}},\ }\href
  {https://doi.org/10.1140/epjc/s10052-022-10113-6} {\bibfield  {journal}
  {\bibinfo  {journal} {Eur. Phys. J. C}\ }\textbf {\bibinfo {volume} {82}},\
  \bibinfo {pages} {164} (\bibinfo {year} {2022})},\ \Eprint
  {https://arxiv.org/abs/2007.05001} {arXiv:2007.05001 [hep-ph]} \BibitemShut
  {NoStop}%
\bibitem [{\citenamefont {Sun}\ \emph {et~al.}(2023)\citenamefont {Sun},
  \citenamefont {Yan}, \citenamefont {Zhao},\ and\ \citenamefont
  {Zhao}}]{Sun:2023cuf}%
  \BibitemOpen
  \bibfield  {author} {\bibinfo {author} {\bibfnamefont {S.}~\bibnamefont
  {Sun}}, \bibinfo {author} {\bibfnamefont {Q.-S.}\ \bibnamefont {Yan}},
  \bibinfo {author} {\bibfnamefont {X.}~\bibnamefont {Zhao}},\ and\ \bibinfo
  {author} {\bibfnamefont {Z.}~\bibnamefont {Zhao}},\ }\bibfield  {title}
  {\bibinfo {title} {{Constraining rare B decays by $\mu^+\mu^-\to tc$ at
  future lepton colliders}},\ }\href@noop {} {\  (\bibinfo {year} {2023})},\
  \Eprint {https://arxiv.org/abs/2302.01143} {arXiv:2302.01143 [hep-ph]}
  \BibitemShut {NoStop}%
\bibitem [{Ale(2001)}]{Aleph:2001dzz}%
  \BibitemOpen
  \bibfield  {title} {\bibinfo {title} {{Search for single top production via
  ﬂavour changing neutral currents: preliminary combined results of the LEP
  experiments}},\ }\href@noop {} {\  (\bibinfo {year} {2001})},\ \bibinfo
  {note} {{DELPHI-2001-119 CONF 542}}\BibitemShut {NoStop}%
\bibitem [{\citenamefont {Achard}\ \emph {et~al.}(2002)\citenamefont {Achard}
  \emph {et~al.}}]{L3:2002hbp}%
  \BibitemOpen
  \bibfield  {author} {\bibinfo {author} {\bibfnamefont {P.}~\bibnamefont
  {Achard}} \emph {et~al.} (\bibinfo {collaboration} {L3}),\ }\bibfield
  {title} {\bibinfo {title} {{Search for single top production at LEP}},\
  }\href {https://doi.org/10.1016/S0370-2693(02)02933-7} {\bibfield  {journal}
  {\bibinfo  {journal} {Phys. Lett. B}\ }\textbf {\bibinfo {volume} {549}},\
  \bibinfo {pages} {290} (\bibinfo {year} {2002})},\ \Eprint
  {https://arxiv.org/abs/hep-ex/0210041} {arXiv:hep-ex/0210041} \BibitemShut
  {NoStop}%
\bibitem [{\citenamefont {Abdallah}\ \emph {et~al.}(2011)\citenamefont
  {Abdallah} \emph {et~al.}}]{DELPHI:2011ab}%
  \BibitemOpen
  \bibfield  {author} {\bibinfo {author} {\bibfnamefont {J.}~\bibnamefont
  {Abdallah}} \emph {et~al.} (\bibinfo {collaboration} {DELPHI}),\ }\bibfield
  {title} {\bibinfo {title} {{Search for single top quark production via
  contact interactions at LEP2}},\ }\href
  {https://doi.org/10.1140/epjc/s10052-011-1555-z} {\bibfield  {journal}
  {\bibinfo  {journal} {Eur. Phys. J. C}\ }\textbf {\bibinfo {volume} {71}},\
  \bibinfo {pages} {1555} (\bibinfo {year} {2011})},\ \Eprint
  {https://arxiv.org/abs/1102.4455} {arXiv:1102.4455 [hep-ex]} \BibitemShut
  {NoStop}%
\bibitem [{\citenamefont {Cheng}\ \emph {et~al.}(2022)\citenamefont {Cheng}
  \emph {et~al.}}]{CEPCPhysicsStudyGroup:2022uwl}%
  \BibitemOpen
  \bibfield  {author} {\bibinfo {author} {\bibfnamefont {H.}~\bibnamefont
  {Cheng}} \emph {et~al.} (\bibinfo {collaboration} {CEPC Physics Study
  Group}),\ }\bibfield  {title} {\bibinfo {title} {{The Physics potential of
  the CEPC. Prepared for the US Snowmass Community Planning Exercise (Snowmass
  2021)}},\ }in\ \href@noop {} {\emph {\bibinfo {booktitle} {{2022 Snowmass
  Summer Study}}}}\ (\bibinfo {year} {2022})\ \Eprint
  {https://arxiv.org/abs/2205.08553} {arXiv:2205.08553 [hep-ph]} \BibitemShut
  {NoStop}%
\bibitem [{\citenamefont {Bernardi}\ \emph {et~al.}(2022)\citenamefont
  {Bernardi} \emph {et~al.}}]{Bernardi:2022hny}%
  \BibitemOpen
  \bibfield  {author} {\bibinfo {author} {\bibfnamefont {G.}~\bibnamefont
  {Bernardi}} \emph {et~al.},\ }\bibfield  {title} {\bibinfo {title} {{The
  Future Circular Collider: a Summary for the US 2021 Snowmass Process}},\
  }\href@noop {} {\  (\bibinfo {year} {2022})},\ \Eprint
  {https://arxiv.org/abs/2203.06520} {arXiv:2203.06520 [hep-ex]} \BibitemShut
  {NoStop}%
\bibitem [{\citenamefont {Aryshev}\ \emph {et~al.}(2022)\citenamefont {Aryshev}
  \emph {et~al.}}]{ILCInternationalDevelopmentTeam:2022izu}%
  \BibitemOpen
  \bibfield  {author} {\bibinfo {author} {\bibfnamefont {A.}~\bibnamefont
  {Aryshev}} \emph {et~al.} (\bibinfo {collaboration} {ILC International
  Development Team}),\ }\bibfield  {title} {\bibinfo {title} {{The
  International Linear Collider: Report to Snowmass 2021}},\ }\href@noop {} {\
  (\bibinfo {year} {2022})},\ \Eprint {https://arxiv.org/abs/2203.07622}
  {arXiv:2203.07622 [physics.acc-ph]} \BibitemShut {NoStop}%
\bibitem [{\citenamefont {Al~Ali}\ \emph {et~al.}(2022)\citenamefont {Al~Ali}
  \emph {et~al.}}]{AlAli:2021let}%
  \BibitemOpen
  \bibfield  {author} {\bibinfo {author} {\bibfnamefont {H.}~\bibnamefont
  {Al~Ali}} \emph {et~al.},\ }\bibfield  {title} {\bibinfo {title} {{The muon
  Smasher\textquoteright{}s guide}},\ }\href
  {https://doi.org/10.1088/1361-6633/ac6678} {\bibfield  {journal} {\bibinfo
  {journal} {Rept. Prog. Phys.}\ }\textbf {\bibinfo {volume} {85}},\ \bibinfo
  {pages} {084201} (\bibinfo {year} {2022})},\ \Eprint
  {https://arxiv.org/abs/2103.14043} {arXiv:2103.14043 [hep-ph]} \BibitemShut
  {NoStop}%
\bibitem [{\citenamefont {Aaron}\ \emph {et~al.}(2009)\citenamefont {Aaron}
  \emph {et~al.}}]{H1:2009yuy}%
  \BibitemOpen
  \bibfield  {author} {\bibinfo {author} {\bibfnamefont {F.~D.}\ \bibnamefont
  {Aaron}} \emph {et~al.} (\bibinfo {collaboration} {H1}),\ }\bibfield  {title}
  {\bibinfo {title} {{Search for Single Top Quark Production at HERA}},\ }\href
  {https://doi.org/10.1016/j.physletb.2009.06.057} {\bibfield  {journal}
  {\bibinfo  {journal} {Phys. Lett. B}\ }\textbf {\bibinfo {volume} {678}},\
  \bibinfo {pages} {450} (\bibinfo {year} {2009})},\ \Eprint
  {https://arxiv.org/abs/0904.3876} {arXiv:0904.3876 [hep-ex]} \BibitemShut
  {NoStop}%
\bibitem [{\citenamefont {Abramowicz}\ \emph {et~al.}(2012)\citenamefont
  {Abramowicz} \emph {et~al.}}]{ZEUS:2011mya}%
  \BibitemOpen
  \bibfield  {author} {\bibinfo {author} {\bibfnamefont {H.}~\bibnamefont
  {Abramowicz}} \emph {et~al.} (\bibinfo {collaboration} {ZEUS}),\ }\bibfield
  {title} {\bibinfo {title} {{Search for single-top production in $ep$
  collisions at HERA}},\ }\href
  {https://doi.org/10.1016/j.physletb.2012.01.025} {\bibfield  {journal}
  {\bibinfo  {journal} {Phys. Lett. B}\ }\textbf {\bibinfo {volume} {708}},\
  \bibinfo {pages} {27} (\bibinfo {year} {2012})},\ \Eprint
  {https://arxiv.org/abs/1111.3901} {arXiv:1111.3901 [hep-ex]} \BibitemShut
  {NoStop}%
\bibitem [{\citenamefont {Greljo}\ and\ \citenamefont
  {Marzocca}(2017)}]{Greljo:2017vvb}%
  \BibitemOpen
  \bibfield  {author} {\bibinfo {author} {\bibfnamefont {A.}~\bibnamefont
  {Greljo}}\ and\ \bibinfo {author} {\bibfnamefont {D.}~\bibnamefont
  {Marzocca}},\ }\bibfield  {title} {\bibinfo {title} {{High-$p_T$ dilepton
  tails and flavor physics}},\ }\href
  {https://doi.org/10.1140/epjc/s10052-017-5119-8} {\bibfield  {journal}
  {\bibinfo  {journal} {Eur. Phys. J. C}\ }\textbf {\bibinfo {volume} {77}},\
  \bibinfo {pages} {548} (\bibinfo {year} {2017})},\ \Eprint
  {https://arxiv.org/abs/1704.09015} {arXiv:1704.09015 [hep-ph]} \BibitemShut
  {NoStop}%
\bibitem [{\citenamefont {Aad}\ \emph {et~al.}(2020)\citenamefont {Aad} \emph
  {et~al.}}]{Aad:2020otl}%
  \BibitemOpen
  \bibfield  {author} {\bibinfo {author} {\bibfnamefont {G.}~\bibnamefont
  {Aad}} \emph {et~al.} (\bibinfo {collaboration} {ATLAS}),\ }\bibfield
  {title} {\bibinfo {title} {{Search for new non-resonant phenomena in
  high-mass dilepton final states with the ATLAS detector}},\ }\href
  {https://doi.org/10.1007/JHEP11(2020)005} {\bibfield  {journal} {\bibinfo
  {journal} {JHEP}\ }\textbf {\bibinfo {volume} {11}},\ \bibinfo {pages}
  {005}},\ \bibinfo {note} {[Erratum: JHEP 04, 142 (2021)]},\ \Eprint
  {https://arxiv.org/abs/2006.12946} {arXiv:2006.12946 [hep-ex]} \BibitemShut
  {NoStop}%
\bibitem [{\citenamefont {Sirunyan}\ \emph
  {et~al.}(2021{\natexlab{b}})\citenamefont {Sirunyan} \emph
  {et~al.}}]{CMS:2021ctt}%
  \BibitemOpen
  \bibfield  {author} {\bibinfo {author} {\bibfnamefont {A.~M.}\ \bibnamefont
  {Sirunyan}} \emph {et~al.} (\bibinfo {collaboration} {CMS}),\ }\bibfield
  {title} {\bibinfo {title} {{Search for resonant and nonresonant new phenomena
  in high-mass dilepton final states at $ \sqrt{s} $ = 13 TeV}},\ }\href
  {https://doi.org/10.1007/JHEP07(2021)208} {\bibfield  {journal} {\bibinfo
  {journal} {JHEP}\ }\textbf {\bibinfo {volume} {07}},\ \bibinfo {pages}
  {208}},\ \Eprint {https://arxiv.org/abs/2103.02708} {arXiv:2103.02708
  [hep-ex]} \BibitemShut {NoStop}%
\bibitem [{\citenamefont {Allwicher}\ \emph
  {et~al.}(2022{\natexlab{a}})\citenamefont {Allwicher}, \citenamefont
  {Faroughy}, \citenamefont {Jaffredo}, \citenamefont {Sumensari},\ and\
  \citenamefont {Wilsch}}]{Allwicher:2022gkm}%
  \BibitemOpen
  \bibfield  {author} {\bibinfo {author} {\bibfnamefont {L.}~\bibnamefont
  {Allwicher}}, \bibinfo {author} {\bibfnamefont {D.~A.}\ \bibnamefont
  {Faroughy}}, \bibinfo {author} {\bibfnamefont {F.}~\bibnamefont {Jaffredo}},
  \bibinfo {author} {\bibfnamefont {O.}~\bibnamefont {Sumensari}},\ and\
  \bibinfo {author} {\bibfnamefont {F.}~\bibnamefont {Wilsch}},\ }\bibfield
  {title} {\bibinfo {title} {{Drell-Yan Tails Beyond the Standard Model}},\
  }\href@noop {} {\  (\bibinfo {year} {2022}{\natexlab{a}})},\ \Eprint
  {https://arxiv.org/abs/2207.10714} {arXiv:2207.10714 [hep-ph]} \BibitemShut
  {NoStop}%
\bibitem [{\citenamefont {Allwicher}\ \emph
  {et~al.}(2022{\natexlab{b}})\citenamefont {Allwicher}, \citenamefont
  {Faroughy}, \citenamefont {Jaffredo}, \citenamefont {Sumensari},\ and\
  \citenamefont {Wilsch}}]{Allwicher:2022mcg}%
  \BibitemOpen
  \bibfield  {author} {\bibinfo {author} {\bibfnamefont {L.}~\bibnamefont
  {Allwicher}}, \bibinfo {author} {\bibfnamefont {D.~A.}\ \bibnamefont
  {Faroughy}}, \bibinfo {author} {\bibfnamefont {F.}~\bibnamefont {Jaffredo}},
  \bibinfo {author} {\bibfnamefont {O.}~\bibnamefont {Sumensari}},\ and\
  \bibinfo {author} {\bibfnamefont {F.}~\bibnamefont {Wilsch}},\ }\bibfield
  {title} {\bibinfo {title} {{HighPT: A Tool for high-$p_T$ Drell-Yan Tails
  Beyond the Standard Model}},\ }\href@noop {} {\  (\bibinfo {year}
  {2022}{\natexlab{b}})},\ \Eprint {https://arxiv.org/abs/2207.10756}
  {arXiv:2207.10756 [hep-ph]} \BibitemShut {NoStop}%
\bibitem [{\citenamefont {Greljo}\ \emph {et~al.}(2022)\citenamefont {Greljo},
  \citenamefont {Salko}, \citenamefont {Smolkovi\v{c}},\ and\ \citenamefont
  {Stangl}}]{Greljo:2022jac}%
  \BibitemOpen
  \bibfield  {author} {\bibinfo {author} {\bibfnamefont {A.}~\bibnamefont
  {Greljo}}, \bibinfo {author} {\bibfnamefont {J.}~\bibnamefont {Salko}},
  \bibinfo {author} {\bibfnamefont {A.}~\bibnamefont {Smolkovi\v{c}}},\ and\
  \bibinfo {author} {\bibfnamefont {P.}~\bibnamefont {Stangl}},\ }\bibfield
  {title} {\bibinfo {title} {{Rare $b$ decays meet high-mass Drell-Yan}},\
  }\href@noop {} {\  (\bibinfo {year} {2022})},\ \Eprint
  {https://arxiv.org/abs/2212.10497} {arXiv:2212.10497 [hep-ph]} \BibitemShut
  {NoStop}%
\bibitem [{\citenamefont {Harland-Lang}\ \emph {et~al.}(2015)\citenamefont
  {Harland-Lang}, \citenamefont {Martin}, \citenamefont {Motylinski},\ and\
  \citenamefont {Thorne}}]{Harland-Lang:2014zoa}%
  \BibitemOpen
  \bibfield  {author} {\bibinfo {author} {\bibfnamefont {L.~A.}\ \bibnamefont
  {Harland-Lang}}, \bibinfo {author} {\bibfnamefont {A.~D.}\ \bibnamefont
  {Martin}}, \bibinfo {author} {\bibfnamefont {P.}~\bibnamefont {Motylinski}},\
  and\ \bibinfo {author} {\bibfnamefont {R.~S.}\ \bibnamefont {Thorne}},\
  }\bibfield  {title} {\bibinfo {title} {{Parton distributions in the LHC era:
  MMHT 2014 PDFs}},\ }\href {https://doi.org/10.1140/epjc/s10052-015-3397-6}
  {\bibfield  {journal} {\bibinfo  {journal} {Eur. Phys. J. C}\ }\textbf
  {\bibinfo {volume} {75}},\ \bibinfo {pages} {204} (\bibinfo {year} {2015})},\
  \Eprint {https://arxiv.org/abs/1412.3989} {arXiv:1412.3989 [hep-ph]}
  \BibitemShut {NoStop}%
\bibitem [{\citenamefont {Alwall}\ \emph {et~al.}(2014)\citenamefont {Alwall},
  \citenamefont {Frederix}, \citenamefont {Frixione}, \citenamefont {Hirschi},
  \citenamefont {Maltoni}, \citenamefont {Mattelaer}, \citenamefont {Shao},
  \citenamefont {Stelzer}, \citenamefont {Torrielli},\ and\ \citenamefont
  {Zaro}}]{Alwall:2014hca}%
  \BibitemOpen
  \bibfield  {author} {\bibinfo {author} {\bibfnamefont {J.}~\bibnamefont
  {Alwall}}, \bibinfo {author} {\bibfnamefont {R.}~\bibnamefont {Frederix}},
  \bibinfo {author} {\bibfnamefont {S.}~\bibnamefont {Frixione}}, \bibinfo
  {author} {\bibfnamefont {V.}~\bibnamefont {Hirschi}}, \bibinfo {author}
  {\bibfnamefont {F.}~\bibnamefont {Maltoni}}, \bibinfo {author} {\bibfnamefont
  {O.}~\bibnamefont {Mattelaer}}, \bibinfo {author} {\bibfnamefont {H.~S.}\
  \bibnamefont {Shao}}, \bibinfo {author} {\bibfnamefont {T.}~\bibnamefont
  {Stelzer}}, \bibinfo {author} {\bibfnamefont {P.}~\bibnamefont {Torrielli}},\
  and\ \bibinfo {author} {\bibfnamefont {M.}~\bibnamefont {Zaro}},\ }\bibfield
  {title} {\bibinfo {title} {{The automated computation of tree-level and
  next-to-leading order differential cross sections, and their matching to
  parton shower simulations}},\ }\href
  {https://doi.org/10.1007/JHEP07(2014)079} {\bibfield  {journal} {\bibinfo
  {journal} {JHEP}\ }\textbf {\bibinfo {volume} {07}},\ \bibinfo {pages}
  {079}},\ \Eprint {https://arxiv.org/abs/1405.0301} {arXiv:1405.0301 [hep-ph]}
  \BibitemShut {NoStop}%
\bibitem [{CMS(2022{\natexlab{a}})}]{CMS:2022gho}%
  \BibitemOpen
  \bibfield  {title} {\bibinfo {title} {{Sensitivity projections for a search
  for new phenomena at high dilepton mass for the LHC Run 3 and the HL-LHC}},\
  }\href@noop {} {\  (\bibinfo {year} {2022}{\natexlab{a}})}\BibitemShut
  {NoStop}%
\bibitem [{\citenamefont {Schael}\ \emph {et~al.}(2013)\citenamefont {Schael}
  \emph {et~al.}}]{ALEPH:2013dgf}%
  \BibitemOpen
  \bibfield  {author} {\bibinfo {author} {\bibfnamefont {S.}~\bibnamefont
  {Schael}} \emph {et~al.} (\bibinfo {collaboration} {ALEPH, DELPHI, L3, OPAL,
  LEP Electroweak}),\ }\bibfield  {title} {\bibinfo {title} {{Electroweak
  Measurements in Electron-Positron Collisions at W-Boson-Pair Energies at
  LEP}},\ }\href {https://doi.org/10.1016/j.physrep.2013.07.004} {\bibfield
  {journal} {\bibinfo  {journal} {Phys. Rept.}\ }\textbf {\bibinfo {volume}
  {532}},\ \bibinfo {pages} {119} (\bibinfo {year} {2013})},\ \Eprint
  {https://arxiv.org/abs/1302.3415} {arXiv:1302.3415 [hep-ex]} \BibitemShut
  {NoStop}%
\bibitem [{\citenamefont {Angelescu}\ \emph {et~al.}(2020)\citenamefont
  {Angelescu}, \citenamefont {Faroughy},\ and\ \citenamefont
  {Sumensari}}]{Angelescu:2020uug}%
  \BibitemOpen
  \bibfield  {author} {\bibinfo {author} {\bibfnamefont {A.}~\bibnamefont
  {Angelescu}}, \bibinfo {author} {\bibfnamefont {D.~A.}\ \bibnamefont
  {Faroughy}},\ and\ \bibinfo {author} {\bibfnamefont {O.}~\bibnamefont
  {Sumensari}},\ }\bibfield  {title} {\bibinfo {title} {{Lepton Flavor
  Violation and Dilepton Tails at the LHC}},\ }\href
  {https://doi.org/10.1140/epjc/s10052-020-8210-5} {\bibfield  {journal}
  {\bibinfo  {journal} {Eur. Phys. J. C}\ }\textbf {\bibinfo {volume} {80}},\
  \bibinfo {pages} {641} (\bibinfo {year} {2020})},\ \Eprint
  {https://arxiv.org/abs/2002.05684} {arXiv:2002.05684 [hep-ph]} \BibitemShut
  {NoStop}%
\bibitem [{\citenamefont {Safronova}\ \emph {et~al.}(2018)\citenamefont
  {Safronova}, \citenamefont {Budker}, \citenamefont {DeMille}, \citenamefont
  {Kimball}, \citenamefont {Derevianko},\ and\ \citenamefont
  {Clark}}]{Safronova:2017xyt}%
  \BibitemOpen
  \bibfield  {author} {\bibinfo {author} {\bibfnamefont {M.~S.}\ \bibnamefont
  {Safronova}}, \bibinfo {author} {\bibfnamefont {D.}~\bibnamefont {Budker}},
  \bibinfo {author} {\bibfnamefont {D.}~\bibnamefont {DeMille}}, \bibinfo
  {author} {\bibfnamefont {D.~F.~J.}\ \bibnamefont {Kimball}}, \bibinfo
  {author} {\bibfnamefont {A.}~\bibnamefont {Derevianko}},\ and\ \bibinfo
  {author} {\bibfnamefont {C.~W.}\ \bibnamefont {Clark}},\ }\bibfield  {title}
  {\bibinfo {title} {{Search for New Physics with Atoms and Molecules}},\
  }\href {https://doi.org/10.1103/RevModPhys.90.025008} {\bibfield  {journal}
  {\bibinfo  {journal} {Rev. Mod. Phys.}\ }\textbf {\bibinfo {volume} {90}},\
  \bibinfo {pages} {025008} (\bibinfo {year} {2018})},\ \Eprint
  {https://arxiv.org/abs/1710.01833} {arXiv:1710.01833 [physics.atom-ph]}
  \BibitemShut {NoStop}%
\bibitem [{\citenamefont {Wood}\ \emph {et~al.}(1997)\citenamefont {Wood},
  \citenamefont {Bennett}, \citenamefont {Cho}, \citenamefont {Masterson},
  \citenamefont {Roberts}, \citenamefont {Tanner},\ and\ \citenamefont
  {Wieman}}]{Wood:1997zq}%
  \BibitemOpen
  \bibfield  {author} {\bibinfo {author} {\bibfnamefont {C.~S.}\ \bibnamefont
  {Wood}}, \bibinfo {author} {\bibfnamefont {S.~C.}\ \bibnamefont {Bennett}},
  \bibinfo {author} {\bibfnamefont {D.}~\bibnamefont {Cho}}, \bibinfo {author}
  {\bibfnamefont {B.~P.}\ \bibnamefont {Masterson}}, \bibinfo {author}
  {\bibfnamefont {J.~L.}\ \bibnamefont {Roberts}}, \bibinfo {author}
  {\bibfnamefont {C.~E.}\ \bibnamefont {Tanner}},\ and\ \bibinfo {author}
  {\bibfnamefont {C.~E.}\ \bibnamefont {Wieman}},\ }\bibfield  {title}
  {\bibinfo {title} {{Measurement of parity nonconservation and an anapole
  moment in cesium}},\ }\href {https://doi.org/10.1126/science.275.5307.1759}
  {\bibfield  {journal} {\bibinfo  {journal} {Science}\ }\textbf {\bibinfo
  {volume} {275}},\ \bibinfo {pages} {1759} (\bibinfo {year}
  {1997})}\BibitemShut {NoStop}%
\bibitem [{\citenamefont {Sahoo}\ \emph {et~al.}(2021)\citenamefont {Sahoo},
  \citenamefont {Das},\ and\ \citenamefont {Spiesberger}}]{Sahoo:2021thl}%
  \BibitemOpen
  \bibfield  {author} {\bibinfo {author} {\bibfnamefont {B.~K.}\ \bibnamefont
  {Sahoo}}, \bibinfo {author} {\bibfnamefont {B.~P.}\ \bibnamefont {Das}},\
  and\ \bibinfo {author} {\bibfnamefont {H.}~\bibnamefont {Spiesberger}},\
  }\bibfield  {title} {\bibinfo {title} {{New physics constraints from atomic
  parity violation in Cs133}},\ }\href
  {https://doi.org/10.1103/PhysRevD.103.L111303} {\bibfield  {journal}
  {\bibinfo  {journal} {Phys. Rev. D}\ }\textbf {\bibinfo {volume} {103}},\
  \bibinfo {pages} {L111303} (\bibinfo {year} {2021})},\ \Eprint
  {https://arxiv.org/abs/2101.10095} {arXiv:2101.10095 [hep-ph]} \BibitemShut
  {NoStop}%
\bibitem [{\citenamefont {Becker}\ \emph {et~al.}(2018)\citenamefont {Becker}
  \emph {et~al.}}]{Becker:2018ggl}%
  \BibitemOpen
  \bibfield  {author} {\bibinfo {author} {\bibfnamefont {D.}~\bibnamefont
  {Becker}} \emph {et~al.},\ }\bibfield  {title} {\bibinfo {title} {{The P2
  experiment}},\ }\href {https://doi.org/10.1140/epja/i2018-12611-6} {\bibfield
   {journal} {\bibinfo  {journal} {Eur. Phys. J. A}\ }\textbf {\bibinfo
  {volume} {54}},\ \bibinfo {pages} {208} (\bibinfo {year} {2018})},\ \Eprint
  {https://arxiv.org/abs/1802.04759} {arXiv:1802.04759 [nucl-ex]} \BibitemShut
  {NoStop}%
\bibitem [{\citenamefont {Dev}\ \emph {et~al.}(2021)\citenamefont {Dev},
  \citenamefont {Rodejohann}, \citenamefont {Xu},\ and\ \citenamefont
  {Zhang}}]{Dev:2021otb}%
  \BibitemOpen
  \bibfield  {author} {\bibinfo {author} {\bibfnamefont {P.~S.~B.}\
  \bibnamefont {Dev}}, \bibinfo {author} {\bibfnamefont {W.}~\bibnamefont
  {Rodejohann}}, \bibinfo {author} {\bibfnamefont {X.-J.}\ \bibnamefont {Xu}},\
  and\ \bibinfo {author} {\bibfnamefont {Y.}~\bibnamefont {Zhang}},\ }\bibfield
   {title} {\bibinfo {title} {{Searching for Z' bosons at the P2 experiment}},\
  }\href {https://doi.org/10.1007/JHEP06(2021)039} {\bibfield  {journal}
  {\bibinfo  {journal} {JHEP}\ }\textbf {\bibinfo {volume} {06}},\ \bibinfo
  {pages} {039}},\ \Eprint {https://arxiv.org/abs/2103.09067} {arXiv:2103.09067
  [hep-ph]} \BibitemShut {NoStop}%
\bibitem [{\citenamefont {Bischer}\ \emph {et~al.}(2022)\citenamefont
  {Bischer}, \citenamefont {Dev}, \citenamefont {Rodejohann}, \citenamefont
  {Xu},\ and\ \citenamefont {Zhang}}]{Bischer:2021jqn}%
  \BibitemOpen
  \bibfield  {author} {\bibinfo {author} {\bibfnamefont {I.}~\bibnamefont
  {Bischer}}, \bibinfo {author} {\bibfnamefont {P.~S.~B.}\ \bibnamefont {Dev}},
  \bibinfo {author} {\bibfnamefont {W.}~\bibnamefont {Rodejohann}}, \bibinfo
  {author} {\bibfnamefont {X.-J.}\ \bibnamefont {Xu}},\ and\ \bibinfo {author}
  {\bibfnamefont {Y.}~\bibnamefont {Zhang}},\ }\bibfield  {title} {\bibinfo
  {title} {{Searching for new physics from SMEFT and leptoquarks at the P2
  experiment}},\ }\href {https://doi.org/10.1103/PhysRevD.105.095016}
  {\bibfield  {journal} {\bibinfo  {journal} {Phys. Rev. D}\ }\textbf {\bibinfo
  {volume} {105}},\ \bibinfo {pages} {095016} (\bibinfo {year} {2022})},\
  \Eprint {https://arxiv.org/abs/2112.12051} {arXiv:2112.12051 [hep-ph]}
  \BibitemShut {NoStop}%
\bibitem [{\citenamefont {Bause}\ \emph {et~al.}(2021)\citenamefont {Bause},
  \citenamefont {Gisbert}, \citenamefont {Golz},\ and\ \citenamefont
  {Hiller}}]{Bause:2021cna}%
  \BibitemOpen
  \bibfield  {author} {\bibinfo {author} {\bibfnamefont {R.}~\bibnamefont
  {Bause}}, \bibinfo {author} {\bibfnamefont {H.}~\bibnamefont {Gisbert}},
  \bibinfo {author} {\bibfnamefont {M.}~\bibnamefont {Golz}},\ and\ \bibinfo
  {author} {\bibfnamefont {G.}~\bibnamefont {Hiller}},\ }\bibfield  {title}
  {\bibinfo {title} {{Interplay of dineutrino modes with semileptonic rare
  B-decays}},\ }\href {https://doi.org/10.1007/JHEP12(2021)061} {\bibfield
  {journal} {\bibinfo  {journal} {JHEP}\ }\textbf {\bibinfo {volume} {12}},\
  \bibinfo {pages} {061}},\ \Eprint {https://arxiv.org/abs/2109.01675}
  {arXiv:2109.01675 [hep-ph]} \BibitemShut {NoStop}%
\bibitem [{\citenamefont {Altmannshofer}\ and\ \citenamefont
  {Archilli}(2022)}]{Altmannshofer:2022hfs}%
  \BibitemOpen
  \bibfield  {author} {\bibinfo {author} {\bibfnamefont {W.}~\bibnamefont
  {Altmannshofer}}\ and\ \bibinfo {author} {\bibfnamefont {F.}~\bibnamefont
  {Archilli}},\ }\bibfield  {title} {\bibinfo {title} {{Rare decays of b and c
  hadrons}},\ }in\ \href@noop {} {\emph {\bibinfo {booktitle} {{2022 Snowmass
  Summer Study}}}}\ (\bibinfo {year} {2022})\ \Eprint
  {https://arxiv.org/abs/2206.11331} {arXiv:2206.11331 [hep-ph]} \BibitemShut
  {NoStop}%
\bibitem [{\citenamefont {Aebischer}\ \emph {et~al.}(2016)\citenamefont
  {Aebischer}, \citenamefont {Crivellin}, \citenamefont {Fael},\ and\
  \citenamefont {Greub}}]{Aebischer:2015fzz}%
  \BibitemOpen
  \bibfield  {author} {\bibinfo {author} {\bibfnamefont {J.}~\bibnamefont
  {Aebischer}}, \bibinfo {author} {\bibfnamefont {A.}~\bibnamefont
  {Crivellin}}, \bibinfo {author} {\bibfnamefont {M.}~\bibnamefont {Fael}},\
  and\ \bibinfo {author} {\bibfnamefont {C.}~\bibnamefont {Greub}},\ }\bibfield
   {title} {\bibinfo {title} {{Matching of gauge invariant dimension-six
  operators for $b\to s$ and $b\to c$ transitions}},\ }\href
  {https://doi.org/10.1007/JHEP05(2016)037} {\bibfield  {journal} {\bibinfo
  {journal} {JHEP}\ }\textbf {\bibinfo {volume} {05}},\ \bibinfo {pages}
  {037}},\ \Eprint {https://arxiv.org/abs/1512.02830} {arXiv:1512.02830
  [hep-ph]} \BibitemShut {NoStop}%
\bibitem [{\citenamefont {Celis}\ \emph {et~al.}(2017)\citenamefont {Celis},
  \citenamefont {Fuentes-Martin}, \citenamefont {Vicente},\ and\ \citenamefont
  {Virto}}]{Celis:2017doq}%
  \BibitemOpen
  \bibfield  {author} {\bibinfo {author} {\bibfnamefont {A.}~\bibnamefont
  {Celis}}, \bibinfo {author} {\bibfnamefont {J.}~\bibnamefont
  {Fuentes-Martin}}, \bibinfo {author} {\bibfnamefont {A.}~\bibnamefont
  {Vicente}},\ and\ \bibinfo {author} {\bibfnamefont {J.}~\bibnamefont
  {Virto}},\ }\bibfield  {title} {\bibinfo {title} {{Gauge-invariant
  implications of the LHCb measurements on lepton-flavor nonuniversality}},\
  }\href {https://doi.org/10.1103/PhysRevD.96.035026} {\bibfield  {journal}
  {\bibinfo  {journal} {Phys. Rev. D}\ }\textbf {\bibinfo {volume} {96}},\
  \bibinfo {pages} {035026} (\bibinfo {year} {2017})},\ \Eprint
  {https://arxiv.org/abs/1704.05672} {arXiv:1704.05672 [hep-ph]} \BibitemShut
  {NoStop}%
\bibitem [{\citenamefont {Camargo-Molina}\ \emph {et~al.}(2018)\citenamefont
  {Camargo-Molina}, \citenamefont {Celis},\ and\ \citenamefont
  {Faroughy}}]{Camargo-Molina:2018cwu}%
  \BibitemOpen
  \bibfield  {author} {\bibinfo {author} {\bibfnamefont {J.~E.}\ \bibnamefont
  {Camargo-Molina}}, \bibinfo {author} {\bibfnamefont {A.}~\bibnamefont
  {Celis}},\ and\ \bibinfo {author} {\bibfnamefont {D.~A.}\ \bibnamefont
  {Faroughy}},\ }\bibfield  {title} {\bibinfo {title} {{Anomalies in Bottom
  from new physics in Top}},\ }\href
  {https://doi.org/10.1016/j.physletb.2018.07.051} {\bibfield  {journal}
  {\bibinfo  {journal} {Phys. Lett. B}\ }\textbf {\bibinfo {volume} {784}},\
  \bibinfo {pages} {284} (\bibinfo {year} {2018})},\ \Eprint
  {https://arxiv.org/abs/1805.04917} {arXiv:1805.04917 [hep-ph]} \BibitemShut
  {NoStop}%
\bibitem [{\citenamefont {Bi\ss{}mann}\ \emph {et~al.}(2020)\citenamefont
  {Bi\ss{}mann}, \citenamefont {Erdmann}, \citenamefont {Grunwald},
  \citenamefont {Hiller},\ and\ \citenamefont
  {Kr\"oninger}}]{Bissmann:2019gfc}%
  \BibitemOpen
  \bibfield  {author} {\bibinfo {author} {\bibfnamefont {S.}~\bibnamefont
  {Bi\ss{}mann}}, \bibinfo {author} {\bibfnamefont {J.}~\bibnamefont
  {Erdmann}}, \bibinfo {author} {\bibfnamefont {C.}~\bibnamefont {Grunwald}},
  \bibinfo {author} {\bibfnamefont {G.}~\bibnamefont {Hiller}},\ and\ \bibinfo
  {author} {\bibfnamefont {K.}~\bibnamefont {Kr\"oninger}},\ }\bibfield
  {title} {\bibinfo {title} {{Constraining top-quark couplings combining
  top-quark and $\boldsymbol{B}$ decay observables}},\ }\href
  {https://doi.org/10.1140/epjc/s10052-020-7680-9} {\bibfield  {journal}
  {\bibinfo  {journal} {Eur. Phys. J. C}\ }\textbf {\bibinfo {volume} {80}},\
  \bibinfo {pages} {136} (\bibinfo {year} {2020})},\ \Eprint
  {https://arxiv.org/abs/1909.13632} {arXiv:1909.13632 [hep-ph]} \BibitemShut
  {NoStop}%
\bibitem [{\citenamefont {Aoude}\ \emph {et~al.}(2020)\citenamefont {Aoude},
  \citenamefont {Hurth}, \citenamefont {Renner},\ and\ \citenamefont
  {Shepherd}}]{Aoude:2020dwv}%
  \BibitemOpen
  \bibfield  {author} {\bibinfo {author} {\bibfnamefont {R.}~\bibnamefont
  {Aoude}}, \bibinfo {author} {\bibfnamefont {T.}~\bibnamefont {Hurth}},
  \bibinfo {author} {\bibfnamefont {S.}~\bibnamefont {Renner}},\ and\ \bibinfo
  {author} {\bibfnamefont {W.}~\bibnamefont {Shepherd}},\ }\bibfield  {title}
  {\bibinfo {title} {{The impact of flavour data on global fits of the MFV
  SMEFT}},\ }\href {https://doi.org/10.1007/JHEP12(2020)113} {\bibfield
  {journal} {\bibinfo  {journal} {JHEP}\ }\textbf {\bibinfo {volume} {12}},\
  \bibinfo {pages} {113}},\ \Eprint {https://arxiv.org/abs/2003.05432}
  {arXiv:2003.05432 [hep-ph]} \BibitemShut {NoStop}%
\bibitem [{\citenamefont {Bi\ss{}mann}\ \emph {et~al.}(2021)\citenamefont
  {Bi\ss{}mann}, \citenamefont {Grunwald}, \citenamefont {Hiller},\ and\
  \citenamefont {Kr\"oninger}}]{Bissmann:2020mfi}%
  \BibitemOpen
  \bibfield  {author} {\bibinfo {author} {\bibfnamefont {S.}~\bibnamefont
  {Bi\ss{}mann}}, \bibinfo {author} {\bibfnamefont {C.}~\bibnamefont
  {Grunwald}}, \bibinfo {author} {\bibfnamefont {G.}~\bibnamefont {Hiller}},\
  and\ \bibinfo {author} {\bibfnamefont {K.}~\bibnamefont {Kr\"oninger}},\
  }\bibfield  {title} {\bibinfo {title} {{Top and Beauty synergies in
  SMEFT-fits at present and future colliders}},\ }\href
  {https://doi.org/10.1007/JHEP06(2021)010} {\bibfield  {journal} {\bibinfo
  {journal} {JHEP}\ }\textbf {\bibinfo {volume} {06}},\ \bibinfo {pages}
  {010}},\ \Eprint {https://arxiv.org/abs/2012.10456} {arXiv:2012.10456
  [hep-ph]} \BibitemShut {NoStop}%
\bibitem [{\citenamefont {Bruggisser}\ \emph {et~al.}(2021)\citenamefont
  {Bruggisser}, \citenamefont {Sch\"afer}, \citenamefont {van Dyk},\ and\
  \citenamefont {Westhoff}}]{Bruggisser:2021duo}%
  \BibitemOpen
  \bibfield  {author} {\bibinfo {author} {\bibfnamefont {S.}~\bibnamefont
  {Bruggisser}}, \bibinfo {author} {\bibfnamefont {R.}~\bibnamefont
  {Sch\"afer}}, \bibinfo {author} {\bibfnamefont {D.}~\bibnamefont {van Dyk}},\
  and\ \bibinfo {author} {\bibfnamefont {S.}~\bibnamefont {Westhoff}},\
  }\bibfield  {title} {\bibinfo {title} {{The Flavor of UV Physics}},\ }\href
  {https://doi.org/10.1007/JHEP05(2021)257} {\bibfield  {journal} {\bibinfo
  {journal} {JHEP}\ }\textbf {\bibinfo {volume} {05}},\ \bibinfo {pages}
  {257}},\ \Eprint {https://arxiv.org/abs/2101.07273} {arXiv:2101.07273
  [hep-ph]} \BibitemShut {NoStop}%
\bibitem [{\citenamefont {Altmannshofer}\ and\ \citenamefont
  {Stangl}(2021)}]{Altmannshofer:2021qrr}%
  \BibitemOpen
  \bibfield  {author} {\bibinfo {author} {\bibfnamefont {W.}~\bibnamefont
  {Altmannshofer}}\ and\ \bibinfo {author} {\bibfnamefont {P.}~\bibnamefont
  {Stangl}},\ }\bibfield  {title} {\bibinfo {title} {{New physics in rare B
  decays after Moriond 2021}},\ }\href
  {https://doi.org/10.1140/epjc/s10052-021-09725-1} {\bibfield  {journal}
  {\bibinfo  {journal} {Eur. Phys. J. C}\ }\textbf {\bibinfo {volume} {81}},\
  \bibinfo {pages} {952} (\bibinfo {year} {2021})},\ \Eprint
  {https://arxiv.org/abs/2103.13370} {arXiv:2103.13370 [hep-ph]} \BibitemShut
  {NoStop}%
\bibitem [{\citenamefont {Aaij}\ \emph {et~al.}(2017)\citenamefont {Aaij} \emph
  {et~al.}}]{LHCb:2017avl}%
  \BibitemOpen
  \bibfield  {author} {\bibinfo {author} {\bibfnamefont {R.}~\bibnamefont
  {Aaij}} \emph {et~al.} (\bibinfo {collaboration} {LHCb}),\ }\bibfield
  {title} {\bibinfo {title} {{Test of lepton universality with $B^{0}
  \rightarrow K^{*0}\ell^{+}\ell^{-}$ decays}},\ }\href
  {https://doi.org/10.1007/JHEP08(2017)055} {\bibfield  {journal} {\bibinfo
  {journal} {JHEP}\ }\textbf {\bibinfo {volume} {08}},\ \bibinfo {pages}
  {055}},\ \Eprint {https://arxiv.org/abs/1705.05802} {arXiv:1705.05802
  [hep-ex]} \BibitemShut {NoStop}%
\bibitem [{\citenamefont {Aaij}\ \emph
  {et~al.}(2020{\natexlab{a}})\citenamefont {Aaij} \emph
  {et~al.}}]{LHCb:2019efc}%
  \BibitemOpen
  \bibfield  {author} {\bibinfo {author} {\bibfnamefont {R.}~\bibnamefont
  {Aaij}} \emph {et~al.} (\bibinfo {collaboration} {LHCb}),\ }\bibfield
  {title} {\bibinfo {title} {{Test of lepton universality with $
  {\Lambda}_b^0\to {pK}^{-}{\mathrm{\ell}}^{+}{\mathrm{\ell}}^{-} $ decays}},\
  }\href {https://doi.org/10.1007/JHEP05(2020)040} {\bibfield  {journal}
  {\bibinfo  {journal} {JHEP}\ }\textbf {\bibinfo {volume} {05}},\ \bibinfo
  {pages} {040}},\ \Eprint {https://arxiv.org/abs/1912.08139} {arXiv:1912.08139
  [hep-ex]} \BibitemShut {NoStop}%
\bibitem [{\citenamefont {Aaij}\ \emph
  {et~al.}(2022{\natexlab{a}})\citenamefont {Aaij} \emph
  {et~al.}}]{LHCb:2021lvy}%
  \BibitemOpen
  \bibfield  {author} {\bibinfo {author} {\bibfnamefont {R.}~\bibnamefont
  {Aaij}} \emph {et~al.} (\bibinfo {collaboration} {LHCb}),\ }\bibfield
  {title} {\bibinfo {title} {{Tests of lepton universality using $B^0\to K^0_S
  \ell^+ \ell^-$ and $B^+\to K^{*+} \ell^+ \ell^-$ decays}},\ }\href
  {https://doi.org/10.1103/PhysRevLett.128.191802} {\bibfield  {journal}
  {\bibinfo  {journal} {Phys. Rev. Lett.}\ }\textbf {\bibinfo {volume} {128}},\
  \bibinfo {pages} {191802} (\bibinfo {year} {2022}{\natexlab{a}})},\ \Eprint
  {https://arxiv.org/abs/2110.09501} {arXiv:2110.09501 [hep-ex]} \BibitemShut
  {NoStop}%
\bibitem [{\citenamefont {Aaij}\ \emph
  {et~al.}(2022{\natexlab{b}})\citenamefont {Aaij} \emph
  {et~al.}}]{LHCb:2021trn}%
  \BibitemOpen
  \bibfield  {author} {\bibinfo {author} {\bibfnamefont {R.}~\bibnamefont
  {Aaij}} \emph {et~al.} (\bibinfo {collaboration} {LHCb}),\ }\bibfield
  {title} {\bibinfo {title} {{Test of lepton universality in beauty-quark
  decays}},\ }\href {https://doi.org/10.1038/s41567-021-01478-8} {\bibfield
  {journal} {\bibinfo  {journal} {Nature Phys.}\ }\textbf {\bibinfo {volume}
  {18}},\ \bibinfo {pages} {277} (\bibinfo {year} {2022}{\natexlab{b}})},\
  \Eprint {https://arxiv.org/abs/2103.11769} {arXiv:2103.11769 [hep-ex]}
  \BibitemShut {NoStop}%
\bibitem [{\citenamefont {Aaij}\ \emph
  {et~al.}(2020{\natexlab{b}})\citenamefont {Aaij} \emph
  {et~al.}}]{LHCb:2020lmf}%
  \BibitemOpen
  \bibfield  {author} {\bibinfo {author} {\bibfnamefont {R.}~\bibnamefont
  {Aaij}} \emph {et~al.} (\bibinfo {collaboration} {LHCb}),\ }\bibfield
  {title} {\bibinfo {title} {{Measurement of $CP$-Averaged Observables in the
  $B^{0}\rightarrow K^{*0}\mu^{+}\mu^{-}$ Decay}},\ }\href
  {https://doi.org/10.1103/PhysRevLett.125.011802} {\bibfield  {journal}
  {\bibinfo  {journal} {Phys. Rev. Lett.}\ }\textbf {\bibinfo {volume} {125}},\
  \bibinfo {pages} {011802} (\bibinfo {year} {2020}{\natexlab{b}})},\ \Eprint
  {https://arxiv.org/abs/2003.04831} {arXiv:2003.04831 [hep-ex]} \BibitemShut
  {NoStop}%
\bibitem [{\citenamefont {Aaij}\ \emph {et~al.}(2014)\citenamefont {Aaij} \emph
  {et~al.}}]{LHCb:2014cxe}%
  \BibitemOpen
  \bibfield  {author} {\bibinfo {author} {\bibfnamefont {R.}~\bibnamefont
  {Aaij}} \emph {et~al.} (\bibinfo {collaboration} {LHCb}),\ }\bibfield
  {title} {\bibinfo {title} {{Differential branching fractions and isospin
  asymmetries of $B \to K^{(*)} \mu^+ \mu^-$ decays}},\ }\href
  {https://doi.org/10.1007/JHEP06(2014)133} {\bibfield  {journal} {\bibinfo
  {journal} {JHEP}\ }\textbf {\bibinfo {volume} {06}},\ \bibinfo {pages}
  {133}},\ \Eprint {https://arxiv.org/abs/1403.8044} {arXiv:1403.8044 [hep-ex]}
  \BibitemShut {NoStop}%
\bibitem [{\citenamefont {Aaij}\ \emph {et~al.}(2016)\citenamefont {Aaij} \emph
  {et~al.}}]{LHCb:2016ykl}%
  \BibitemOpen
  \bibfield  {author} {\bibinfo {author} {\bibfnamefont {R.}~\bibnamefont
  {Aaij}} \emph {et~al.} (\bibinfo {collaboration} {LHCb}),\ }\bibfield
  {title} {\bibinfo {title} {{Measurements of the S-wave fraction in
  $B^{0}\rightarrow K^{+}\pi^{-}\mu^{+}\mu^{-}$ decays and the
  $B^{0}\rightarrow K^{\ast}(892)^{0}\mu^{+}\mu^{-}$ differential branching
  fraction}},\ }\href {https://doi.org/10.1007/JHEP11(2016)047} {\bibfield
  {journal} {\bibinfo  {journal} {JHEP}\ }\textbf {\bibinfo {volume} {11}},\
  \bibinfo {pages} {047}},\ \bibinfo {note} {[Erratum: JHEP 04, 142 (2017)]},\
  \Eprint {https://arxiv.org/abs/1606.04731} {arXiv:1606.04731 [hep-ex]}
  \BibitemShut {NoStop}%
\bibitem [{\citenamefont {Aaij}\ \emph {et~al.}(2021)\citenamefont {Aaij} \emph
  {et~al.}}]{LHCb:2021zwz}%
  \BibitemOpen
  \bibfield  {author} {\bibinfo {author} {\bibfnamefont {R.}~\bibnamefont
  {Aaij}} \emph {et~al.} (\bibinfo {collaboration} {LHCb}),\ }\bibfield
  {title} {\bibinfo {title} {{Branching Fraction Measurements of the Rare
  $B^0_s\rightarrow\phi\mu^+\mu^-$ and $B^0_s\rightarrow
  f_2^\prime(1525)\mu^+\mu^-$ Decays}},\ }\href
  {https://doi.org/10.1103/PhysRevLett.127.151801} {\bibfield  {journal}
  {\bibinfo  {journal} {Phys. Rev. Lett.}\ }\textbf {\bibinfo {volume} {127}},\
  \bibinfo {pages} {151801} (\bibinfo {year} {2021})},\ \Eprint
  {https://arxiv.org/abs/2105.14007} {arXiv:2105.14007 [hep-ex]} \BibitemShut
  {NoStop}%
\bibitem [{\citenamefont {Be\v{c}irevi\'c}\ and\ \citenamefont
  {Sumensari}(2017)}]{Becirevic:2017jtw}%
  \BibitemOpen
  \bibfield  {author} {\bibinfo {author} {\bibfnamefont {D.}~\bibnamefont
  {Be\v{c}irevi\'c}}\ and\ \bibinfo {author} {\bibfnamefont {O.}~\bibnamefont
  {Sumensari}},\ }\bibfield  {title} {\bibinfo {title} {{A leptoquark model to
  accommodate $R_K^\mathrm{exp} < R_K^\mathrm{SM}$ and $R_{K^\ast}^\mathrm{exp}
  < R_{K^\ast}^\mathrm{SM}$}},\ }\href
  {https://doi.org/10.1007/JHEP08(2017)104} {\bibfield  {journal} {\bibinfo
  {journal} {JHEP}\ }\textbf {\bibinfo {volume} {08}},\ \bibinfo {pages}
  {104}},\ \Eprint {https://arxiv.org/abs/1704.05835} {arXiv:1704.05835
  [hep-ph]} \BibitemShut {NoStop}%
\bibitem [{\citenamefont {Kamenik}\ \emph {et~al.}(2018)\citenamefont
  {Kamenik}, \citenamefont {Soreq},\ and\ \citenamefont
  {Zupan}}]{Kamenik:2017tnu}%
  \BibitemOpen
  \bibfield  {author} {\bibinfo {author} {\bibfnamefont {J.~F.}\ \bibnamefont
  {Kamenik}}, \bibinfo {author} {\bibfnamefont {Y.}~\bibnamefont {Soreq}},\
  and\ \bibinfo {author} {\bibfnamefont {J.}~\bibnamefont {Zupan}},\ }\bibfield
   {title} {\bibinfo {title} {{Lepton flavor universality violation without new
  sources of quark flavor violation}},\ }\href
  {https://doi.org/10.1103/PhysRevD.97.035002} {\bibfield  {journal} {\bibinfo
  {journal} {Phys. Rev. D}\ }\textbf {\bibinfo {volume} {97}},\ \bibinfo
  {pages} {035002} (\bibinfo {year} {2018})},\ \Eprint
  {https://arxiv.org/abs/1704.06005} {arXiv:1704.06005 [hep-ph]} \BibitemShut
  {NoStop}%
\bibitem [{\citenamefont {Fox}\ \emph {et~al.}(2018)\citenamefont {Fox},
  \citenamefont {Low},\ and\ \citenamefont {Zhang}}]{Fox:2018ldq}%
  \BibitemOpen
  \bibfield  {author} {\bibinfo {author} {\bibfnamefont {P.~J.}\ \bibnamefont
  {Fox}}, \bibinfo {author} {\bibfnamefont {I.}~\bibnamefont {Low}},\ and\
  \bibinfo {author} {\bibfnamefont {Y.}~\bibnamefont {Zhang}},\ }\bibfield
  {title} {\bibinfo {title} {{Top-philic $Z'$ forces at the LHC}},\ }\href
  {https://doi.org/10.1007/JHEP03(2018)074} {\bibfield  {journal} {\bibinfo
  {journal} {JHEP}\ }\textbf {\bibinfo {volume} {03}},\ \bibinfo {pages}
  {074}},\ \Eprint {https://arxiv.org/abs/1801.03505} {arXiv:1801.03505
  [hep-ph]} \BibitemShut {NoStop}%
\bibitem [{\citenamefont {Coy}\ \emph {et~al.}(2020)\citenamefont {Coy},
  \citenamefont {Frigerio}, \citenamefont {Mescia},\ and\ \citenamefont
  {Sumensari}}]{Coy:2019rfr}%
  \BibitemOpen
  \bibfield  {author} {\bibinfo {author} {\bibfnamefont {R.}~\bibnamefont
  {Coy}}, \bibinfo {author} {\bibfnamefont {M.}~\bibnamefont {Frigerio}},
  \bibinfo {author} {\bibfnamefont {F.}~\bibnamefont {Mescia}},\ and\ \bibinfo
  {author} {\bibfnamefont {O.}~\bibnamefont {Sumensari}},\ }\bibfield  {title}
  {\bibinfo {title} {{New physics in $b\to s\ell\ell$ transitions at one
  loop}},\ }\href {https://doi.org/10.1140/epjc/s10052-019-7581-y} {\bibfield
  {journal} {\bibinfo  {journal} {Eur. Phys. J. C}\ }\textbf {\bibinfo {volume}
  {80}},\ \bibinfo {pages} {52} (\bibinfo {year} {2020})},\ \Eprint
  {https://arxiv.org/abs/1909.08567} {arXiv:1909.08567 [hep-ph]} \BibitemShut
  {NoStop}%
\bibitem [{\citenamefont {Li}\ \emph {et~al.}(2022{\natexlab{b}})\citenamefont
  {Li}, \citenamefont {Shen}, \citenamefont {Wang}, \citenamefont {Yang},\ and\
  \citenamefont {Yuan}}]{Li:2021cty}%
  \BibitemOpen
  \bibfield  {author} {\bibinfo {author} {\bibfnamefont {X.-Q.}\ \bibnamefont
  {Li}}, \bibinfo {author} {\bibfnamefont {M.}~\bibnamefont {Shen}}, \bibinfo
  {author} {\bibfnamefont {D.-Y.}\ \bibnamefont {Wang}}, \bibinfo {author}
  {\bibfnamefont {Y.-D.}\ \bibnamefont {Yang}},\ and\ \bibinfo {author}
  {\bibfnamefont {X.-B.}\ \bibnamefont {Yuan}},\ }\bibfield  {title} {\bibinfo
  {title} {{Explaining the $b \to s \ell^+ \ell^-$ anomalies in $Z^\prime$
  scenarios with top-FCNC couplings}},\ }\href
  {https://doi.org/10.1016/j.nuclphysb.2022.115828} {\bibfield  {journal}
  {\bibinfo  {journal} {Nucl. Phys. B}\ }\textbf {\bibinfo {volume} {980}},\
  \bibinfo {pages} {115828} (\bibinfo {year} {2022}{\natexlab{b}})},\ \Eprint
  {https://arxiv.org/abs/2112.14215} {arXiv:2112.14215 [hep-ph]} \BibitemShut
  {NoStop}%
\bibitem [{LHC(2022{\natexlab{a}})}]{LHCb:2022qnv}%
  \BibitemOpen
  \bibfield  {title} {\bibinfo {title} {{Test of lepton universality in $b
  \rightarrow s \ell^+ \ell^-$ decays}},\ }\href@noop {} {\  (\bibinfo {year}
  {2022}{\natexlab{a}})},\ \Eprint {https://arxiv.org/abs/2212.09152}
  {arXiv:2212.09152 [hep-ex]} \BibitemShut {NoStop}%
\bibitem [{LHC(2022{\natexlab{b}})}]{LHCb:2022zom}%
  \BibitemOpen
  \bibfield  {title} {\bibinfo {title} {{Measurement of lepton universality
  parameters in $B^+\to K^+\ell^+\ell^-$ and $B^0\to K^{*0}\ell^+\ell^-$
  decays}},\ }\href@noop {} {\  (\bibinfo {year} {2022}{\natexlab{b}})},\
  \Eprint {https://arxiv.org/abs/2212.09153} {arXiv:2212.09153 [hep-ex]}
  \BibitemShut {NoStop}%
\bibitem [{CMS(2022{\natexlab{b}})}]{CMS:2022dbz}%
  \BibitemOpen
  \bibfield  {title} {\bibinfo {title} {{Measurement of $B^0_s\to\mu^+\mu^-$
  decay properties and search for the $B^0\to\mu\mu$ decay in proton-proton
  collisions at $\sqrt{s}=13~\rm{TeV}$}},\ }\href@noop {} {\  (\bibinfo {year}
  {2022}{\natexlab{b}})}\BibitemShut {NoStop}%
\bibitem [{\citenamefont {Ciuchini}\ \emph {et~al.}(2022)\citenamefont
  {Ciuchini}, \citenamefont {Fedele}, \citenamefont {Franco}, \citenamefont
  {Paul}, \citenamefont {Silvestrini},\ and\ \citenamefont
  {Valli}}]{Ciuchini:2022wbq}%
  \BibitemOpen
  \bibfield  {author} {\bibinfo {author} {\bibfnamefont {M.}~\bibnamefont
  {Ciuchini}}, \bibinfo {author} {\bibfnamefont {M.}~\bibnamefont {Fedele}},
  \bibinfo {author} {\bibfnamefont {E.}~\bibnamefont {Franco}}, \bibinfo
  {author} {\bibfnamefont {A.}~\bibnamefont {Paul}}, \bibinfo {author}
  {\bibfnamefont {L.}~\bibnamefont {Silvestrini}},\ and\ \bibinfo {author}
  {\bibfnamefont {M.}~\bibnamefont {Valli}},\ }\bibfield  {title} {\bibinfo
  {title} {{Constraints on Lepton Universality Violation from Rare $B$
  Decays}},\ }\href@noop {} {\  (\bibinfo {year} {2022})},\ \Eprint
  {https://arxiv.org/abs/2212.10516} {arXiv:2212.10516 [hep-ph]} \BibitemShut
  {NoStop}%
\bibitem [{\citenamefont {Amhis}\ \emph {et~al.}(2022)\citenamefont {Amhis}
  \emph {et~al.}}]{HFLAV:2022pwe}%
  \BibitemOpen
  \bibfield  {author} {\bibinfo {author} {\bibfnamefont {Y.}~\bibnamefont
  {Amhis}} \emph {et~al.} (\bibinfo {collaboration} {HFLAV}),\ }\bibfield
  {title} {\bibinfo {title} {{Averages of $b$-hadron, $c$-hadron, and
  $\tau$-lepton properties as of 2021}},\ }\href@noop {} {\  (\bibinfo {year}
  {2022})},\ \Eprint {https://arxiv.org/abs/2206.07501} {arXiv:2206.07501
  [hep-ex]} \BibitemShut {NoStop}%
\bibitem [{\citenamefont {Wehle}\ \emph {et~al.}(2017)\citenamefont {Wehle}
  \emph {et~al.}}]{Belle:2016fev}%
  \BibitemOpen
  \bibfield  {author} {\bibinfo {author} {\bibfnamefont {S.}~\bibnamefont
  {Wehle}} \emph {et~al.} (\bibinfo {collaboration} {Belle}),\ }\bibfield
  {title} {\bibinfo {title} {{Lepton-Flavor-Dependent Angular Analysis of $B\to
  K^\ast \ell^+\ell^-$}},\ }\href
  {https://doi.org/10.1103/PhysRevLett.118.111801} {\bibfield  {journal}
  {\bibinfo  {journal} {Phys. Rev. Lett.}\ }\textbf {\bibinfo {volume} {118}},\
  \bibinfo {pages} {111801} (\bibinfo {year} {2017})},\ \Eprint
  {https://arxiv.org/abs/1612.05014} {arXiv:1612.05014 [hep-ex]} \BibitemShut
  {NoStop}%
\bibitem [{\citenamefont {Choudhury}\ \emph {et~al.}(2021)\citenamefont
  {Choudhury} \emph {et~al.}}]{BELLE:2019xld}%
  \BibitemOpen
  \bibfield  {author} {\bibinfo {author} {\bibfnamefont {S.}~\bibnamefont
  {Choudhury}} \emph {et~al.} (\bibinfo {collaboration} {BELLE}),\ }\bibfield
  {title} {\bibinfo {title} {{Test of lepton flavor universality and search for
  lepton flavor violation in $B \rightarrow K\ell \ell$ decays}},\ }\href
  {https://doi.org/10.1007/JHEP03(2021)105} {\bibfield  {journal} {\bibinfo
  {journal} {JHEP}\ }\textbf {\bibinfo {volume} {03}},\ \bibinfo {pages}
  {105}},\ \Eprint {https://arxiv.org/abs/1908.01848} {arXiv:1908.01848
  [hep-ex]} \BibitemShut {NoStop}%
\bibitem [{\citenamefont {Abdesselam}\ \emph {et~al.}(2021)\citenamefont
  {Abdesselam} \emph {et~al.}}]{Belle:2019oag}%
  \BibitemOpen
  \bibfield  {author} {\bibinfo {author} {\bibfnamefont {A.}~\bibnamefont
  {Abdesselam}} \emph {et~al.} (\bibinfo {collaboration} {Belle}),\ }\bibfield
  {title} {\bibinfo {title} {{Test of Lepton-Flavor Universality in ${B\to
  K^\ast\ell^+\ell^-}$ Decays at Belle}},\ }\href
  {https://doi.org/10.1103/PhysRevLett.126.161801} {\bibfield  {journal}
  {\bibinfo  {journal} {Phys. Rev. Lett.}\ }\textbf {\bibinfo {volume} {126}},\
  \bibinfo {pages} {161801} (\bibinfo {year} {2021})},\ \Eprint
  {https://arxiv.org/abs/1904.02440} {arXiv:1904.02440 [hep-ex]} \BibitemShut
  {NoStop}%
\bibitem [{\citenamefont {Straub}(2018)}]{Straub:2018kue}%
  \BibitemOpen
  \bibfield  {author} {\bibinfo {author} {\bibfnamefont {D.~M.}\ \bibnamefont
  {Straub}},\ }\bibfield  {title} {\bibinfo {title} {{flavio: a Python package
  for flavour and precision phenomenology in the Standard Model and beyond}},\
  }\href@noop {} {\  (\bibinfo {year} {2018})},\ \Eprint
  {https://arxiv.org/abs/1810.08132} {arXiv:1810.08132 [hep-ph]} \BibitemShut
  {NoStop}%
\bibitem [{\citenamefont {Workman}\ \emph {et~al.}(2022)\citenamefont {Workman}
  \emph {et~al.}}]{ParticleDataGroup:2022pth}%
  \BibitemOpen
  \bibfield  {author} {\bibinfo {author} {\bibfnamefont {R.~L.}\ \bibnamefont
  {Workman}} \emph {et~al.} (\bibinfo {collaboration} {Particle Data Group}),\
  }\bibfield  {title} {\bibinfo {title} {{Review of Particle Physics}},\ }\href
  {https://doi.org/10.1093/ptep/ptac097} {\bibfield  {journal} {\bibinfo
  {journal} {PTEP}\ }\textbf {\bibinfo {volume} {2022}},\ \bibinfo {pages}
  {083C01} (\bibinfo {year} {2022})}\BibitemShut {NoStop}%
\bibitem [{\citenamefont {Davidson}\ and\ \citenamefont
  {Saporta}(2019)}]{Davidson:2018rqt}%
  \BibitemOpen
  \bibfield  {author} {\bibinfo {author} {\bibfnamefont {S.}~\bibnamefont
  {Davidson}}\ and\ \bibinfo {author} {\bibfnamefont {A.}~\bibnamefont
  {Saporta}},\ }\bibfield  {title} {\bibinfo {title} {{Constraints on $2\ell
  2q$ operators from $\mu - e$ flavour-changing meson decays}},\ }\href
  {https://doi.org/10.1103/PhysRevD.99.015032} {\bibfield  {journal} {\bibinfo
  {journal} {Phys. Rev. D}\ }\textbf {\bibinfo {volume} {99}},\ \bibinfo
  {pages} {015032} (\bibinfo {year} {2019})},\ \Eprint
  {https://arxiv.org/abs/1807.10288} {arXiv:1807.10288 [hep-ph]} \BibitemShut
  {NoStop}%
\bibitem [{\citenamefont {Carpentier}\ and\ \citenamefont
  {Davidson}(2010)}]{Carpentier:2010ue}%
  \BibitemOpen
  \bibfield  {author} {\bibinfo {author} {\bibfnamefont {M.}~\bibnamefont
  {Carpentier}}\ and\ \bibinfo {author} {\bibfnamefont {S.}~\bibnamefont
  {Davidson}},\ }\bibfield  {title} {\bibinfo {title} {{Constraints on
  two-lepton, two quark operators}},\ }\href
  {https://doi.org/10.1140/epjc/s10052-010-1482-4} {\bibfield  {journal}
  {\bibinfo  {journal} {Eur. Phys. J. C}\ }\textbf {\bibinfo {volume} {70}},\
  \bibinfo {pages} {1071} (\bibinfo {year} {2010})},\ \Eprint
  {https://arxiv.org/abs/1008.0280} {arXiv:1008.0280 [hep-ph]} \BibitemShut
  {NoStop}%
\bibitem [{\citenamefont {Dawson}\ and\ \citenamefont
  {Giardino}(2022)}]{Dawson:2022bxd}%
  \BibitemOpen
  \bibfield  {author} {\bibinfo {author} {\bibfnamefont {S.}~\bibnamefont
  {Dawson}}\ and\ \bibinfo {author} {\bibfnamefont {P.~P.}\ \bibnamefont
  {Giardino}},\ }\bibfield  {title} {\bibinfo {title} {{Flavorful electroweak
  precision observables in the Standard Model effective field theory}},\ }\href
  {https://doi.org/10.1103/PhysRevD.105.073006} {\bibfield  {journal} {\bibinfo
   {journal} {Phys. Rev. D}\ }\textbf {\bibinfo {volume} {105}},\ \bibinfo
  {pages} {073006} (\bibinfo {year} {2022})},\ \Eprint
  {https://arxiv.org/abs/2201.09887} {arXiv:2201.09887 [hep-ph]} \BibitemShut
  {NoStop}%
\bibitem [{\citenamefont {Schael}\ \emph {et~al.}(2006)\citenamefont {Schael}
  \emph {et~al.}}]{ALEPH:2005ab}%
  \BibitemOpen
  \bibfield  {author} {\bibinfo {author} {\bibfnamefont {S.}~\bibnamefont
  {Schael}} \emph {et~al.} (\bibinfo {collaboration} {ALEPH, DELPHI, L3, OPAL,
  SLD, LEP Electroweak Working Group, SLD Electroweak Group, SLD Heavy Flavour
  Group}),\ }\bibfield  {title} {\bibinfo {title} {{Precision electroweak
  measurements on the $Z$ resonance}},\ }\href
  {https://doi.org/10.1016/j.physrep.2005.12.006} {\bibfield  {journal}
  {\bibinfo  {journal} {Phys. Rept.}\ }\textbf {\bibinfo {volume} {427}},\
  \bibinfo {pages} {257–454} (\bibinfo {year} {2006})},\ \Eprint
  {https://arxiv.org/abs/hep-ex/0509008} {arXiv:hep-ex/0509008 [hep-ex]}
  \BibitemShut {NoStop}%
\bibitem [{\citenamefont {Benesch}\ \emph {et~al.}(2014)\citenamefont {Benesch}
  \emph {et~al.}}]{MOLLER:2014iki}%
  \BibitemOpen
  \bibfield  {author} {\bibinfo {author} {\bibfnamefont {J.}~\bibnamefont
  {Benesch}} \emph {et~al.} (\bibinfo {collaboration} {MOLLER}),\ }\bibfield
  {title} {\bibinfo {title} {{The MOLLER Experiment: An Ultra-Precise
  Measurement of the Weak Mixing Angle Using M\o ller Scattering}},\
  }\href@noop {} {\  (\bibinfo {year} {2014})},\ \Eprint
  {https://arxiv.org/abs/1411.4088} {arXiv:1411.4088 [nucl-ex]} \BibitemShut
  {NoStop}%
\bibitem [{\citenamefont {Du}\ \emph {et~al.}(2021)\citenamefont {Du},
  \citenamefont {Freitas}, \citenamefont {Patel},\ and\ \citenamefont
  {Ramsey-Musolf}}]{Du:2019evk}%
  \BibitemOpen
  \bibfield  {author} {\bibinfo {author} {\bibfnamefont {Y.}~\bibnamefont
  {Du}}, \bibinfo {author} {\bibfnamefont {A.}~\bibnamefont {Freitas}},
  \bibinfo {author} {\bibfnamefont {H.~H.}\ \bibnamefont {Patel}},\ and\
  \bibinfo {author} {\bibfnamefont {M.~J.}\ \bibnamefont {Ramsey-Musolf}},\
  }\bibfield  {title} {\bibinfo {title} {{Parity-Violating M\o{}ller Scattering
  at Next-to-Next-to-Leading Order: Closed Fermion Loops}},\ }\href
  {https://doi.org/10.1103/PhysRevLett.126.131801} {\bibfield  {journal}
  {\bibinfo  {journal} {Phys. Rev. Lett.}\ }\textbf {\bibinfo {volume} {126}},\
  \bibinfo {pages} {131801} (\bibinfo {year} {2021})},\ \Eprint
  {https://arxiv.org/abs/1912.08220} {arXiv:1912.08220 [hep-ph]} \BibitemShut
  {NoStop}%
\bibitem [{\citenamefont {Jenkins}\ \emph {et~al.}(2014)\citenamefont
  {Jenkins}, \citenamefont {Manohar},\ and\ \citenamefont
  {Trott}}]{Jenkins:2013wua}%
  \BibitemOpen
  \bibfield  {author} {\bibinfo {author} {\bibfnamefont {E.~E.}\ \bibnamefont
  {Jenkins}}, \bibinfo {author} {\bibfnamefont {A.~V.}\ \bibnamefont
  {Manohar}},\ and\ \bibinfo {author} {\bibfnamefont {M.}~\bibnamefont
  {Trott}},\ }\bibfield  {title} {\bibinfo {title} {{Renormalization Group
  Evolution of the Standard Model Dimension Six Operators II: Yukawa
  Dependence}},\ }\href {https://doi.org/10.1007/JHEP01(2014)035} {\bibfield
  {journal} {\bibinfo  {journal} {JHEP}\ }\textbf {\bibinfo {volume} {01}},\
  \bibinfo {pages} {035}},\ \Eprint {https://arxiv.org/abs/1310.4838}
  {arXiv:1310.4838 [hep-ph]} \BibitemShut {NoStop}%
\bibitem [{\citenamefont {Alonso}\ \emph {et~al.}(2014)\citenamefont {Alonso},
  \citenamefont {Jenkins}, \citenamefont {Manohar},\ and\ \citenamefont
  {Trott}}]{Alonso:2013hga}%
  \BibitemOpen
  \bibfield  {author} {\bibinfo {author} {\bibfnamefont {R.}~\bibnamefont
  {Alonso}}, \bibinfo {author} {\bibfnamefont {E.~E.}\ \bibnamefont {Jenkins}},
  \bibinfo {author} {\bibfnamefont {A.~V.}\ \bibnamefont {Manohar}},\ and\
  \bibinfo {author} {\bibfnamefont {M.}~\bibnamefont {Trott}},\ }\bibfield
  {title} {\bibinfo {title} {{Renormalization Group Evolution of the Standard
  Model Dimension Six Operators III: Gauge Coupling Dependence and
  Phenomenology}},\ }\href {https://doi.org/10.1007/JHEP04(2014)159} {\bibfield
   {journal} {\bibinfo  {journal} {JHEP}\ }\textbf {\bibinfo {volume} {04}},\
  \bibinfo {pages} {159}},\ \Eprint {https://arxiv.org/abs/1312.2014}
  {arXiv:1312.2014 [hep-ph]} \BibitemShut {NoStop}%
\bibitem [{\citenamefont {Jenkins}\ \emph
  {et~al.}(2018{\natexlab{a}})\citenamefont {Jenkins}, \citenamefont
  {Manohar},\ and\ \citenamefont {Stoffer}}]{Jenkins:2017jig}%
  \BibitemOpen
  \bibfield  {author} {\bibinfo {author} {\bibfnamefont {E.~E.}\ \bibnamefont
  {Jenkins}}, \bibinfo {author} {\bibfnamefont {A.~V.}\ \bibnamefont
  {Manohar}},\ and\ \bibinfo {author} {\bibfnamefont {P.}~\bibnamefont
  {Stoffer}},\ }\bibfield  {title} {\bibinfo {title} {{Low-Energy Effective
  Field Theory below the Electroweak Scale: Operators and Matching}},\ }\href
  {https://doi.org/10.1007/JHEP03(2018)016} {\bibfield  {journal} {\bibinfo
  {journal} {JHEP}\ }\textbf {\bibinfo {volume} {03}},\ \bibinfo {pages}
  {016}},\ \Eprint {https://arxiv.org/abs/1709.04486} {arXiv:1709.04486
  [hep-ph]} \BibitemShut {NoStop}%
\bibitem [{\citenamefont {Jenkins}\ \emph
  {et~al.}(2018{\natexlab{b}})\citenamefont {Jenkins}, \citenamefont
  {Manohar},\ and\ \citenamefont {Stoffer}}]{Jenkins:2017dyc}%
  \BibitemOpen
  \bibfield  {author} {\bibinfo {author} {\bibfnamefont {E.~E.}\ \bibnamefont
  {Jenkins}}, \bibinfo {author} {\bibfnamefont {A.~V.}\ \bibnamefont
  {Manohar}},\ and\ \bibinfo {author} {\bibfnamefont {P.}~\bibnamefont
  {Stoffer}},\ }\bibfield  {title} {\bibinfo {title} {{Low-Energy Effective
  Field Theory below the Electroweak Scale: Anomalous Dimensions}},\ }\href
  {https://doi.org/10.1007/JHEP01(2018)084} {\bibfield  {journal} {\bibinfo
  {journal} {JHEP}\ }\textbf {\bibinfo {volume} {01}},\ \bibinfo {pages}
  {084}},\ \Eprint {https://arxiv.org/abs/1711.05270} {arXiv:1711.05270
  [hep-ph]} \BibitemShut {NoStop}%
\bibitem [{\citenamefont {Czarnecki}\ and\ \citenamefont
  {Marciano}(1996)}]{Czarnecki:1995fw}%
  \BibitemOpen
  \bibfield  {author} {\bibinfo {author} {\bibfnamefont {A.}~\bibnamefont
  {Czarnecki}}\ and\ \bibinfo {author} {\bibfnamefont {W.~J.}\ \bibnamefont
  {Marciano}},\ }\bibfield  {title} {\bibinfo {title} {{Electroweak radiative
  corrections to polarized Moller scattering asymmetries}},\ }\href
  {https://doi.org/10.1103/PhysRevD.53.1066} {\bibfield  {journal} {\bibinfo
  {journal} {Phys. Rev. D}\ }\textbf {\bibinfo {volume} {53}},\ \bibinfo
  {pages} {1066} (\bibinfo {year} {1996})},\ \Eprint
  {https://arxiv.org/abs/hep-ph/9507420} {arXiv:hep-ph/9507420} \BibitemShut
  {NoStop}%
\end{thebibliography}%

\end{document}